\documentclass[12pt]{article}
\pdfoutput=1
\usepackage{jheppub}
\usepackage[utf8]{inputenc}
\usepackage{amsmath}
\usepackage{amssymb}
\usepackage{amsthm}
\usepackage{amsfonts}
\usepackage[dvipsnames]{xcolor}
\usepackage{enumerate}
\usepackage{ifthen}
\usepackage[mathscr]{eucal}
\usepackage[errorshow]{tracefnt}
\usepackage{slashed}
\usepackage{array}
\usepackage{epsf}
\usepackage{epsfig}
\usepackage{psfrag}
\usepackage{multirow}
\usepackage{wasysym} 
\usepackage{tikz-cd} 
\usepgfmodule{shapes}
\usetikzlibrary{backgrounds, arrows,calc,shapes,decorations.pathreplacing, decorations.markings, automata,positioning}
\usepackage{relsize}
\usepackage{ytableau} 
\usepackage{dirtytalk} 
\usepackage{cancel} 

\tikzset{
  arrow/.pic={\path[tips,every arrow/.try,->,>=#1] (0,0) -- +(.1pt,0);},
  pics/arrow/.default={latex,very thick}
}

\newcommand\bovermat[2]{%
\makebox[0pt][l]{$\smash{\overbrace{\phantom{%
\begin{matrix}#2\end{matrix}}}^{\text{#1}}}$}#2}

\newcommand{\be}{\begin{equation}}
\newcommand{\ee}{\end{equation}}
\newcommand{\bea}{\begin{eqnarray}}
\newcommand{\eea}{\end{eqnarray}}
\newcommand{\ba}{\begin{array}}
\newcommand{\ea}{\end{array}}
\newcommand{\bpic}{\begin{tikzpicture}}
\newcommand{\epic}{\end{tikzpicture}}

\newcommand{\nn}{\nonumber}

\newcommand{\tr}{\text{tr}\,}
\newcommand{\Pf}{\text{Pfaff}\,}

\newcommand{\Llra}{\Longleftrightarrow}

\newcommand{\lra}{\leftrightarrow}
\newcommand{\llra}{\longleftrightarrow}

\newcommand\bt{\tilde b}
\newcommand\ct{\tilde c}
\newcommand\dt{\tilde d}
\newcommand\lt{\tilde l}
\newcommand\nt{\tilde n}

\newcommand\At{\tilde A}
\newcommand\Bt{\tilde B}
\newcommand\Ht{\tilde H}
\newcommand\Qt{\tilde Q}

\newcommand\Xt{\tilde X}
\newcommand\Yt{\tilde Y}

\newcommand{\CT}{{\mathcal T}}

\newcommand{\CW}{{\mathcal W}}

\newcommand{\cN}{\mathcal{N}}

\newcommand{\cW}{\mathcal{W}}

\newcommand{\bZ}{\mathbb{Z}}

\def\a{\alpha} 
\def\b{\beta} 
 
\def\de{\delta} 
\def\e{\varepsilon}

\def\P{\Phi}

\def\Bt{\tilde{B}}

\title{S-confinements from deconfinements}

\begin{document}

\author[1]{Stephane Bajeot}
\author[2]{Sergio Benvenuti}

\affiliation[1]{SISSA, Via Bonomea 265, 34136 Trieste, Italy}
\affiliation[2]{INFN, Sezione di Trieste, Via Valerio 2, 34127 Trieste, Italy}
\emailAdd{sbajeot@sissa.it, benve79@gmail.com}

\abstract{We consider four dimensional $\mathcal{N}=1$ gauge theories that are S-confining, that is they are dual to a Wess-Zumino model. S-confining theories with a simple gauge group  have been classified. We prove all the S-confining dualities in the list, when the matter fields transform in rank-$1$ and/or rank-$2$ representations. Our only assumptions are the S-confining dualities for $SU(N)$ with $N+1$ flavors and for $Usp(2N)$ with $2N+4$ fundamentals. The strategy consists in a sequence of deconfinements and re-confinements. We pay special attention to the explicit superpotential at each step.}

\maketitle

\section{Introduction and summary}
The infrared (IR) behavior of four-dimensional gauge theories is a problem of central importance in theoretical physics. When minimal supersymmetry is present, substantial progress can be made \cite{Seiberg:1994bz,Seiberg:1994pq,Intriligator:1995au}.
One possible IR behaviour of a gauge theory is \emph{S-confinement}, where $S$ stands for \emph{smooth}. More precisely this terminology has been introduced in \cite{Csaki:1996sm} for \say{smooth confinement without chiral symmetry breaking and with a non-vanishing confining superpotential}. The IR behavior of an S-confining gauge theory, by definition, can be captured by a theory with trivial gauge dynamics, that is a Wess-Zumino (WZ) model. The elementary fields of the IR WZ description map with the gauge invariant operators of the ultraviolet (UV) gauge theory, more precisely they are in one-to-one correspondence with the generators of the chiral ring of the UV gauge theory. In addition, we require that this WZ description is valid everywhere on the moduli space including the origin where therefore all global symmetries are unbroken. The two paradigmatic, and simplest, examples of S-confining gauge theories are
\begin{itemize}
\item $SU(N)$ with $N+1$ flavors, described by a theory of mesons and baryons.
\item $Usp(2N)$ with $N+2$ flavors\footnote{For $Usp$ gauge group, we call a flavor two chiral fields in the fundamental representation.}, described by a theory of mesons.
\end{itemize}
There are also other examples of S-confining gauge theories. \cite{Csaki:1996zb} classified all the S-confining gauge theories with a simple gauge group and vanishing tree-level superpotential. These theories were argued to be S-confining by proposing a WZ description and checking that all the 't Hooft anomalies of the UV gauge theory match the 't Hooft anomalies of the WZ description.

Our aim in this paper is to \emph{prove} these S-confining dualities. By proof we mean that we find a sequence of dualities that takes us from the gauge theory to the WZ model. At each step, we write the explicit superpotential. The dualities we use are only the two basic S-confining dualities discussed above, that is $SU(N+1)/Usp(2N)$ with $N+2$ flavors. Our strategy, especially for the case of $Usp(2N)$ with antisymmetric and $6$ fundamentals, is similar to the strategy of \cite{Pasquetti:2019tix, Pasquetti:2019uop, Benvenuti:2021nwt}, implemented in $3$ dimensions.\footnote{Indeed, dimensional reducing $Usp(2N)$ with antisymmetric and $6$ fundamentals on a circle and turning on appropriate real mass deformations as in \cite{Benini:2017dud, Benvenuti:2018bav}, it is posssible to flow to the S-confining duality of $3d$ $\cN=2$ $U(N)$ with adjoint and $(1,1)$ fundamentals discussed in \cite{Pasquetti:2019tix, Pasquetti:2019uop}. See  \cite{Nii:2019ebv} for a study of $3d$ $\cN=2$ S-confining single node quivers.}

We are able to do this for all S-confining \emph{one node quivers}, that is gauge theories with a simple gauge group and matter in rank-$1$ and/or rank-$2$ representations. The set of theories includes $3$ infinite series: $Usp(2N)$ with antisymmetric and $3$ flavors, $SU(N)$ with antisymmetric and $(4, N)$ flavors, $SU(N)$ with anti-symmetric, conjugate antisymmetric and $(3,3)$ flavors. Moreover there are $4$ exceptional cases, with $SU(5), SU(6), SU(7)$ gauge group and $2$ or $3$ anti-symmetrics plus flavors.\footnote{\cite{Csaki:1996zb} also found non-quivers S-confining models, with $Spin(N)$ gauge group and spinorial matter, and an $SU(6)$ model with a rank-$3$ antisymmetric representation. Our techniques cannot tackle such cases. We leave this problem to future work.} In the two sporadic $SU(5)$ cases, we will be confronted to extra subtleties in the computation of the superpotential associated to degenerate operators, see also \cite{BAJEOTBENVENUTI2}. 

The main tool we are going to use is the deconfinement method \cite{Berkooz:1995km, Luty:1996cg} that allows us to trade theories with antisymmetric field for theories involving only fields in the fundamental representation of simple gauge factor. More precisely, we use the deconfinement method of \cite{Garcia-Etxebarria:2012ypj,Garcia-Etxebarria:2013tba,BOTTINITESI}. This deconfinement method in $4d \, \cN=1$ has been used recently \cite{Etxebarria:2021lmq} to construct Lagrangians for families of $\cN=2$ superconformal field theories (SCFT).

One immediate lesson that can be drawn from these results is that the basic Seiberg dualities, involving only fundamental matter, seem to be strong enough to prove dualities involving more general matter content. A recent results that corroborates this expectation appeared recently in \cite{Bottini:2021vms, Hwang:2021ulb}, where it is shown that $4d$ mirror symmetry dualities \cite{Pasquetti:2019hxf, Hwang:2020wpd} (and hence $3d$ mirror symmetry dualities \cite{Intriligator:1996ex, Hanany:1996ie})  can be proven using only Intriligator-Pouliot (IP) duality, relating theories with $Usp$ gauge group and fundamental matter. Moreover, in \cite{BAJEOTBENVENUTI2} we will use the sequential deconfinement techniques of \cite{Benvenuti:2020wpc, Benvenuti:2020gvy} to find a quiver dual of $Usp(2N)$ with $F$ flavors, and use this deconfined version to prove a knwon self-duality of $Usp(2N)$ with $4$ flavors \cite{Csaki:1996eu}.

\vspace{0.8cm}
This paper is organized as follows. 

In section \ref{sec:deconfmethod} we describe the way we deconfine the anti-symmetric fields which appear in the S-confining theories with one node and vanishing superpotential.

In sections \ref{Usp2N}, \ref{SUaa} and \ref{SUa} we consider the infinite series. In order to have a simple close formula for the superpotential at generic $N$ (for these infinite series, \cite{Csaki:1996zb} only gave the superpotential for very small values of $N$), we flip some operators in the gauge theory. The analysis of the superpotential at all intermediate steps of our procedure is quite involved, so we relegate many details in Appendix \ref{SuperpotInTk}.

In section \ref{sporadic} we discuss the four sporadic S-confining models. 

In appendix \ref{unitary} we give a unitary presentation of the S-confining dualities discussed in this paper, flipping the operators that violate the unitarity bounds.

\vspace{1cm}
\noindent Note added: while completing this work, we became aware of \cite{BOTTINIetAL}, which has some overlap with the results of Section 3. We thank the authors of \cite{BOTTINIetAL} for coordinating the submission.

\section{Elementary S-confining theories and deconfinement method}\label{sec:deconfmethod}

In this section we present the elementary S-confining theories that will be the building blocks to obtain the more complicated cases of the next sections.

The first one is the original Seiberg S-confining theory\footnote{Before \cite{Seiberg:1994bz}, it has been already remarked in \cite{Amati:1988ft} that the 't Hooft anomalies were matched by the mesons and baryons but the confining superpotential was not found.} \cite{Seiberg:1994bz}, $SU(N)$ SQCD with $F=N+1$ flavors. In quivers notation it reads 

\noindent \textbf{$SU$ building block}
\be \label{SUbuildingBlock} \bpic[node distance=2cm,gSUnode/.style={circle,red,draw,minimum size=8mm},gUSpnode/.style={circle,blue,draw,minimum size=8mm},fnode/.style={rectangle,draw,minimum size=8mm}]  
\node[gSUnode] (G1) at (0,0) {$N$};
\node[fnode] (F1) at (-1,-2) {$N+1$};
\node[fnode] (F2) at (1,-2) {$N+1$};
\draw (G1) -- pic[pos=0.4,sloped,very thick]{arrow=latex reversed} (F1.north);
\draw (G1) -- pic[pos=0.5,sloped,very thick]{arrow=latex reversed} (F2.north);
\node[right] at (-0.8,-3) {$ \CW= 0$};
\node at (-1,-0.7) {$\Qt$};
\node at (1,-0.7) {$Q$};
\node at (3,-0.7) {$\llra$};
\node[fnode] (F3) at (4.5,-2) {$N+1$};
\node[fnode] (F4) at (7.5,-2) {$N+1$};
\node[fnode] (F5) at (6,0) {$1$};
\draw (F5) -- pic[pos=0.4,sloped]{arrow} (F3);
\draw (F5) -- pic[pos=0.6,sloped]{arrow} (F4);
\draw (F3) -- pic[pos=0.5,sloped,very thick]{arrow=latex reversed} (F4);
\node[right] at (4,-3) {$ \CW= B M \Bt + \det M$};
\node at (5,-0.7) {$\Bt$};
\node at (6,-2.4) {$M$};
\node at (7.1,-0.9) {$B$};
\epic \ee
The red circle means a $SU$ gauge node and the square node is for the $SU$ flavor symmetry. An oriented arrow between two nodes means a bi-fundamental field (transforming in the fundamental/antifundamental representation for the ingoing/outgoing arrow). We have also set the dynamical scale $\Lambda$ that should appear in the superpotential in the r.h.s to $1$ since it won't play any role in our work. 

The second building block is the IP S-confining theory \cite{Intriligator:1995ne}, $Usp(2N)$ with $2N+4$ fundamental fields.

\noindent \textbf{$Usp$ building block}
\be \label{UspbuildingBlock} \bpic[node distance=2cm,gSUnode/.style={circle,red,draw,minimum size=8mm},gUSpnode/.style={circle,blue,draw,minimum size=8mm},fnode/.style={rectangle,draw,minimum size=8mm}]  
\node[gUSpnode] (G1) at (-3,0) {$2N$};
\node[fnode] (F1) at (-0.5,0) {$2N+4$};
\draw (G1) -- pic[pos=0.7,sloped]{arrow} (F1);
\node[right] at (-2.7,-1.2) {$ \CW= 0$};
\node at (1.5,0) {$\longleftrightarrow$};
\node[fnode] (F2) at (3.5,0) {$2N+4$};
\draw (4,0.4) to[out=90,in=0] pic[pos=0.1,sloped]{arrow} (3.5,1) to[out=180,in=90] pic[pos=0.7,sloped,very thick]{arrow=latex reversed} (3,0.4);
\node[right] at (2.3,-1.2) {$ \CW= \Pf( \mu)$};
\node at (4.2,1) {$\mu$};
\epic \ee
The blue circle means a $Usp(2N)$ gauge group and the half-circle with arrows represents an antisymmetric field.

The examples that we are going to discuss in the next sections are gauge theories with rank-2 matter, more precisely antisymmetric fields. Therefore if we want to use only the building block confining theories, we need a way to get a situation in which only fundamental fields are present. This is possible and it is called deconfinement. The price we have to pay is an additional gauge group. We trade the rank-2 field by a confining gauge group. This method was introduced by Berkooz \cite{Berkooz:1995km} and further developed in \cite{Luty:1996cg}. \\
The original deconfinement method of \cite{Berkooz:1995km} was for $SU(2N)$ with an antisymmetric field and it reads
\be \label{BerkoozDeconfinement} \scalebox{0.9}{\bpic[node distance=2cm,gSUnode/.style={circle,red,draw,minimum size=8mm},gUSpnode/.style={circle,blue,draw,minimum size=8mm},fnode/.style={rectangle,draw,minimum size=8mm}]  
\node[gSUnode] (G1) at (0,0) {$2N$};
\node[fnode] (F1) at (-1,-2) {$2N+F-4$};
\node[fnode] (F2) at (1,-2) {$F$};
\draw (G1) -- pic[pos=0.4,sloped,very thick]{arrow=latex reversed} (F1.north);
\draw (G1) -- pic[pos=0.5,sloped,very thick]{arrow=latex reversed} (F2.north);
\draw (0.3,0.4) to[out=90,in=0] pic[pos=0.1,sloped]{arrow} (0,0.8) to[out=180,in=90] pic[pos=0.5,sloped,very thick]{arrow=latex reversed} (-0.3,0.4);
\node[right] at (-1.4,-3) {$ \CW= \Pf(A)$};
\node at (0.6,0.9) {$A$};
\node at (-1,-0.7) {$\Qt$};
\node at (1,-0.7) {$Q$};
\node at (3,0) {$\llra$};
\node[gUSpnode] (G2) at (6,0) {$2N-4$};
\node[gSUnode] (G3) at (9,0) {$2N$};
\node[fnode] (F3) at (8,-2) {$2N+F-4$};
\node[fnode] (F4) at (10,-2) {$F$};
\draw (G2) -- pic[pos=0.6,sloped]{arrow} (G3);
\draw (G3) -- pic[pos=0.4,sloped,very thick]{arrow=latex reversed} (F3);
\draw (G3) -- pic[pos=0.5,sloped,very thick]{arrow=latex reversed} (F4);
\node[right] at (8.2,-3) {$ \CW= 0$};
\node at (8.1,-0.9) {$\Qt$};
\node at (9.9,-0.9) {$Q$};
\node at (7.8,0.3) {$b$};
\epic} \ee 
The justification is straightforward, we start on the r.h.s, we notice that the $Usp(2N-4)$ is coupled to 2N fields, so we apply the $Usp$ building block \eqref{UspbuildingBlock} and we get the l.h.s.

We would prefer a situation where the superpotential is zero on the side with the antisymmetric field. A way has been found by Luty, Schmaltz and Terning \cite{Luty:1996cg}. It applies for any group G and it says
\be \label{LSTDeconfinement} \hspace{2cm} \scalebox{0.9}{\bpic[node distance=2cm,gSUnode/.style={circle,red,draw,minimum size=8mm},gUSpnode/.style={circle,blue,draw,minimum size=8mm},fnode/.style={rectangle,draw,minimum size=8mm}]  
\node[circle,draw,minimum size=8mm] (G1) at (1,0) {$G$};
\draw (1.3,0.3) to[out=90,in=0] pic[pos=0.1,sloped]{arrow} (1,0.8) to[out=180,in=90] pic[pos=0.5,sloped,very thick]{arrow=latex reversed} (0.7,0.3);
\node[right] at (0.2,-1) {$ \CW= 0$};
\node at (1.6,0.9) {$X$};
\node at (3,0) {$\llra$};
\node[gUSpnode] (G2) at (6,0) {\scalebox{0.9}{$N+K-4$}};
\node[circle,draw,minimum size=8mm] (G3) at (9,0) {$G$};
\node[fnode] (F1) at (6,-3) {$K$};
\draw (G2) -- pic[pos=0.6,sloped]{arrow} (G3);
\draw (G2) -- pic[pos=0.5,sloped,very thick]{arrow=latex reversed} (F1);
\draw (G3) -- pic[pos=0.4,sloped,very thick]{arrow=latex reversed} (F1);
\draw (6.3,-3.4) to[out=-90,in=0] pic[pos=0.1,sloped]{arrow} (6,-3.9) to[out=-180,in=-90] pic[pos=0.6,sloped,very thick]{arrow=latex reversed} (5.7,-3.4);  
\node[right] at (8,-2.6) {$ \CW=$ Planar Triangle $\,+ \b l l$};
\node at (5.7,-1.8) {$l$};
\node at (8,-1.5) {$\ct$};
\node at (7.8,0.3) {$b$};
\node at (6.5,-3.9) {$\b$};
\epic} \ee 
On the r.h.s, N is the dimension of the fundamental representation of G (the antisymmetric field has indices $X_{ij}$ with  $i,j =1,\dots,N$), $K$ is the smallest integer such that $N+K-4$ is even and $\b$ is an antisymmetric field of the flavor group. Few remarks
\begin{itemize}
\item Depending on the group $G$, there is additional matter, that contributes to cancel the gauge anomaly. However $G$ is a spectator in this deconfinement.  
\item Planar Triangle corresponds to the term $l b \ct$ with the obvious contraction of indices.
\item \say{$X_{ij} \llra b_i \, b_j \equiv b^c_i \, b^d_j \, J^{Usp}_{cd} \,$}. We put the mapping into quotation mark because it does not correspond to gauge invariant operators.
\item If $K>1$ we have a fictitious $SU(K)$ global symmetry. It is fictitious because the only fields that transform under this symmetry are not present in the low-energy theory.
\end{itemize}
The justification of \eqref{LSTDeconfinement} goes as follows. We start once again from the r.h.s and notice that $Usp(N+K-4)$ is connected to $N+K$ so we can use our $Usp$ building block \eqref{UspbuildingBlock}. The result is
\be \bpic[node distance=2cm,gSUnode/.style={circle,red,draw,minimum size=8mm},gUSpnode/.style={circle,blue,draw,minimum size=8mm},fnode/.style={rectangle,draw,minimum size=8mm}]  
\node[circle,draw,minimum size=8mm] (G1) at (0,0) {$G$};
\node[fnode] (F1) at (0,-1.8) {$K$};
\draw (0.3,0.3) to[out=90,in=0] pic[pos=0.1,sloped]{arrow} (0,0.8) to[out=180,in=90] pic[pos=0.5,sloped,very thick]{arrow=latex reversed} (-0.3,0.3);
\draw (-0.4,-1.8) to[out=180,in=90] pic[pos=0.1,sloped]{arrow} (-0.8,-2.2) to[out=270,in=180] (-0.4,-2.6) to[out=0,in=-90] pic[pos=0.9,sloped]{arrow} (-0.1,-2.2);  
\draw (0.4,-1.8) to[out=0,in=90] pic[pos=0.4,sloped,very thick]{arrow=latex reversed} (0.8,-2.2) to[out=270,in=0] (0.4,-2.6) to[out=180,in=270] pic[pos=0.6,sloped,very thick]{arrow=latex reversed} (0.1,-2.2);
\draw (0.2,-0.4) --pic[pos=0.5,sloped,very thick]{arrow=latex reversed} (0.2,-1.4);
\draw (-0.2,-0.4) --pic[pos=0.6,sloped,]{arrow} (-0.2,-1.4);
\node[right] at (2,-1) {$ \CW= \ct p + \a \b + \Pf\begin{pmatrix}
& X & \vdots & p \\
\dots & \dots & \dots & \dots \\
& & \vdots & \a
\end{pmatrix}$};
\node at (0.6,0.9) {$X$};
\node at (0.6,-1) {$p$};
\node at (-0.6,-0.9) {$\ct$};
\node at (1,-2.5) {$\a$};
\node at (-1,-2.5) {$\b$};
\epic \ee 
Then since $\ct$ and $\b$ are massive we can integrate them out which put to zero $p$ and $\a$. Now if we look at the Pfaffian term, we see columns of zero and so it vanishes. Therefore we obtain the l.h.s of \eqref{LSTDeconfinement}.

\noindent For $G=SU(2N)/Usp(2N)/SO(2N)$ $K$ is an even integer greater or equal to $2$, so using this method there is a fictitious global symmetry (this $SU(2N)$ deconfinement appears in Terning \cite{Terning:1997jj}). For our purposes this is not enough, we need a deconfinement without this additional fake symmetry group. When the theories have at least one matter fields in the fundamental representation, we will use the following deconfinement
\be \label{SU2NDeconfinement} \scalebox{0.9}{\bpic[node distance=2cm,gSUnode/.style={circle,red,draw,minimum size=8mm},gUSpnode/.style={circle,blue,draw,minimum size=8mm},fnode/.style={rectangle,draw,minimum size=8mm}]  
\node[gSUnode] (G1) at (1,0) {$2N$};
\node[fnode] (F1) at (0,-2) {$2N+F-4$};
\node[fnode] (F2) at (2,-2) {$F$};
\draw (G1) -- pic[pos=0.4,sloped,very thick]{arrow=latex reversed} (F1.north);
\draw (G1) -- pic[pos=0.5,sloped,very thick]{arrow=latex reversed} (F2.north);
\draw (1.3,0.4) to[out=90,in=0] pic[pos=0.1,sloped]{arrow} (1,0.8) to[out=180,in=90] pic[pos=0.5,sloped,very thick]{arrow=latex reversed} (0.7,0.4);
\node at (1.6,0.9) {$A$};
\node at (0,-0.7) {$\Qt$};
\node at (2,-0.7) {$Q$};
\node at (3,0) {$\equiv$};
\node[right] at (2,-3) {$ \CW= 0$};
\node[gSUnode] (G2) at (6,0) {$2N$};
\node[fnode] (F3) at (4.3,-2) {$2N+F-4$};
\node[fnode] (F4) at (6.6,-2) {$F-1$};
\node[fnode,violet] (F5) at (7.9,-2) {$1$};
\draw (G2) -- pic[pos=0.4,sloped,very thick]{arrow=latex reversed} (F3);
\draw (G2) -- pic[pos=0.5,sloped,very thick]{arrow=latex reversed} (F4);
\draw (G2) -- pic[pos=0.5,sloped,very thick]{arrow=latex reversed} (F5);
\draw (6.3,0.4) to[out=90,in=0] pic[pos=0.1,sloped]{arrow} (6,0.8) to[out=180,in=90] pic[pos=0.5,sloped,very thick]{arrow=latex reversed} (5.7,0.4);
\node at (6.6,0.9) {$A$};
\node at (4.9,-0.7) {$\Qt$};
\node at (6,-0.9) {$q$};
\node at (7.3,-0.9) {$F$};
\node at (8,0) {$\llra$};
\node[gSUnode] (G3) at (13,0) {$2N$};
\node[gUSpnode] (G4) at (10,0) {$2N-2$};
\node[fnode] (F6) at (12.5,-2.5) {$2N+F-4$};
\node[fnode] (F7) at (14.5,-2.5) {$F-1$};
\node[fnode,violet] (F8) at (9,-2.5) {$1$};
\node[fnode,blue] (F9) at (10.5,-2.5) {$1$};
\draw (G3) -- pic[pos=0.4,sloped]{arrow} (G4);
\draw (G3) -- pic[pos=0.5,sloped,very thick]{arrow=latex reversed} (F6);
\draw (G3) -- pic[pos=0.5,sloped,very thick]{arrow=latex reversed} (F7.north);
\draw (G3) -- pic[pos=0.4,sloped,very thick]{arrow=latex reversed} (F9);
\draw (G4) -- (F8);
\draw (G4) -- (F9);
\draw (F8) -- (F9);
\node[right] at (10,-3.5) {$ \CW=$ 2 Planar Triangles};
\node at (13.1,-1.2) {$\Qt$};
\node at (14.2,-1.2) {$q$};
\node at (11.7,0.4) {$b$};
\node at (9.2,-1.4) {$l$};
\node at (11.5,-1) {$\ct$};
\node at (10.6,-1.4) {$d$};
\node at (9.8,-2.8) {$h$};
\epic} \ee
This type of $4d$ $\cN=1$ deconfinement reduces to the $3d$ $\cN=2$ deconfinement used in \cite{Benvenuti:2020gvy} and also appeared in \cite{BOTTINITESI}. The mapping of the chiral ring generators is
\be \label{mapSU2NDeconfinement} 
\ba{c}
q \, \Qt \\
F \, \Qt \\
A \, \Qt^2 \\
\e_{2N} \, \Qt^{2N} \\
\e_{2N} \, A^N \\
\e_{2N} \, (A^{N-J} \, q^{2J}) \\
\e_{2N} \, (A^{N-1} \, q \, F) \\
\e_{2N} \, (A^{N-J} \, q^{2J-1} \, F) 
\ea
\quad \Longleftrightarrow \quad
\ba{c}
q \, \Qt \\
l \, b \, \Qt \\
b \, b \, \Qt^2 \\
\e_{2N} \, \Qt^{2N} \\
h \\
\e_{2N} \, (b^{2N-2J} \, q^{2J}) \\
\ct \, q \\
\e_{2N} \, (b^{2N-2K} \, l \, b \, q^{2K-1})  
\ea
\quad
\ba{c}
\\
\\
\\
\\
\\
J = 1,\dots,\lfloor\frac{F-1}{2}\rfloor \\
\\
K= 2,\dots,\lfloor\frac{F}{2}\rfloor
\ea
\ee
The trick was to split the $F$ fundamental fields into $F-1$ and $1$ and use this extra $1$ to deconfine without introducing any extra "fake" global symmetry. The proof of \eqref{SU2NDeconfinement} is similar to the previous ones. We start on the r.h.s by confining the $Usp(2N-2)$ gauge group. The initial superpotential terms become mass terms of the unwanted mesons which therefore are set to 0. This kills the would-be Pfaffian term because, as before, we get a vanishing column.

\noindent The $Usp(2N)$ case is similar
\be \label{UspDeconfinement} \scalebox{0.9}{\bpic[node distance=2cm,gSUnode/.style={circle,red,draw,minimum size=8mm},gUSpnode/.style={circle,blue,draw,minimum size=8mm},fnode/.style={rectangle,draw,minimum size=8mm}]  
\node[gUSpnode] (G1) at (-3,0) {$2N$};
\node[fnode] (F1) at (-1,0) {$2F$};
\draw (G1) -- pic[pos=0.7,sloped]{arrow} (F1);
\draw (-2.7,0.4) to[out=90,in=0]  (-3,0.8) to[out=180,in=90] (-3.3,0.4);
\node[right] at (-2,-1.2) {$ \CW= 0$};
\node at (-2.5,0.9) {$A$};
\node at (0,0) {$\equiv$};
\node[gUSpnode] (G2) at (1.5,0) {$2N$};
\node[fnode] (F2) at (4,0) {$2F-1$};
\node[fnode,violet] (F3) at (1.5,-1.8) {$1$};
\draw (G2) -- pic[pos=0.7,sloped]{arrow} (F2) node[midway,above] {$Q_0$};
\draw (G2) -- (F3) node[midway,left] {$F_0$};
\draw (1.8,0.4) to[out=90,in=0]  (1.5,0.8) to[out=180,in=90] (1.2,0.4);
\node at (2.1,0.9) {$A_0$};
\node at (5.5,0) {$\llra$};
\node[gUSpnode] (G3) at (7,0) {$2N$};
\node[gUSpnode] (G4) at (10.5,0) {$2N-2$};
\node[fnode] (F4) at (8.3,1.8) {$2F-1$};
\node[fnode] (F5) at (7,-2.5) {$1$};
\node[fnode,violet] (F6) at (10.5,-2.5) {$1$};
\draw (G3) -- (G4) node[midway] {$\times$};
\draw (G3) -- pic[pos=0.7,sloped]{arrow} (F4.south west);
\draw (G3) -- (F5) node[midway,left] {$v_1$};
\draw (G4) -- (F6) node[midway,right] {$F_1$};
\draw (F5) -- (F6) node[midway,below] {$h_1$};
\draw (F5.north east) -- (G4);
\node[right] at (5,-4) {$ \CW= $ 2 Planar Triangles $+ \b_1 \, \tr(b_1 b_1) $};
\node at (8.7,0.3) {$b_1$};
\node at (8.5,-1) {$d_1$};
\node at (6.9,1.1) {$Q_0$};
\epic} \ee
The cross, $\times$, on the r.h.s quiver stands for the singlet $\b_1$. This singlet is called a flipper. In this situation we should add this flipper because we want the antisymmetric $A_0$ on the l.h.s to be traceless\footnote{Meaning $A_0^{\a \b} J_{\a \b} = 0$ with $J_{\a \b}$ the invariant tensor of $Usp(2N)$}.
  
\noindent In this case the mapping is 
\be \label{mapUspDeconfinement} 
\ba{c}
Q_0 \, A_0^i \, F_0 \\
Q_0 \, A^{N-1} \, F_0 \\
Q_0 \, A^j \, Q_0 \\
\tr A^k \\
\tr A^N
\ea
\quad \Longleftrightarrow \quad
\ba{c}
Q_0 \, b_1 \, (b_1 \, b_1)^i \, F_1 \\
Q_0 \, v_1 \\
Q_0 \, (b_1 \, b_1)^j \, Q_0 \\
\tr (b_1 \, b_1)^k \\
h_1
\ea
\quad
\ba{c}
i=0,\dots,N-2 \\
\\
j=0,\dots,N-1 \\
k=2,\dots,N-1 \\
\\
\ea
\ee
Let's write more explicitly the indices of the operators.
\begin{itemize}
\item $Q_0 \, A_0^j \, Q_0 \equiv (Q_0)_{i_1}^{\a_1} \, A_0^{\a_2 \a_3} \, \cdots \, A_0^{\a_{2j} \a_{2j+1}} \, (Q_0)_{i_2}^{\a_{2j+2}} \, J^{2N}_{\a_1 \a_2} \, \cdots \, J^{2N}_{\a_{2j+1} \a_{2j+2}}$ that transforms in the antisymmetric representation of the $SU(2F-1)$ global symmetry.
\item \scalebox{0.9}{$Q_0 \, (b_1 \, b_1)^j \, Q_0 \equiv$} 

\noindent \scalebox{0.9}{$(Q_0)_{i_1}^{\a_1} \, ((b_1)_{\b_1}^{\a_2} \, (b_1)_{\b_2}^{\a_3} \, J^{\b_1 \b_2}_{2N-2}) \, \cdots \, ((b_1)_{\b_{2J-1}}^{\a_{2J}} \, (b_1)_{\b_{2J}}^{\a_{2J+1}} \, J^{\b_{2J-1} \b_{2J}}_{2N-2}) \, (Q_0)_{i_2}^{\a_{2j+2}} \, J^{2N}_{\a_1 \a_2} \, \cdots \, J^{2N}_{\a_{2j+1} \a_{2j+2}}$}
\item $\tr A_0^k \equiv A_0^{\a_1 \a_2} \cdots A_0^{\a_{2k-1} \a_{2k}} \, J^{2N}_{\a_2 \a_3} \cdots J^{2N}_{\a_{2k-2} \a_{2k-1}} \, J^{2N}_{\a_{2k} \a_1}$
\item $\tr( b_1 \, b_1)^k \equiv$ 

\noindent $((b_1)_{\b_1}^{\a_1} \, (b_1)_{\b_2}^{\a_2} \, J^{\b_1 \b_2}_{2N-2}) \, \cdots \, ((b_1)_{\b_{2J-1}}^{\a_{2J-1}} \, (b_1)_{\b_{2J}}^{\a_{2J}} \, J^{\b_{2J-1} \b_{2J}}_{2N-2}) \, J^{2N}_{\a_2 \a_3} \cdots J^{2N}_{\a_{2k-2} \a_{2k-1}} \, J^{2N}_{\a_{2k} \a_1}$
\end{itemize}

To summarize with these two ways of deconfining. The advantages are
\begin{itemize}
\item $\cW=0$
\item No additional matter fields in the deconfined frame that introduces fictitious global symmetry
\end{itemize}
The disadvantage is the apparent breaking of the global symmetry
\begin{itemize}
\item $SU(2N)$ case: $SU(2N+F-4) \times SU(F) \times U(1)^2 \longrightarrow SU(2N+F-4) \times SU(F-1) \times U(1)^3$
\item $Usp(2N)$ case: $SU(2F) \times U(1) \longrightarrow SU(2F-1) \times U(1)^2$
\end{itemize}

\noindent For $G=SU(2N+1)$ with $F$ fundamentals we have $K=1$ and the deconfinement reads
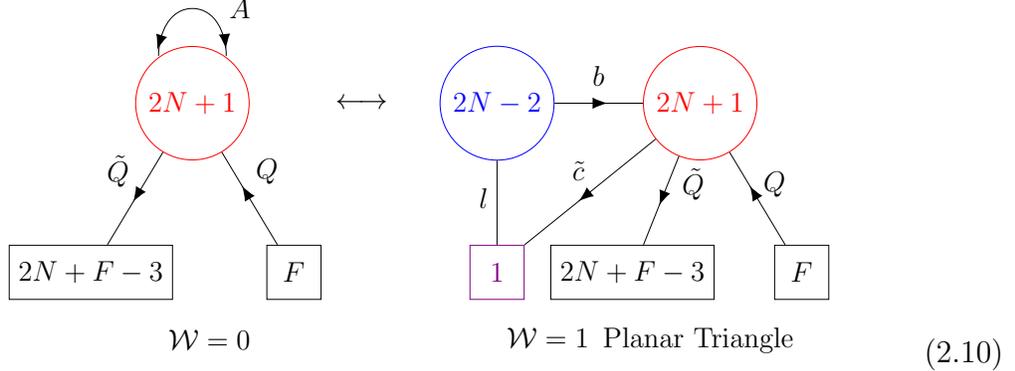
\be \label{SU2N+1Deconfinement} \scalebox{0.9}{\bpic[node distance=2cm,gSUnode/.style={circle,red,draw,minimum size=8mm},gUSpnode/.style={circle,blue,draw,minimum size=8mm},fnode/.style={rectangle,draw,minimum size=8mm}]  
\node[gSUnode] (G1) at (-0.5,0) {$2N+1$};
\node[fnode] (F1) at (-2,-2.5) {$2N+F-3$};
\node[fnode] (F2) at (1,-2.5) {$F$};
\draw (G1) -- pic[pos=0.4,sloped,very thick]{arrow=latex reversed} (F1);
\draw (G1) -- pic[pos=0.5,sloped,very thick]{arrow=latex reversed} (F2);
\draw (0,0.7) to[out=90,in=0] pic[pos=0.1,sloped]{arrow} (-0.5,1.4) to[out=180,in=90] pic[pos=0.7,sloped,very thick]{arrow=latex reversed} (-1,0.7);
\node[right] at (-1,-3.5) {$ \CW= 0$};
\node at (0.2,1.4) {$A$};
\node at (-1.6,-1) {$\Qt$};
\node at (0.6,-1) {$Q$};
\node at (2,0) {$\llra$};
\node[gUSpnode] (G2) at (4,0) {$2N-2$}; 
\node[gSUnode] (G3) at (7,0) {$2N+1$};
\node[fnode] (F3) at (6,-2.5) {$2N+F-3$};
\node[fnode] (F4) at (8.5,-2.5) {$F$};
\node[fnode,violet] (F5) at (4,-2.5) {$1$};
\draw (G3) -- pic[pos=0.4,sloped]{arrow} (G2);
\draw (G3) -- pic[pos=0.4,sloped,very thick]{arrow=latex reversed} (F3);
\draw (G3) -- pic[pos=0.5,sloped,very thick]{arrow=latex reversed} (F4);
\draw (G3) -- pic[pos=0.5,sloped,very thick]{arrow=latex reversed} (F5.north east);
\draw (G2) -- (F5);
\node[right] at (4,-3.5) {$ \CW=1 \,$ Planar Triangle};
\node at (6.9,-1.2) {$\Qt$};
\node at (8.1,-1.2) {$Q$};
\node at (5.5,0.4) {$b$};
\node at (3.8,-1.4) {$l$};
\node at (5.2,-1) {$\ct$};
\epic} \ee
The mapping of the chiral ring generators is the following:
\be \label{mapSU2N+1Deconfinement} 
\ba{c}
Q \, \Qt  \\
A \, \Qt^2  \\
\e_{2N+1} \, \Qt^{2N+1} \\
\e_{2N+1} \, (A^N \, Q)  \\
\e_{2N+1} \, (A^{N-J} \, Q^{2J+1}) 
\ea
\quad \Longleftrightarrow \quad
\ba{c}
Q \, \Qt \\
b \, b \, \Qt^2 \\
\e_{2N+1} \, \Qt^{2N+1} \\
\ct \, Q \\
\e_{2N+1} \, (b^{2N-2J} \, Q^{2J+1})  
\ea
\quad
\ba{c}
\\
\\
\\
\\
J = 1,\dots,\lfloor\frac{F}{2}\rfloor-1
\ea
\ee 
This form of deconfinement appears first in Pouliot \cite{Pouliot:1995me}.

In the following sections, we will apply  the two S-confinements \eqref{SUbuildingBlock}, \eqref{UspbuildingBlock} and the deconfinements \eqref{SU2NDeconfinement} \eqref{UspDeconfinement}
to prove all S-confining dualities for single node quivers. S-confining theories with a single gauge group were classified by \cite{Csaki:1996zb}. By "proof" we mean using only the basic building blocks \eqref{SUbuildingBlock} and \eqref{UspbuildingBlock}. The strategy is to first deconfine the rank-$2$ matter, then confine one by one all the gauge groups.

\section{$Usp(2N)$ with ${\tiny\ydiagram{1,1}} $ + $6 \, {\tiny\ydiagram{1}}$ series} \label{Usp2N}
Let us start from $Usp(2N)$ gauge theory with one antisymmetric and 6 fundamental fields. This theory has a continuous global symmetry $SU(6) \times U(1)$ (on top of the $U(1)_R$ symmetry). We also turn on a superpotential   
\be 
\label{UspQuantumChiralRing} \bpic[node distance=2cm,gSUnode/.style={circle,red,draw,minimum size=8mm},gUSpnode/.style={circle,blue,draw,minimum size=8mm},fnode/.style={rectangle,draw,minimum size=8mm}]  
\node[gUSpnode] (G1) at (-3,0) {$2N$};
\node[fnode] (F1) at (-0.5,0) {$6$};
\draw (G1) -- pic[pos=0.7,sloped]{arrow} (F1);
\draw (-2.7,0.4) to[out=90,in=0]  (-3,0.8) to[out=180,in=90] (-3.3,0.4);
\node[right] at (-4,-1.2) {$ \CW= \sum_{i=2}^{N} \, \a_i \, \tr(A^i)$};
\node at (-2.5,0.9) {$A$};
\node at (-1.7,0.3) {$q$};
\epic \ee  
where $\a_i$ are gauge singlets. The F-term equations of these flippers set the $\tr(A^i)$ to $0$ on the chiral ring. In addition, on the quantum chiral ring these flippers $\a_i$ are not generators. We can understand it as follows: Start with the theory with $\cW =0$. After a-maximization \cite{Intriligator:2003jj}, we discover that the operators $\tr(A^i)$ violate the unitary bound, see Appendix~\ref{unitary}. Therefore, we expect that in the IR the theory breaks into a free and an interacting part. If we want to focus on the interacting part, the procedure is to flip these operators \cite{Benvenuti:2017lle} and the flippers are not generators of the quantum chiral ring. The conclusion is that the quantum chiral ring generators of \eqref{UspQuantumChiralRing} are the dressed mesons: $\tr(q \, A^a \, q), \, a=0, \dots, N-1$ and these are the operators that we have to map. We turned on this superpotential because it will be much easier to keep track of the superpotential when doing the sequence of confinement/deconfinement. 

Now we start by using the trick of splitting the $6$ fundamentals into $5+1$ as in \cite{Pasquetti:2019tix, Pasquetti:2019uop, Benvenuti:2021nwt}
\be \label{UspT0} \bpic[node distance=2cm,gSUnode/.style={circle,red,draw,minimum size=8mm},gUSpnode/.style={circle,blue,draw,minimum size=8mm},fnode/.style={rectangle,draw,minimum size=8mm}]  
\node at (-5,1) {$\CT_0:$};
\node[gUSpnode] (G1) at (-3,0) {$2N$};
\node[fnode] (F1) at (-0.5,0) {$6$};
\draw (G1) -- pic[pos=0.7,sloped]{arrow} (F1);
\draw (-2.7,0.4) to[out=90,in=0]  (-3,0.8) to[out=180,in=90] (-3.3,0.4);
\node[right] at (-4,-1.2) {$ \CW= \sum_{i=2}^{N} \, \a_i \, \tr(A^i)$};
\node at (-2.5,0.9) {$A$};
\node at (-1.7,0.3) {$q$};
\node at (1,0) {$\equiv$};
\node[gUSpnode] (G2) at (2.5,0) {$2N$};
\node[fnode] (F2) at (5,0) {$5$};
\node[fnode,red] (F3) at (2.5,-1.8) {$1$};
\draw (G2) -- pic[pos=0.7,sloped]{arrow} (F2) node[midway,above] {$Q_0$};
\draw (G2) -- (F3) node[midway,left] {$F_0$};
\draw (2.8,0.4) to[out=90,in=0]  (2.5,0.8) to[out=180,in=90] (2.2,0.4);
\node[right] at (1.2,-2.9) {$ \CW= \sum_{i=2}^{N} \, \a_i \, \tr(A_0^i)$};
\node at (3.1,0.9) {$A_0$};
\epic \ee  

\noindent The first step is the use of the deconfinement \eqref{UspDeconfinement}.
\be \label{UspT0'} \scalebox{0.9}{\bpic[node distance=2cm,gSUnode/.style={circle,red,draw,minimum size=8mm},gUSpnode/.style={circle,blue,draw,minimum size=8mm},fnode/.style={rectangle,draw,minimum size=8mm}] 
\node at (-1.5,2.5) {$\CT_{0'}:$};
\node[gUSpnode] (G1) at (0,0) {$2N$};
\node[gUSpnode] (G2) at (3.5,0) {$2N-2$};
\node[fnode] (F1) at (1.3,1.8) {$5$};
\node[fnode] (F2) at (0,-2.5) {$1$};
\node[fnode,red] (F3) at (3.5,-2.5) {$1$};
\draw (G1) -- (G2) node[midway] {$\times$};
\draw (G1) -- pic[pos=0.7,sloped]{arrow} (F1.south west);
\draw (G1) -- (F2) node[midway,left] {$v_1$};
\draw (G2) -- (F3) node[midway,right] {$F_1$};
\draw (F2.north east) -- (G2);
\node[right] at (5.5,0) {$ \CW=$ Planar Triangle};
\node[right] at (5.5,-1) {$+ \b_1 \, \tr(b_1 b_1) + \sum_{i=2}^{N-1} \a_i \, \tr(b_1 b_1)^i$};
\node at (1.7,0.3) {$b_1$};
\node at (1.5,-1) {$d_1$};
\node at (0.1,1.1) {$Q_0$};
\epic} \ee 
Let us remark that the $h_1$ field has been integrated out using the equation of motion (E.O.M) of the massive $\a_N$. Now we see that the $Usp(2N)$ is coupled to $2N-2+5+1=2N+4$ fundamental fields. Therefore we can apply the basic S-confining result \eqref{UspbuildingBlock}. We get  
\be \label{UspT0'Intermediate} \scalebox{0.9}{\bpic[node distance=2cm,gSUnode/.style={circle,red,draw,minimum size=8mm},gUSpnode/.style={circle,blue,draw,minimum size=8mm},fnode/.style={rectangle,draw,minimum size=8mm}]  
\node[gUSpnode] (G1) at (0,0) {$2N-2$};
\node[fnode] (F1) at (3,0) {$5$};
\node[fnode,red] (F2) at (0,-2.5) {$1$};
\node[fnode] (F3) at (3,-2.5) {$1$};
\draw (G1) -- pic[pos=0.7,sloped]{arrow} (F1) node[midway,above] {$Q_1$};
\draw (G1) -- (F2) node[midway,left] {$F_1$};
\draw (F1) -- pic[pos=0.4,sloped,very thick]{arrow=latex reversed} (F3) node[midway,right] {$O_1$};
\draw (0.6,-0.6) -- (2.8,-2.1);
\draw (0.4,-0.8) -- (2.6,-2.3);
\draw (0.4,0.7) to[out=90,in=0]  (0,1.3) to[out=180,in=90] (-0.4,0.7);
\draw (3.4,0.3) to[out=0,in=90] pic[pos=0.4,sloped,very thick]{arrow=latex reversed} (3.8,0) to[out=-90,in=0] pic[pos=0.6,sloped,very thick]{arrow=latex reversed} (3.4,-0.3);
\node[right] at (5,0) {$ \CW= p_1 d_1 + \b_1 \, \tr( A_1) + \sum_{i=2}^{N-1} \, \a_i \, \tr(A_1^i)$};
\node[right] at (5.8,-1.5) {$+ \Pf
\begin{pmatrix}
A_1 & Q_1 & p_1 \\
& M_1 & O_1 \\
& & 0
\end{pmatrix}$};
\node at (0.7,1.3) {$A_1$};
\node at (4,0.5) {$M_1$};
\node at (2,-1.1) {$p_1$};
\node at (1.2,-1.8) {$d_1$};
\epic} \ee
The E.O.M of $\b_1$ and $d_1$ set $\tr(A_1) \equiv (A_1)^{\a_1 \a_2} J^{2N-2}_{\a_1 \a_2} =0$ and $p_1 =0$. Therefore the Pfaffian term becomes
\begin{multline} 
\Pf \mu = \Pf
\begin{pmatrix}
A_1 & Q_1 & 0 \\
& M_1 & O_1 \\
& & 0
\end{pmatrix} \sim \e_{a_1 \dots a_{2N+4}} \, \mu^{a_1 a_2} \dots \mu^{a_{2N+3} a_{2N+4}} \\
=\e_{2N-2} \, \e_5 \, \left[ A_1^{N-1} \, M_1^2 O_1 + A_1^{N-2} \, Q_1^2 M_1 O_1 + A_1^{N-3} \, Q_1^4 O_1 \right] \label{ChiralRingStab}
\end{multline} 
Let us focus on the first term: $\e_{2N-2} \, \e_5 \, \left[ A_1^{N-1} \, M_1^2 O_1 \right] = \e_{2N-2} \, \left[ A_1^{N-1} \right] \, \e_5 \, \left[M_1^2 O_1 \right]$. We claim that we can drop this term. Indeed the part $\e_{2N-2} \, \left[ A_1^{N-1} \right]$ can be written, by linear algebra, as the product of all traces of $A_i$ (which are all set to 0 on the chiral ring by the F-term equations of $\a_i$). Then using the chiral ring stability argument\footnote{The chiral ring stability argument says the following \cite{Benvenuti:2017lle}: Start with a theory with superpotential $\cW_{\CT} = \sum_i \cW_i$. For each $i$, consider the modified theory $\CT_i$ where the term $\cW_i$ is removed. Then, check if the operator $\cW_i$ is in the chiral ring of $\CT_i$. If it is not, drop $\cW_i$ from the full superpotential $\cW$.} \cite{Benvenuti:2017lle} on $\e_{2N-2} \, \e_5 \, \left[ A_1^{N-1} \, M_1^2 O_1 \right]$ we conclude that we can remove this term from the full superpotential in \eqref{UspT0'Intermediate}. More generally, the chiral ring stability allows us to drop terms of the form: $\e_{2c} \, \left[ A^{c} \right] \, \e_5 \,  \left[ M_i \, M_j \, O_k \right]$ with $c>0$\footnote{Even if we don't want to use the chiral ring stability criterion, these terms will disappear once we reach the final frame. Indeed every power of the antisymmetric field will at some point be mapped to a singlet $h_a$ field. This singlet will enter in the superpotential as a mass term with one flipper $\a$. Therefore when we integrate out the massive flipper $\a_{N-k}$, its E.O.M set $h_{k+1}$ to $0$. Conclusion, if we are doing enough confinement/deconfinement steps, these terms $\e_{2c} \, \left[ A^{c} \right] \, \e_5 \,  \left[ M_i \, M_j \, O_k \right]$ with $c>0$ are set to 0.}. Therefore from now on we will discard these terms. This is the reason why we turn on the superpotential in \eqref{UspT0}, to avoid the proliferation of this kind of term.
\be \label{UspT1} \scalebox{0.9}{\bpic[node distance=2cm,gSUnode/.style={circle,red,draw,minimum size=8mm},gUSpnode/.style={circle,blue,draw,minimum size=8mm},fnode/.style={rectangle,draw,minimum size=8mm}]  
\node at (-1.5,2) {$\CT_1:$};
\node[gUSpnode] (G1) at (0,0) {$2N-2$};
\node[fnode] (F1) at (3,0) {$5$};
\node[fnode,red] (F2) at (0,-2.5) {$1$};
\node[fnode] (F3) at (3,-2.5) {$1$};
\draw (G1) -- pic[pos=0.7,sloped]{arrow} (F1) node[midway,above] {$Q_1$};
\draw (G1) -- (F2) node[midway,left] {$F_1$};
\draw (F1) -- pic[pos=0.4,sloped,very thick]{arrow=latex reversed} (F3) node[midway,right] {$O_1$};
\draw (0.4,0.7) to[out=90,in=0]  (0,1.3) to[out=180,in=90] (-0.4,0.7);
\draw (3.4,0.3) to[out=0,in=90] pic[pos=0.4,sloped,very thick]{arrow=latex reversed} (3.8,0) to[out=-90,in=0] pic[pos=0.6,sloped,very thick]{arrow=latex reversed} (3.4,-0.3);
\node[right] at (5,0) {$ \CW= \sum_{i=2}^{N-1} \, \a_i \, \tr(A_1^i)$};
\node[right] at (5.8,-1) {$+ \, \e_{2N-2} \, \e_5 \left[A_1^{N-2} \, Q_1^2 M_1 O_1 \right.$};
\node[right] at (5.8,-2) {$\left.+  A_1^{N-3} \, Q_1^4 O_1 \right]$};
\node at (0.7,1.3) {$A_1$};
\node at (4,0.5) {$M_1$};
\epic} \ee
The next step is to deconfine again the antisymmetric field. We get 
\be \label{UspT1'} \scalebox{0.9}{\bpic[node distance=2cm,gSUnode/.style={circle,red,draw,minimum size=8mm},gUSpnode/.style={circle,blue,draw,minimum size=8mm},fnode/.style={rectangle,draw,minimum size=8mm}]  
\node at (-1,3.2) {$\CT_{1'}:$};
\node[gUSpnode] (G1) at (0,0) {$2N-2$};
\node[gUSpnode] (G2) at (3.5,0) {$2N-4$};
\node[fnode] (F1) at (1.3,2.2) {$5$};
\node[fnode] (F2) at (0,-2.5) {$1$};
\node[fnode,red] (F3) at (3.5,-2.5) {$1$};
\node[fnode] (F4) at (3.3,2.2) {$1$};
\draw (G1) -- (G2) node[midway] {$\times$};
\draw (G1) -- pic[pos=0.7,sloped]{arrow} (F1.south west);
\draw (G1) -- (F2) node[midway,left] {$v_2$};
\draw (G2) -- (F3) node[midway,right] {$F_2$};
\draw (F2.north east) -- (G2);
\draw (F1) -- pic[pos=0.4,sloped,very thick]{arrow=latex reversed} (F4);
\draw (1.6,2.6) to[out=90,in=0] pic[pos=0.1,sloped]{arrow} (1.3,3) to[out=180,in=90] pic[pos=0.5,sloped,very thick]{arrow=latex reversed} (1,2.6);
\node[right] at (4.7,1) {$\CW= \tr(v_2 b_2 d_2) + \b_2 \, \tr(b_2 b_2) + \sum_{i=2}^{N-2} \a_2 \, \tr(b_2 b_2)^i$};
\node[right] at (5.3,0) {$+ \e_{2N-2} \, \e_5 \left[ b_2^{2N-4} Q_1^2 M_1 O_1 + b_2^{2N-6} Q_1^4 O_1\right]$};
\node at (1.7,0.3) {$b_2$};
\node at (1.5,-1) {$d_2$};
\node at (0.2,1.3) {$Q_1$};
\node at (2.4,1.8) {$O_1$};
\node at (2,3.1) {$M_1$};
\epic} \ee 
Now we confine the $Usp(2N-2)$. It is similar to the previous step between the frames $\CT_{0'}$ and $\CT_1$. There is once again a confining superpotential given by a Pfaffian term. What is interesting and non-trivial is the mapping of the existing superpotential terms in \eqref{UspT1'}.

\noindent Let's start with $\e_{2N-2} \, \e_5 \left[ b_2^{2N-4} Q_1^2 M_1 O_1 \right]$. In principle when we deconfine the antisymmetric field $A_1$ with the $Usp(2N-4)$ gauge group, the $b_2$ fields are contracted pairwise with the invariant tensor of the Usp group $J^{2N-4}$. Therefore a term like $\e_{2N-2} \, \e_5 \left[ b_2^{2N-4} Q_1^2 M_1 O_1 \right]$ contain implicitly $N-2$ invariant tensor $J^{2N-4}$ that are contracted with the $2N-4$ $b_2$ fields. But we can do something else. Since the number of $b_2$ fields is smaller than the order of $\e_{2N-2}$, the epsilon forces to antisymmetrize all the $2N-4$ indices. So we can trade, modulo an irrelevant numerical factor, the bunch of $J^{2N-4}$ for $\e^{2n-4}$. Putting indices we get
\begin{multline}
\e_{2N-2} \, \e_5 \, \left[ b_2^{2N-4} \, Q_1^2 M_1 O_1 \right] \sim \e^{2N-4} \, \e_5 \, \e_{2N-2} \, \left[b_2^{2N-4} \, Q_1^2 M_1 O_1 \right] \\
= \textcolor{magenta}{\e^{2N-4}} \, \e_5 \, \textcolor{ForestGreen}{\e_{2N-2}} \, \left[ (b_2)_{\textcolor{magenta}{\b_1}}^{\textcolor{ForestGreen}{\a_1}} \dots (b_2)_{\textcolor{magenta}{\b_{2N-4}}}^{\textcolor{ForestGreen}{\a_{2N-4}}} \, (Q_1)_{i_1}^{\textcolor{ForestGreen}{\a_{2N-3}}} \, (Q_1)_{i_2}^{\textcolor{ForestGreen}{\a_{2N-2}}} (M_1)_{i_3 i_4} (O_1)_{i_5}\right]
\end{multline}
To improve the readability, we have colored the $Usp(2N-4)$ indices in \textcolor{magenta}{magenta} and the $Usp(2N-2)$ indices in \textcolor{ForestGreen}{green}. Using this form, it is easier to see where this term is mapped to. Indeed, when we confine the $Usp(2N-2)$, we have two different possibilities. Either we contract the two $Q_1$ together which gives the meson $M_2$ or we contract each $Q_1$ with a $b_2$ which gives a new fundamental $Q_2$.

\noindent So after the confinement of $Usp(2N-2)$, this term is mapped to
\begin{itemize}
\item $\e^{2N-4} \, \e_5 \, \e_{2N-2} \, \left[b_2^{2N-4} \, Q_1^2 M_1 O_1 \right] \longrightarrow \e^{2N-4} \, \e_5 \, \left[ A_2^{N-2} \, M_2 M_1 O_1 + A_2^{N-3} \, Q_2^2 M_1 O_1\right]$
\end{itemize}
Again we use the chiral ring stability argument to remove the first term.

\noindent The second term we have to map is $\e_{2N-2} \, \e_5 \left[b_2^{2N-6} Q_1^4 O_1 \right]$. The strategy is the same as for the previous term. We also trade the bunch of $J^{2N-4}$ with $\e^{2N-6}$ tensor\footnote{$\e^{2N-6}$ is an abuse of terminology. What we mean is a totally antisymmetric tensor with $2N-6$ indices but each index goes from $1$ to $2N-4$.}. Now there are $4$ $Q_1$ to play with. We can form $3$ different terms. The first one is contracting the $4$ $Q_1$ among themselves. The second is contracting $2$ $Q_1$ with $2$ $b_2$ and the remaining $2$ $Q_1$ together. The last one is contracting the $4$ $Q_1$ with $4$ $b_2$.

\noindent So after the confinement of $Usp(2N-2)$ we get
\begin{itemize}
\item \scalebox{0.9}{$\e^{2N-6} \, \e_5 \, \e_{2N-2} \, \left[b_2^{2N-6} Q_1^4 O_1 \right] \longrightarrow \e^{2N-6} \, \e_5 \, \left[ A_2^{N-3} \, M_2^2 O_1 + A_2^{N-4} \, Q_2^2 M_2 O_1 + A_2^{N-5} \, Q_2^4 O_1\right]$}
\end{itemize}
After eliminating the first term using the chiral ring stability we get 
\be \label{UspT2} \scalebox{0.9}{\bpic[node distance=2cm,gSUnode/.style={circle,red,draw,minimum size=8mm},gUSpnode/.style={circle,blue,draw,minimum size=8mm},fnode/.style={rectangle,draw,minimum size=8mm}]   
\node at (-0.8,2) {$\CT_2:$};
\node[gUSpnode] (G1) at (0,0) {$2N-4$};
\node[fnode,minimum size=1cm] (F1) at (3,0) {$5$};
\node[fnode,red] (F2) at (0,-2.5) {$1$};
\node[fnode] (F3) at (2.3,-2.5) {$1$};
\node[fnode] (F4) at (3.7,-2.5) {$1$};
\draw (G1) -- pic[pos=0.7,sloped]{arrow} (F1) node[midway,above] {$Q_2$};
\draw (G1) -- (F2) node[midway,left] {$F_2$};
\draw (F1) -- pic[pos=0.3,sloped]{arrow} (F3) node[midway,left] {$O_1$};
\draw (F1) -- pic[pos=0.4,sloped,very thick]{arrow=latex reversed} (F4) node[midway,right] {$O_2$};
\draw (0.4,0.7) to[out=90,in=0]  (0,1.3) to[out=180,in=90] (-0.4,0.7);
\draw (3.5,0.2) to[out=0,in=90] pic[pos=0.4,sloped,very thick]{arrow=latex reversed} (3.9,0) to[out=-90,in=0] pic[pos=0.6,sloped,very thick]{arrow=latex reversed} (3.5,-0.2);
\draw (3.5,0.4) to[out=0,in=90] pic[pos=0.4,sloped,very thick]{arrow=latex reversed} (4.1,0) to[out=-90,in=0] pic[pos=0.6,sloped,very thick]{arrow=latex reversed} (3.5,-0.4);
\node[right] at (5,0) {$\CW = \sum_{i=2}^{N-2} \, \a_i \, \tr(A_2^i) + \e^{2N-4} \, \e_5 \, \left[A_2^{N-3} \, Q_2^2 M_1 O_1 \right]$};
\node[right] at (5.7,-1) {$+\e^{2N-6} \, \e_5 \, \left[A_2^{N-4} \, Q_2^2 M_2 O_1 + A_2^{N-5} \, Q_2^4 O_1 \right]$};
\node[right] at (5.7,-2) {$+ \e^{2N-4} \, \e_5 \, \left[A_2^{N-3} \, Q_2^2 M_2 O_2 + A_2^{N-4} \, Q_2^4 O_2 \right]$};
\node at (0.7,1.3) {$A_2$};
\node at (4,0.8) {$M_1, M_2$};
\epic} \ee
We iterate this procedure of deconfinement/deconfinement. In the appendix \ref{SuperpotInTk} we present the theory in the arbitrary $\CT_k$ frame with the full superpotential. 

\noindent After $N-1$ iterations we get 
\be \label{UspTN-1} \scalebox{0.9}{\bpic[node distance=2cm,gSUnode/.style={circle,red,draw,minimum size=8mm},gUSpnode/.style={circle,blue,draw,minimum size=8mm},fnode/.style={rectangle,draw,minimum size=8mm}]   
\node at (-1,1.5) {$\CT_{N-1}:$};
\node[gUSpnode] (G1) at (0,0) {$2$};
\node[fnode,minimum size=1cm] (F1) at (3,0) {$5$};
\node[fnode,red] (F2) at (0,-2) {$1$};
\node[fnode] (F3) at (2.1,-2) {$1$};
\node at (3.1,-2) {$\dots$};
\node[fnode] (F4) at (3.9,-2) {$1$};
\draw (G1) -- pic[pos=0.7,sloped]{arrow} (F1) node[midway,above] {$Q_{N-1}$};
\draw (G1) -- (F2) node[midway,left] {$F_{N-1}$};
\draw (F1) -- pic[pos=0.3,sloped]{arrow} (F3) node[midway,left] {$O_1$};
\draw (F1) -- pic[pos=0.4,sloped,very thick]{arrow=latex reversed} (F4) node[midway,right] {$O_{N-1}$};
\draw (3.5,0.2) to[out=0,in=90] pic[pos=0.4,sloped,very thick]{arrow=latex reversed} (3.9,0) to[out=-90,in=0] pic[pos=0.6,sloped,very thick]{arrow=latex reversed} (3.5,-0.2);
\node at (4.3,0) {$\dots$};
\draw (3.5,-0.4) to[out=0,in=-90] pic[pos=0.4,sloped,very thick]{arrow=latex reversed} (4.7,0) to[out=90,in=0] pic[pos=0.6,sloped,very thick]{arrow=latex reversed} (3.5,0.4);
\node[right] at (5.7,0) {$\CW = \sum_{i=1}^{N-1} \e_2 \, \e_5 \, \left[ Q_{N-1}^2 M_i O_i \right]$};
\node[right] at (5.3,-1) {$+\sum_{k=1}^{N-2} \sum_{k+1 \le i \le j \le N-1} \, \e_5 \, \left[ M_i M_j O_{k} \right] \de_{i+j-k,N} $};
\node at (4.6,0.9) {$M_1, \dots, M_{N-1}$};
\epic} \ee
Now since we reach $Usp(2) \simeq SU(2)$ the traceless antisymmetric field simply does not exist anymore. Therefore there is nothing to deconfine and we can directly apply the building block once more. We reach the final \say{Deconfined} frame
\be \label{UspTDec} \scalebox{0.9}{\bpic[node distance=2cm,gSUnode/.style={circle,red,draw,minimum size=8mm},gUSpnode/.style={circle,blue,draw,minimum size=8mm},fnode/.style={rectangle,draw,minimum size=8mm}]   
\node at (0,1.8) {$\CT_{Dec}:$};
\node[fnode,minimum size=1cm] (F1) at (2,0) {$5$};
\node[fnode] (F2) at (1.1,-2) {$1$};
\node at (2.1,-2) {$\dots$};
\node[fnode,red] (F3) at (2.9,-2) {$1$};
\draw (F1) -- pic[pos=0.3,sloped]{arrow} (F2) node[midway,left] {$O_1$};
\draw (F1) -- pic[pos=0.4,sloped,very thick]{arrow=latex reversed} (F3) node[midway,right] {$O_{N}$};
\draw (2.2,0.5) to[out=90,in=0] pic[pos=0.1,sloped]{arrow} (2,0.9) to[out=180,in=90] pic[pos=0.6,sloped,very thick]{arrow=latex reversed} (1.8,0.5);
\node at (2,1.3) {$\vdots$};
\draw (1.6,0.5) to[out=90,in=-180] pic[pos=0.4,sloped,very thick]{arrow=latex reversed} (2,1.5) to[out=0,in=90] pic[pos=0.8,sloped]{arrow} (2.4,0.5);
\node[right] at (-2,-3.5) {$\CW = \sum_{k=1}^{N} \sum_{k \le i \le j \le N-1} \, \e_5 \, \left[ M_i M_j O_{k} \right] \de_{i+j-k,N}$};
\node at (3.7,1) {$M_1, \dots, M_{N}$};
\node at (5,0) {$\equiv$};
\node[fnode,minimum size=1cm] (F4) at (8,0) {$6$};
\draw (8.2,0.5) to[out=90,in=0] pic[pos=0.1,sloped]{arrow} (8,0.9) to[out=180,in=90] pic[pos=0.6,sloped,very thick]{arrow=latex reversed} (7.8,0.5);
\node at (8,1.3) {$\vdots$};
\draw (7.6,0.5) to[out=90,in=-180] pic[pos=0.4,sloped,very thick]{arrow=latex reversed} (8,1.5) to[out=0,in=90] pic[pos=0.8,sloped]{arrow} (8.4,0.5);
\node at (9.7,1) {$\mu^1, \dots, \mu^{N}$};
\node[right] at (5,-1.5) {$\CW = \sum_{N\ge a\ge b\ge c\ge 1} \, \e_{6} \, \left[ \mu^a \, \mu^b \,  \mu^c \right] \, \de_{a+b+c,2N+1}$};
\node[right] at (7,-3.5) {$\mu^a =
\begin{pmatrix}
\bovermat{5}{\phantom{12}  M_a \phantom{12}} & \vdots & \bovermat{1}{O_{N+1-a}} \\
\dotfill & \vdots \dotfill & \dotfill \\
& \vdots &  0
\end{pmatrix}
\begin{aligned}
&\left. \begin{matrix}
\vphantom{O_N} \\
\\
\end{matrix} \right\}
5 \\
&\left. \begin{matrix}
\\
\end{matrix} \right\}
1\\
\end{aligned} $}; 
\epic} \ee
In the last equality, we have repackage the superpotential in a manifestly $SU(6)$ invariant way. We recover the result of \cite{Csaki:1996zb} and the superpotential for a generic N proposed in \cite{Benvenuti:2018bav}.

\noindent The mapping of the chiral ring generators between the original frame and the final one is the following
\be \label{Uspmap1}
\ba{c}\CT_{0} \\
\tr(Q_0 \, A_0^i \, Q_0) \\
\tr(Q_0 \, A_0^i \, F_0) 
\ea
\qquad \Longleftrightarrow \qquad
\ba{c} \CT_{DEC} \\
M_{i+1}  \\
O_{N-i}
\ea
\qquad
\ba{l}\\
i= 0,\dots,N-1 \\
i= 0,\dots,N-1
\ea
\ee
In the repackaged form, the mapping becomes: $\tr(q \, A^a \, q) \Longleftrightarrow \mu^{a+1} , \,\, a=0, \dots, N-1$.
\section{$SU(M)$ with  ${\tiny\ydiagram{1,1}},  \overline{\tiny\ydiagram{1,1}}  + 3 \, {\tiny\ydiagram{1}} + 3  \, \overline{{\tiny\ydiagram{1}}}$ series}\label{SUaa}
\subsection{Odd rank: $M = 2N+1$} 
The first series is $SU(2N+1)$ gauge theory with fields in the antisymmetric and conjugate antisymmetric representation and $(3,3)$ fundamentals, antifundamentals. The continuous global symmetry is $SU(3)_{Q} \times SU(3)_{\Qt} \times U(1)^3$.

\be \label{SU2N+1antisym'T1} \scalebox{0.9}{\bpic[node distance=2cm,gSUnode/.style={circle,red,draw,minimum size=8mm},gUSpnode/.style={circle,blue,draw,minimum size=8mm},fnode/.style={rectangle,draw,minimum size=8mm}]   
\node at (3,1.8) {$\CT_1:$};
\node[gSUnode] (G1) at (6,0) {$2N+1$};
\node[fnode,orange] (F1) at (7.3,-2.3) {$3$};
\node[fnode] (F2) at (4.7,-2.3) {$3$};
\draw (G1) -- pic[pos=0.6,sloped]{arrow} (F1);
\draw (G1) -- pic[pos=0.4,sloped]{arrow} (F2);
\draw (5.2,0.3) to[out=135,in=-90] pic[pos=0.1,sloped]{arrow} (5,0.8) to[out=90,in=180] (5.3,1.3) to[out=0,in=90] pic[pos=0.9,sloped]{arrow} (5.8,0.8);  
\draw (6.8,0.3) to[out=45,in=-90] pic[pos=1,sloped,very thick]{arrow} (7,0.8) to[out=90,in=0] (6.7,1.3) to[out=180,in=90] pic[pos=0.4,sloped,very thick]{arrow} (6.2,0.8);
\node[right] at (8.5,0) {$ \CW= 0$};
\node at (4.9,1.5) {$A$};
\node at (7.1,1.5) {$\At$};
\node at (7.1,-1.1) {$\Qt$};
\node at (4.9,-1.1) {$Q$};
\epic} \ee
The chiral ring generators are
\begin{itemize}
\item \scalebox{0.95}{$Q (A \, \At)^k \Qt \sim Q_i^{\a_1} (\At_{\a_1 \a_2} \, A^{\a_2 \a_3} \, \At_{\a_3 \a_4} \cdots A^{\a_{2k} \a_{2k+1}}) \, \Qt_{\a_{2k+1}}^I$}\footnote{$\a_i=1,\ldots,2N+1$ are gauge indices. $I,J,\ldots=1,\ldots,3$ are $SU(3)_{\Qt}$ flavor indices. $i,j,k,l\dots=1,\dots,3$ are $SU(3)_{Q}$ flavor indices.} $\, , k=0,\dots, N-1 \, $ 

(transforming in the $(\, \overline{\scalebox{0.5}{\ydiagram{1}}} \, , \scalebox{0.5}{\ydiagram{1}} \,)$ of $SU(3)_Q \times SU(3)_{\Qt}$)
\item \scalebox{0.95}{$\At \, (A \, \At)^k \, Q^2 \sim \At_{\a_1 \a_2}(A^{\a_2 \a_3} \, \At_{\a_3 \a_4} \, A^{\a_4 \a_5} \cdots A^{\a_{2k+1} \a_{2k+2}} \, \At_{\a_{2k+2} \a_{2k+3}}) \, Q_{[i}^{\a_{2k+3}} Q_{j]}^{\a_1} \, ,$}

$k=0,\dots, N-1 \,$ $ \sim (\, \overline{\scalebox{0.5}{\ydiagram{1,1}}} \, , 1)$ 
\item \scalebox{0.95}{$A \, (A \, \At)^k \, \Qt^2 \sim A^{\a_1 \a_2}(\At_{\a_2 \a_3} \, A^{\a_3 \a_4} \, \At_{\a_4 \a_5} \cdots \At_{\a_{2k+1} \a_{2k+2}} \, A^{\a_{2k+2} \a_{2k+3}}) \, \Qt_{\a_{2k+3}}^{[I} \Qt_{\a_1}^{J]} \, ,$}

$k=0,\dots, N-1 \, $ $ \sim (1 , \scalebox{0.5}{\ydiagram{1,1}} \,)$
\item \scalebox{0.95}{$(A \, \At)^m \sim A^{\a_1 \a_2} \, \At_{\a_2 \a_3} \, A^{\a_3 \a_4} \cdots A^{\a_{2m-1} \a_{2m}} \, \At_{\a_{2m} \a_{1}} \, $ $m=1,\dots,N \,$ $\sim (1,1)$} 
\item \scalebox{0.95}{$\e_{2N+1} \, (A^N \, Q) \sim \e_{2N+1} \, (A^{\a_1 \a_2} \cdots A^{\a_{2N-1} \a_{2N}} \, Q_i^{\a_{2N+1}})$ $\sim (\, \overline{\scalebox{0.5}{\ydiagram{1}}} \, , 1)$}
\item \scalebox{0.95}{$\e^{2N+1} \, (\At^N \, \Qt) \sim \e^{2N+1} \, (\At_{\a_1 \a_2} \cdots \At_{\a_{2N-1} \a_{2N}} \, Q_{\a_{2N+1}}^I)$ $\sim (1, \scalebox{0.5}{\ydiagram{1}} \,)$}
\item \scalebox{0.95}{$\e_{2N+1} \, (A^{N-1} \, Q^3) \sim \e_{2N+1} \, \e^{ijk} \, (A^{\a_1 \a_2} \cdots A^{\a_{2N-3} \a_{2N-2}} \, Q_i^{\a_{2N-1}} \, Q_j^{\a_{2N}} \, Q_k^{\a_{2N+1}})$ $\sim (1,1)$}
\item \scalebox{0.95}{$\e^{2N+1} \, (\At^{N-1} \, \Qt^3) \sim \e^{2N+1} \, \e_{IJK} \, (\At_{\a_1 \a_2} \cdots \At_{\a_{2N-3} \a_{2N-2}} \, \Qt_{\a_{2N-1}}^I \, \Qt_{\a_{2N}}^J \, Q_{\a_{2N+1}}^K)$ $\sim (1,1)$}
\end{itemize}
We deconfine the two antisymmetric fields with the help of \eqref{SU2N+1Deconfinement}.
\be \label{SU2N+1antisym'T1'} \scalebox{0.9}{\bpic[node distance=2cm,gSUnode/.style={circle,red,draw,minimum size=8mm},gUSpnode/.style={circle,blue,draw,minimum size=8mm},fnode/.style={rectangle,draw,minimum size=8mm}]   
\node at (1.5,2.8) {$\CT_{1'}:$};
\node[gSUnode] (G1) at (6,0) {$2N+1$};
\node[gUSpnode] (G2) at (3,0) {$2N-2$};
\node[gUSpnode] (G3) at (9,0) {$2N-2$};
\node[fnode,orange] (F1) at (7.3,-2.3) {$3$};
\node[fnode] (F2) at (4.7,-2.3) {$3$};
\node[fnode,violet] (F3) at (4.2,2) {$1$};
\node[fnode,blue] (F4) at (7.8,2) {$1$};
\draw (G1) -- pic[pos=0.4,sloped]{arrow} (G2);
\draw (G1) -- pic[sloped]{arrow} (G3);
\draw (G1) -- pic[pos=0.6,sloped]{arrow} (F1);
\draw (G1) -- pic[pos=0.4,sloped]{arrow} (F2);
\draw (G1) -- pic[sloped,very thick]{arrow=latex reversed} (F3);
\draw (G1) -- pic[sloped,very thick]{arrow=latex reversed} (F4);
\draw (G2) -- (F3);
\draw (G3) -- (F4);
\node[right] at (10.5,0) {$ \CW= 2 \,$ Planar Triangles};
\node at (4.9,-1.1) {$Q$};
\node at (7.1,-1.1) {$\Qt$};
\node at (4.5,-0.4) {$b$};
\node at (7.5,-0.4) {$\bt$};
\node at (5.4,1.2) {$\ct$};
\node at (6.6,1.2) {$c$};
\node at (3.4,1.2) {$l$};
\node at (8.6,1.2) {$\lt$};
\epic} \ee
Now we confine the $SU$ node with \eqref{SUbuildingBlock}.
\be \label{SU2N+1antisym'T2} \scalebox{0.9}{\bpic[node distance=2cm,gSUnode/.style={circle,red,draw,minimum size=8mm},gUSpnode/.style={circle,blue,draw,minimum size=8mm},fnode/.style={rectangle,draw,minimum size=8mm}]     
\node at (-3,3.5) {$\CT_2:$};
\node[fnode,red] (F1) at (0,0) {$1$};
\node[gUSpnode] (G1) at (-1.8,2) {$2N-2$};
\node[gUSpnode] (G2) at (1.8,2) {$2N-2$};
\node[fnode,orange] (F2) at (-2.7,-0.2) {$3$};
\node[fnode] (F3) at (2.7,-0.2) {$3$};
\node[fnode,blue] (F4) at (-1,-2.2) {$1$};
\node[fnode,violet] (F5) at (1,-2.2) {$1$};
\draw (G1) -- (F1);
\draw (G1) -- (G2);
\draw (G1) -- pic[sloped,very thick]{arrow=latex reversed} (F2);
\draw (G2) -- (F1);
\draw (G2) -- pic[sloped,very thick]{arrow=latex reversed} (F3);
\draw (F1) -- pic[pos=0.4,sloped]{arrow} (F2);
\draw (F1) -- pic[pos=0.6,sloped]{arrow} (F3);
\draw (F1) -- (F4);
\draw (F1) -- (F5);
\draw (F2) -- pic[sloped,very thick]{arrow=latex reversed} (F4);
\draw (F3) -- pic[sloped,very thick]{arrow=latex reversed} (F5);
\draw (F4) -- (F5);
\draw (-2.9,-0.6) to[out=-90,in=180] pic[pos=1,sloped,very thick]{arrow=latex reversed} (0,-3.5) to[out=0,in=-90] (2.9,-0.6);
\node[right] at (4,1.7) {$\CW= \e_{2N-2} \, \e^{2N-2} \, \e_3 \, \e^3 \left[K_2^{2N-2} M_0^3 T_N \right. $};
\node[right] at (4,0.7) {$+ K_2^{2N-2} M_0^2 B_1 \Bt_1 + K_2^{2N-3} K_1 R M_0^2 T_N$};
\node[right] at (4,-0.3) {$+ K_2^{2N-3} K_1 R M_0 B_1 \Bt_1$};
\node[right] at (4,-1.3) {$+ K_2^{2N-4} K_1^2 R^2 M_0 T_N + K_2^{2N-4} K_1^2 R^2 B_1 \Bt_1$};
\node[right] at (4,-2.3) {$+ \left. K_2^{2N-5} K_1^3 R^3 T_N \right] + X H_N M_0$};
\node[right] at (4,-3.3) {+ 6 Planar Triangles};
\node at (-2.8,0.9) {$K_1$};
\node at (0,2.4) {$K_2$};
\node at (2.8,0.9) {$R$};
\node at (2.2,-1.4) {$B_1$};
\node at (0,-2.5) {$T_N$};
\node at (-2.1,-1.4) {$\Bt_1$};
\node at (-1.3,-0.4) {$X$};
\node at (-1.1,0.8) {$Y$};
\node at (0.5,1.1) {$\Yt$};
\node at (1.4,0.2) {$H_N$};
\node at (0.8,-1) {$B_3$};
\node at (-0.2,-1.2) {$\Bt_3$};
\node at (0.4,-3.2) {$M_0$};
\epic} \ee 
Let's give more details. The first 7 terms in the superpotential \eqref{SU2N+1antisym'T2} come from the determinant of the meson matrix $\P$
\be 
\det \P \equiv \det
\begin{pmatrix}
K_2 & K_1 & 0 \\
R & M_0 & B_1 \\
0 & \Bt_1 & T_N
\end{pmatrix}
= \e^{a_1 \dots a_{2N+4}} \, \e_{b_1 \dots b_{2N+4}} \, \P_{a_1}^{b_1} \cdots \P_{a_{2N+4}}^{b_{2N+4}} \nn
\ee
\be 
\begin{split}
\longrightarrow &\, \e^{2N-2} \, \e_{2N-2} \, \e^3 \, \e_3 \, \left[ K_2^{2N-2} \left( M_0^3 T_N + M_0^2 B_1 \Bt_1 \right) + K_2^{2N-3} \left( K_1 R M_0^2 T_N \right. \right. \\
&\left. \left. + K_1 R M_0 B_1 \Bt_1 \right) + K_2^{2N-4} \left(K_1^2 R^2 M_0 T_N + K_1^2 R^2 B_1 \Bt_1 \right) + K_2^{2N-5} K_1^3 R^3 T_N \right] \\
\end{split}
\ee

The \say{$X H_N M_0 + 6\,$ Planar Triangles} terms come from the cubic interaction \say{meson $\times$ baryon $\times$ meson} when the $SU$ group confines. We also rescale the fields such that the coefficient in front of each term is $+1$.

The next step is to confine the left $Usp(2N-2)$ using \eqref{UspbuildingBlock}. 
\be \label{SU2N+1antisym'T3} \scalebox{0.9}{\bpic[node distance=2cm,gSUnode/.style={circle,red,draw,minimum size=8mm},gUSpnode/.style={circle,blue,draw,minimum size=8mm},fnode/.style={rectangle,draw,minimum size=8mm}]    
\node at (-3.7,4.2) {$\CT_3:$};
\node[fnode,red] (F1) at (0,0) {$1$};
\node[gUSpnode] (G1) at (0,2.4) {$2N-2$};
\node[fnode,orange] (F2) at (-2.7,-0.2) {$3$};
\node[fnode] (F3) at (2.7,-0.2) {$3$};
\node[fnode,blue] (F4) at (-1,-2.2) {$1$};
\node[fnode,violet] (F5) at (1,-2.2) {$1$};
\draw (F1) -- pic[pos=0.6,sloped]{arrow} (F3);
\draw (F1) -- (F4);
\draw (F1) -- (F5);
\draw (F2) -- pic[sloped,very thick]{arrow=latex reversed} (F4);
\draw (F3) -- pic[sloped,very thick]{arrow=latex reversed} (F5);
\draw (F4) -- (F5);
\draw (-2.9,-0.6) to[out=-90,in=180] pic[pos=1,sloped,very thick]{arrow=latex reversed} (0,-3.5) to[out=0,in=-90] (2.9,-0.6);
\draw (-2.7,0.2) to[out=90,in=180] pic[pos=0.4,sloped,very thick]{arrow=latex reversed} (-0.9,2.4);
\draw (2.7,0.2) to[out=90,in=0] pic[pos=0.3,sloped,very thick]{arrow=latex reversed} (0.9,2.4);
\draw (-3.1,0.1) to[out=180,in=90] pic[pos=0.2,sloped]{arrow} (-3.6,-0.2) to[out=-90,in=180] pic[pos=0.8,sloped]{arrow} (-3.1,-0.5);
\draw (-0.5,3.1) to[out=90,in=180] (0,3.7) to[out=0,in=90] (0.5,3.1);
\node[right] at (3.8,1.6) {$\CW= \e_3 \, \e^3 \, \e_{2N-2} \left[B^{N-1} M_0^3 T_N \right. $};
\node[right] at (3.8,0.6) {$+ B^{N-1} M_0^2 \Bt_1 B_1 + B^{N-2} L R M_0^2 T_N $};
\node[right] at (3.8,-0.4) {$+ B^{N-2} L R M_0 \Bt_1 B_1 + B^{N-2} \Ht_0 R^2 M_0 T_N$};
\node[right] at (3.8,-1.4) {$+ B^{N-3} L^2 R^2 M_0 T_N + B^{N-2} \Ht_0 R^2 B_1 \Bt_1 $};
\node[right] at (3.8,-2.4) {$+ B^{N-3} L^2 R^2 B_1 \Bt_1 + B^{N-3} \Ht_0 L R^3 T_N$};
\node[right] at (3.8,-3.4) {$\left. + B^{N-4} L^3 R^3 T_N \right] + T_N B_3 \Bt_3 + B_1 H_N B_3$};
\node[right] at (0.4,-4.4) {$+ \e_3 \, \e_{2N-2} \left[ B^{N-1} \Ht_0 \Bt_1 \Bt_3 + B^{N-1} \Ht_0 M_0 H_N + B^{N-2} L^2 \Bt_1 \Bt_3 \right.$};
\node[right] at (1.8,-5.4) {$\left.+ B^{N-2} L^2 M_0 H_N + B^{N-2} L \Ht_0 R H_N + B^{N-3} L^3 R H_N \right]$};
\node at (0.8,3.6) {$B$};
\node at (-3.9,-0.2) {$\Ht_0$};
\node at (-2.1,1.3) {$L$};
\node at (2,1.3) {$R$};
\node at (2.2,-1.4) {$B_1$};
\node at (0,-2.5) {$T_N$};
\node at (-2.1,-1.4) {$\Bt_1$};
\node at (1.4,0.2) {$H_N$};
\node at (0.8,-1) {$B_3$};
\node at (-0.8,-1) {$\Bt_3$};
\node at (0.4,-3.2) {$M_0$};
\epic} \ee
Let's explain how to get the superpotential in \eqref{SU2N+1antisym'T3}. The first 7 terms in \eqref{SU2N+1antisym'T2} are mapped to the first 10 terms in \eqref{SU2N+1antisym'T3}. The additional 3 terms come from
\begin{itemize}
\item \scalebox{0.95}{$\e^{2N-2} \, \e_{2N-2} \, \left( K_2^{2N-4} K_1^2 R^2 M_0 T_N \right) \longrightarrow \e_{2N-2} \, \left( B^{N-2} \Ht_0 R^2 M_0 T_N + B^{N-3} L^2 R^2 M_0 T_N \right)$}
\item \scalebox{0.95}{$ \e^{2N-2} \, \e_{2N-2} \, \left( K_2^{2N-4} K_1^2 R^2 B_1 \Bt_1 \right) \longrightarrow \e_{2N-2} \, \left( B^{N-2} \Ht_0 R^2 B_1 \Bt_1 + B^{N-3} L^2 R^2 B_1 \Bt_1 \right)$}
\item \scalebox{0.95}{$\e^{2N-2} \, \e_{2N-2} \, \left(K_2^{2N-5} K_1^3 R^3 T_N \right) \longrightarrow \e_{2N-2} \, \left( B^{N-3} \Ht_0 L R^3 T_N + B^{N-4} L^3 R^3 T_N \right)$}
\end{itemize}
Then $X$ and $\Yt$ of \eqref{SU2N+1antisym'T2} get massive from the planar triangles terms. Indeed we obtain the following mass terms: $P_1 X$ and $P_2 \Yt$ where $P_1$ is the meson $[K_1 Y]$ and $P_2$ the meson $[K_2 Y]$ in the frame $\CT_2$. Therefore after integrating out massive fields, we are left with only two \say{triangles terms}: $T_N B_3 \Bt_3 + B_1 H_N B_3$.\\
The remaining terms come from the Pfaffian confining superpotential \eqref{UspbuildingBlock}
\be 
\Pf \phi \equiv \Pf
\begin{pmatrix}
B & L & P_2 \\
& \Ht_0 & P_1 \\
& & 0
\end{pmatrix}
\sim \e_{a_1 \dots a_{2N+2}} \, \phi^{a_1 a_2} \cdots \phi^{a_{2N+1} a_{2N+2}} \nn
\ee
\be 
\longrightarrow \, \e_3 \, \e_{2N-2} \, \left[B^{N-1} \Ht_0 P_1 + B^{N-2} (L^2 P_1 + L \Ht_0 P_2) + B^{N-3} L^3 P_2 \right] 
\ee

As we said $P_1$ and $P_2$ are massive fields and their expression are obtained by the E.O.M of $X$ and $\Yt$. We get (after rescaling fields)
\begin{align*}
\text{E.O.M: \, from \,} &X: P_1 = \Bt_1 \Bt_3 + M_0 H_N \\
\text{from \,} &\Yt:  P_2 = R H_N
\end{align*}  
Putting all together, we get the superpotential written in \eqref{SU2N+1antisym'T3}.\\
The next step is to confine the $Usp(2N-2)$ node using the result \eqref{UspTDec}. In order to apply it and get a superpotential in a close form, we need to flip the tower of traces of the antisymmetric field $B$. Using the mapping \eqref{mapSU2N+1antisym'1}, it amounts to flip in the frame $\CT_1$ the tower of $(A \, \At)^j$. To simplify even further we also flip the two singlets $\e_{2N+1} \, (A^{N-1} \, Q^3)$ and $\e_{2N+1} \, (\At^{N-1} \, \Qt^3)$. Therefore we kill terms with $T_N, B_3$ and $\Bt_3$.
\be \label{mapSU2N+1antisym'1}
\scalebox{0.74}{$
\ba{c} \CT_1 \\
Q \, \Qt \\
Q \, (A \, \At)^j \, \Qt \\
\At \, (A \, \At)^i \, Q^2 \\
\At \, (A \, \At)^{N-1} \, Q^2 \\
A \, \Qt^2 \\
A \, (A \, \At)^k \, \Qt^2 \\
A \, (A \, \At)^{N-1} \, \Qt^2 \\
(A \, \At)^m  \\
(A \, \At)^N \\
\e_{2N+1} \, (A^N \, Q) \\
\e_{2N+1} \, (\At^N \, \Qt) \\
\e_{2N+1} \, (A^{N-1} \, Q^3) \\
\e_{2N+1} \, (\At^{N-1} \, \Qt^3)
\ea
\Longleftrightarrow
\ba{c} \CT_{1'} \\
Q \, \Qt \\
Q \, (b b \, \bt \bt)^j \, \Qt  \\
\bt \bt \, (b b \, \bt \bt)^i \, Q^2  \\
\e_{2N+1} ( (b b)^{N-1} c \, Q^2) \\
b b \, \Qt^2 \\
b b \, (b b \, \bt \bt)^k \, \Qt^2 \\
\e_{2N+1} ((\bt \bt)^{N-1} \ct \, \Qt^2) \\
(b b \, \bt \bt)^m \\
c \, \ct \\
\ct \, Q \\
c \, \Qt \\
\e_{2N+1} \, ((b b)^{N-1} \, Q^3) \\
\e_{2N+1} \, ((\bt \bt)^{N-1} \, \Qt^3)
\ea
\Longleftrightarrow 
\ba{c}\CT_2 \\
M_0 \\
R K_2 (K_2 K_2)^{j-1} K_1 \\
R (K_2 K_2)^i R \\
H_N \\
K_1 K_1 \\
K_1 (K_2 K_2)^k K_1 \\
X \\
(K_2 K_2)^m \\
T_N \\
B_1 \\
\Bt_1 \\
B_3 \\
\Bt_3  
\ea
\Longleftrightarrow 
\ba{c}\CT_3 \\
M_0 \\
L \, B^{j-1} \, R \\
R \, B^{i} \, R \\
H_N  \\
\Ht_0 \\
L \, B^{k-1} \, L \\
\Ht_0 B^{N-1} + L B^{N-2} L \\
B^m \\
T_N \\
B_1 \\
\Bt_1 \\
B_3 \\
\Bt_3 
\ea
\quad 
\ba{l} 
\\
j=1,\dots,N-1 \\
i=0,\dots,N-2 \\
\\
\\
k=1,\dots,N-2 \\
\\
m=1,\dots,N-1 \\
\\
\\
\\
\\
\ea
$}
\ee
The summary of this flipping procedure is the following duality 
\be \label{SU2N+1antisym'T1Flip} \scalebox{0.9}{\bpic[node distance=2cm,gSUnode/.style={circle,red,draw,minimum size=8mm},gUSpnode/.style={circle,blue,draw,minimum size=8mm},fnode/.style={rectangle,draw,minimum size=8mm}]   
\node at (0.5,1.8) {$\CT_{1,flip}:$};
\node[gSUnode] (G1) at (3,0) {$2N+1$};
\node[fnode,orange] (F1) at (4.3,-2.3) {$3$};
\node[fnode] (F2) at (1.7,-2.3) {$3$};
\draw (G1) -- pic[pos=0.6,sloped]{arrow} (F1);
\draw (G1) -- pic[pos=0.4,sloped]{arrow} (F2);
\draw (2.2,0.3) to[out=135,in=-90] pic[pos=0.1,sloped]{arrow} (2,0.8) to[out=90,in=180] (2.3,1.3) to[out=0,in=90] pic[pos=0.9,sloped]{arrow} (2.8,0.8);  
\draw (3.8,0.3) to[out=45,in=-90] pic[pos=1,sloped,very thick]{arrow} (4,0.8) to[out=90,in=0] (3.7,1.3) to[out=180,in=90] pic[pos=0.4,sloped,very thick]{arrow} (3.2,0.8);
\node[right] at (0.2,-3.5) {$ \CW= \sum_{i=1}^{N} \, \a_i \, \tr(A \, \At)^i +$};
\node[right] at (0.2,-4.5) {$+ \b \left(\e_{2N+1} \, \e^3 \, [A^{N-1} \, Q^3] \right) +$};
\node[right] at (0.2,-5.5) {$+ \tilde \b \left(\e^{2N+1} \, \e_3 \, [\At^{N-1} \, \Qt^3] \right)$};
\node at (1.9,1.5) {$A$};
\node at (4.1,1.5) {$\At$};
\node at (4.1,-1.1) {$\Qt$};
\node at (1.9,-1.1) {$Q$};
\node at (6,0) {$\llra$};
\node at (9,1.8) {$\CT_{3,flip}:$};
\node[fnode,red] (F3) at (11.5,-2) {$1$};
\node[gUSpnode] (G2) at (11.5,0) {$2N-2$};
\node[fnode,orange] (F4) at (-2.7+11.5,-2.6) {$3$};
\node[fnode] (F5) at (2.7+11.5,-2.6) {$3$};
\node[fnode,blue] (F6) at (10.5,-4.2) {$1$};
\node[fnode,violet] (F7) at (12.5,-4.2) {$1$};
\draw (F3) -- pic[pos=0.6,sloped]{arrow} (F5);
\draw (F4) -- pic[sloped,very thick]{arrow=latex reversed} (F6);
\draw (F5) -- pic[sloped,very thick]{arrow=latex reversed} (F7);
\draw (-2.9+11.5,-3) to[out=-90,in=180] pic[pos=1,sloped,very thick]{arrow=latex reversed} (11.5,-5.5) to[out=0,in=-90] (2.9+11.5,-3);
\draw (-2.7+11.5,-2.2) to[out=90,in=180] pic[pos=0.4,sloped,very thick]{arrow=latex reversed} (-0.9+11.5,0);
\draw (2.7+11.5,-2.2) to[out=90,in=0] pic[pos=0.3,sloped,very thick]{arrow=latex reversed} (0.9+11.5,0);
\draw (-3.1+11.5,-2.3) to[out=180,in=90] pic[pos=0.2,sloped]{arrow} (-3.6+11.5,-2.6) to[out=-90,in=180] pic[pos=0.8,sloped]{arrow} (-3.1+11.5,-2.9);
\draw (11,0.7) to[out=90,in=180] (11.5,1.4) to[out=0,in=90] (12,0.7);
\node[right] at (4.8,-6.5) {$\CW= \e_3 \, \e^3 \, \e_{2N-2} \left[B^{N-1} M_0^2 \Bt_1 B_1 + B^{N-2} L R M_0 \Bt_1 B_1 \right. $};
\node[right] at (4.6,-7.5) {$+ B^{N-2} \Ht_0 R^2 B_1 \Bt_1 + B^{N-3} L^2 R^2 B_1 \Bt_1 + B^{N-1} \Ht_0 M_0 H_N$};
\node[right] at (2.1,-8.5) {$\left. + B^{N-2} L^2 M_0 H_N + B^{N-2} L \Ht_0 R H_N + B^{N-3} L^3 R H_N \right] + \sum_{i=2}^{N-1} \, \a_i \tr(B^i)$};
\node at (12.3,1.2) {$B$};
\node at (-3.9+11.5,-2.2) {$\Ht_0$};
\node at (-2.1+11.5,-1.1) {$L$};
\node at (2+11.5,-1.1) {$R$};
\node at (12.8,-3.2) {$B_1$};
\node at (10.2,-3.2) {$\Bt_1$};
\node at (1.4+11.5,-1.9) {$H_N$};
\node at (11.5,-5.1) {$M_0$};
\epic} \ee
We can now use our confining result of Section \ref{Usp2N}. We rewrite \eqref{UspTDec} after splitting the 6 fundamentals into two groups of 3, getting
\be \label{UspTDecSplit} \bpic[node distance=2cm,gSUnode/.style={circle,red,draw,minimum size=8mm},gUSpnode/.style={circle,blue,draw,minimum size=8mm},fnode/.style={rectangle,draw,minimum size=8mm}]   
\node[gUSpnode] (G1) at (3,0) {$2M$};
\node[fnode] (F1) at (4,-2) {$3$};
\node[fnode,orange] (F2) at (2,-2) {$3$};
\draw (G1) -- pic[pos=0.6,sloped,very thick]{arrow=latex reversed} (F1);
\draw (G1) -- pic[pos=0.5,sloped,very thick]{arrow=latex reversed} (F2);
\draw (2.7,0.4) to[out=90,in=180] (3,0.9) to[out=0,in=90] (3.3,0.4);
\node[right] at (1.2,-3.2) {$ \CW= \sum_{i=2}^{M} \a_i \, \tr(B^i)$};
\node at (3.5,1.1) {$B$};
\node at (4,-0.9) {$R$};
\node at (2,-0.9) {$L$};
\node at (6,0) {$\Llra$};
\node[fnode] (F3) at (9,0) {$3$};
\node[fnode,orange] (F4) at (12,0) {$3$};
\draw (9.4,-0.3) -- pic[pos=0.6,sloped]{arrow} (11.6,-0.3);
\node at (10.5,0.1) {$\vdots$};
\draw (9.4,0.3) -- pic[pos=0.6,sloped]{arrow} (11.6,0.3);
\draw (8.6,0.2) to[out=180,in=90] pic[pos=0.2,sloped,very thick]{arrow=latex reversed} (8.2,0) to[out=-90,in=180] pic[pos=0.8,sloped,very thick]{arrow=latex reversed} (8.6,-0.2);
\node at (7.9,0) {$\dots$};
\draw (8.6,0.4) to[out=180,in=90] pic[pos=0.2,sloped,very thick]{arrow=latex reversed} (7.5,0) to[out=-90,in=180] pic[pos=0.8,sloped,very thick]{arrow=latex reversed} (8.6,-0.4);
\draw (12.4,0.2) to[out=0,in=90] pic[pos=0.4,sloped,very thick]{arrow=latex reversed} (12.8,0) to[out=-90,in=0] pic[pos=0.6,sloped,very thick]{arrow=latex reversed} (12.4,-0.2);
\node at (13.2,0) {$\dots$};
\draw (12.4,0.4) to[out=0,in=90] pic[pos=0.6,sloped,very thick]{arrow=latex reversed} (13.5,0) to[out=-90,in=0] pic[pos=0.4,sloped,very thick]{arrow=latex reversed} (12.4,-0.4);
\node[right] at (6.2,-1.7) {$ \CW= \sum_{i,j,k=1}^{M} \, \e_3 \, \e^3 \, \left(\Ht_i \, M_j \, H_k + M_i \, M_j \, M_k \right)\, \de_{i+j+k,2M+1}$};
\node at (10.5,-0.7) {$M_1$};
\node at (10.5,0.7) {$M_M$};
\node at (7.8,-0.8) {$H_1, \dots,H_M$};
\node at (13.2,-0.8) {$\Ht_1, \dots,\Ht_M$};
\node at (10,-3) {$\tr(L \, B^k \, L) \llra \Ht_{k+1}$};
\node at (9.7,-3.7) {$\tr(R \, B^k \, R) \llra H_{k+1}$};
\node at (9.9,-4.4) {$\tr(L \, B^k \, R) \llra M_{k+1}$};
\node at (14,-3.7) {$k=0,\dots,M-1$};
\epic \ee
Apply this result to our frame $\CT_{3,flip}$ in \eqref{SU2N+1antisym'T1Flip} we get the final WZ
\be \label{SU2N+1antisym'T4Flip} \bpic[node distance=2cm,gSUnode/.style={circle,red,draw,minimum size=8mm},gUSpnode/.style={circle,blue,draw,minimum size=8mm},fnode/.style={rectangle,draw,minimum size=8mm}]     
\node at (5.5,2.5) {$\CT_{4,flip}:$};
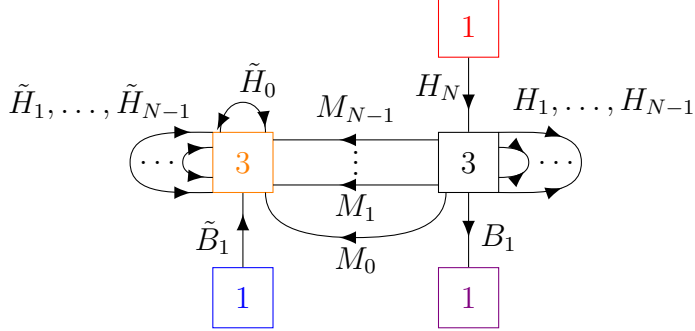
\node[fnode,orange] (F1) at (9,0) {$3$};
\node[fnode] (F2) at (12,0) {$3$};
\node[fnode,blue] (F3) at (9,-1.8) {$1$};
\node[fnode,violet] (F4) at (12,-1.8) {$1$};
\node[fnode,red] (F5) at (12,1.8) {$1$};
\draw (F1) -- pic[sloped,very thick]{arrow=latex reversed} (F3);
\draw (F2) -- pic[pos=0.6,sloped,very thick]{arrow} (F4);
\draw (F2) -- pic[sloped,very thick]{arrow=latex reversed} (F5);
\draw (9.4,-0.3) -- pic[pos=0.5,sloped,very thick]{arrow=latex reversed} (11.6,-0.3);
\node at (10.5,0.1) {$\vdots$};
\draw (9.4,0.3) -- pic[pos=0.5,sloped,very thick]{arrow=latex reversed} (11.6,0.3);
\draw (8.6,0.2) to[out=180,in=90] pic[pos=0.2,sloped,very thick]{arrow} (8.2,0) to[out=-90,in=180] pic[pos=0.8,sloped,very thick]{arrow} (8.6,-0.2);
\node at (7.9,0) {$\dots$};
\draw (8.6,0.4) to[out=180,in=90] pic[pos=0.2,sloped,very thick]{arrow} (7.5,0) to[out=-90,in=180] pic[pos=0.8,sloped,very thick]{arrow} (8.6,-0.4);
\draw (12.4,0.2) to[out=0,in=90] pic[pos=0.7,sloped,very thick]{arrow} (12.8,0) to[out=-90,in=0] pic[pos=0.3,sloped,very thick]{arrow} (12.4,-0.2);
\node at (13.2,0) {$\dots$};
\draw (12.4,0.4) to[out=0,in=90] pic[pos=0.6,sloped,very thick]{arrow} (13.5,0) to[out=-90,in=0] pic[pos=0.4,sloped,very thick]{arrow} (12.4,-0.4);
\draw (9.3,0.4) to[out=90,in=0] pic[pos=0.1,sloped]{arrow} (9,0.8) to[out=180,in=90] pic[pos=0.6,sloped,very thick]{arrow=latex reversed} (8.7,0.4);
\draw (9.3,-0.4) to[out=-90,in=180] pic[pos=1,sloped,very thick]{arrow=latex reversed} (10.5,-1) to[out=0,in=-90] (11.7,-0.4);
\node[right] at (4.5,-3) {$\CW=\e_3 \, \e^3 \left[ M_{N-1} M_0 \Bt_1 B_1 + \sum_{l=0}^{N-3} \, \left( \Ht_{l+1} H_{N-2-l} B_1 \Bt_1 +M_{l+1} M_{N-2-l} B_1 \Bt_1 \right) \right.$};
\node[right] at (5.3,-4) {$+ H_{N-1} \Ht_0 B_1 \Bt_1 + \Ht_{N-1} M_0 H_N + \Ht_0 M_{N-1} H_N + \sum_{l=0}^{N-3} \, \left( \Ht_{l+1} M_{N-2-l} H_N \right)$};
\node[right] at (5.3,-5) {$\left. + \sum_{i,j,k=1}^{N-1} \, \left(\Ht_i \, M_j \, H_k + M_i \, M_j \, M_k \right) \, \de_{i+j+k,2N-1} \right]$};
\node at (10.5,-1.3) {$M_0$};
\node at (10.5,-0.6) {$M_1$};
\node at (10.5,0.7) {$M_{N-1}$};
\node at (7.1,0.8) {$\Ht_1, \dots,\Ht_{N-1}$};
\node at (13.8,0.8) {$H_1, \dots,H_{N-1}$};
\node at (12.4,-1) {$B_1$};
\node at (8.6,-1) {$\Bt_1$};
\node at (9.2,1.1) {$\Ht_0$};
\node at (11.6,1) {$H_N$};
\epic \ee 
As a consistency check, for $N=2$ we recover the superpotential given in \cite{Csaki:1996zb} Section 3.1.4. The mapping of the chiral ring in the flipping case is
\be
\ba{c}\CT_{1,flip} \\
Q \, \Qt \\
Q \, (A \, \At)^j \, \Qt \\
\At \, (A \, \At)^i \, Q^2 \\
\At \, (A \, \At)^{N-1} \, Q^2 \\
A \, \Qt^2 \\
A \, (A \, \At)^k \, \Qt^2 \\
A \, (A \, \At)^{N-1} \, \Qt^2 \\
\e_{2N+1} \, (A^N \, Q) \\
\e_{2N+1} \, (\At^N \, \Qt)
\ea
\Longleftrightarrow
\ba{c} \CT_{3,flip} \\
M_0 \\
L \, B^{j-1} \, R \\
R \, B^{i} \, R \\
H_N  \\
\Ht_0 \\
L \, B^{k-1} \, L \\
L B^{N-2} L \\
B_1 \\
\Bt_1
\ea
\Longleftrightarrow 
\ba{c}\CT_{4,flip} \\
M_0 \\
M_j \\
H_{i+1} \\
H_N \\
\Ht_0 \\
\Ht_k \\
\Ht_{N-1} \\
B_1 \\
\Bt_1  
\ea
\quad \quad
\ba{l} \\
\\
j=1,\dots,N-1 \\
i=0,\dots,N-2 \\
\\
\\
k=1,\dots,N-2 \\
\\
\\
\\  

\ea
\ee

\subsection{Even rank: $M = 2N$}
Let us now study the even rank case of the previous gauge theory.
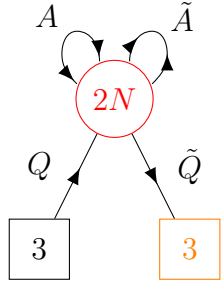
\be \label{SU2Nantisym'T1} \bpic[node distance=2cm,gSUnode/.style={circle,red,draw,minimum size=8mm},gUSpnode/.style={circle,blue,draw,minimum size=8mm},fnode/.style={rectangle,draw,minimum size=8mm}]   
\node at (3.5,1.7) {$\CT_1:$};
\node[gSUnode] (G1) at (6,0) {$2N$};
\node[fnode,orange] (F1) at (7,-2) {$3$};
\node[fnode] (F2) at (5,-2) {$3$};
\draw (G1) -- pic[pos=0.6,sloped]{arrow} (F1);
\draw (G1) -- pic[pos=0.4,sloped]{arrow} (F2);
\draw (5.5,0.2) to[out=135,in=-90] pic[pos=0.1,sloped]{arrow} (5.3,0.6) to[out=90,in=180] (5.5,0.9) to[out=0,in=90] pic[pos=0.9,sloped]{arrow} (5.8,0.5);  
\draw (6.5,0.2) to[out=45,in=-90] pic[pos=1,sloped,very thick]{arrow} (6.7,0.6) to[out=90,in=0] (6.5,0.9) to[out=180,in=90] pic[pos=0.4,sloped,very thick]{arrow} (6.2,0.5);
\node[right] at (8,0) {$ \CW= 0$};
\node at (5.1,1.1) {$A$};
\node at (6.9,1.1) {$\At$};
\node at (7,-0.9) {$\Qt$};
\node at (5,-0.9) {$Q$};
\epic \ee
The chiral ring generators are
\begin{itemize}
\item \scalebox{0.95}{$Q (A \, \At)^k \Qt \sim Q_i^{\a_1} (\At_{\a_1 \a_2} \, A^{\a_2 \a_3} \, \At_{\a_3 \a_4} \cdots A^{\a_{2k} \a_{2k+1}}) \, \Qt_{\a_{2k+1}}^I \, ,$ $k=0,\dots, N-1 \,$} 

(transforming in the $(\, \overline{\scalebox{0.5}{\ydiagram{1}}} \, ,\scalebox{0.5}{\ydiagram{1}} \,)$ of $SU(3)_Q \times SU(3)_{\Qt}$)
\item \scalebox{0.95}{$\At \, (A \, \At)^k \, Q^2 \sim \At_{\a_1 \a_2}(A^{\a_2 \a_3} \, \At_{\a_3 \a_4} \, A^{\a_4 \a_5} \cdots A^{\a_{2k+1} \a_{2k+2}} \, \At_{\a_{2k+2} \a_{2k+3}}) \, Q_{[i}^{\a_{2k+3}} Q_{j]}^{\a_1} \, ,$}

$k=0,\dots, N-1$ $ \sim (\, \overline{\scalebox{0.5}{\ydiagram{1,1}}} \, ,1)$ 
\item \scalebox{0.95}{$A \, (A \, \At)^k \, \Qt^2 \sim A^{\a_1 \a_2}(\At_{\a_2 \a_3} \, A^{\a_3 \a_4} \, \At_{\a_4 \a_5} \cdots \At_{\a_{2k+1} \a_{2k+2}} \, A^{\a_{2k+2} \a_{2k+3}}) \, \Qt_{\a_{2k+3}}^{[I} \Qt_{\a_1}^{J]} \, ,$}

$k=0,\dots, N-1$ $ \sim (1, \scalebox{0.5}{\ydiagram{1,1}} \,)$
\item \scalebox{0.95}{$(A \, \At)^m \sim (A^{\a_1 \a_2} \, \At_{\a_2 \a_3} \cdots A^{\a_{2m-1} \a_{2m}} \, \At_{\a_{2m} \a_{1}}) \, ,$ $n=1,\dots,N-1$ $\sim (1,1)$}  
\item \scalebox{0.95}{$\e_{2N} \, A^N \sim \e_{2N} \, (A^{\a_1 \a_2} \cdots A^{\a_{2N-1} \a_{2N}})$ $\sim (1,1)$}
\item \scalebox{0.95}{$\e_{2N} \, \At^N \sim \e_{2N} \, (\At_{\a_1 \a_2} \cdots \At_{\a_{2N-1} \a_{2N}})$ $\sim (1,1)$}
\item \scalebox{0.95}{$\e_{2N} \, (A^{N-1} \, Q^2) \sim \e_{2N} \, (A^{\a_1 \a_2} \cdots A^{\a_{2N-3} \a_{2N-2}} \, Q_{[i}^{\a_{2N-1}} \, Q_{j]}^{\a_{2N}})$ $\sim (\, \overline{\scalebox{0.5}{\ydiagram{1,1}}} \, ,1)$}
\item \scalebox{0.95}{$\e_{2N} \, (\At^{N-1} \, \Qt^2) \sim \e_{2N} \, (\At_{\a_1 \a_2} \cdots \At_{\a_{2N-3} \a_{2N-2}}) \, \Qt_{\a_{2N-1}}^{[I} \, \Qt_{\a_{2N}}^{J]}$ $\sim (1, , \scalebox{0.5}{\ydiagram{1,1}} \,)$}
\end{itemize}	
The next step is to deconfine the antisymmetric and the conjugate antisymmetric fields. To do so we use a variant of the deconfinement method \eqref{SU2NDeconfinement}, where we don't have to split the flavor symmetry. By doing so we don't have to split the chiral ring generators and it is then easier.  
\be \label{SU2Nantisym'T1'} \bpic[node distance=2cm,gSUnode/.style={circle,red,draw,minimum size=8mm},gUSpnode/.style={circle,blue,draw,minimum size=8mm},fnode/.style={rectangle,draw,minimum size=8mm}]  
\node at (1,1) {$\CT_{1'}:$};
\node[gSUnode] (G1) at (6,0) {$2N$};
\node[gUSpnode] (G2) at (3,0) {$2N$};
\node[gUSpnode] (G3) at (9,0) {$2N$};
\node[fnode,blue] (F1) at (7,-2) {$1$};
\node[fnode,violet] (F2) at (5,-2) {$1$};
\node[fnode] (F3) at (3,-2) {$3$};
\node[fnode,orange] (F4) at (9,-2) {$3$};
\draw (G1) -- pic[pos=0.3,sloped]{arrow} (G2);
\draw (G1) -- pic[pos=0.4,sloped]{arrow} (G3);
\draw (G1) -- pic[pos=0.4,sloped,very thick]{arrow= latex reversed} (F1);
\draw (G1) -- pic[pos=0.3,sloped,very thick]{arrow= latex reversed} (F2);
\draw (G2) -- pic[pos=0.7,sloped,very thick]{arrow= latex reversed} (F3);
\draw (G2) -- (F2);
\draw (F2) -- pic[pos=0.7,sloped,very thick]{arrow= latex reversed} (F3);
\draw (G3) -- pic[pos=0.8,sloped]{arrow} (F4);
\draw (G3) -- (F1);
\draw (F1) -- pic[pos=0.8,sloped,very thick]{arrow= latex reversed} (F4);
\draw (2.6,-1.7) to[out=180,in=90] pic[pos=0.2,sloped]{arrow} (2.2,-2) to[out=-90,in=180] pic[pos=0.8,sloped]{arrow} (2.6,-2.3);
\draw (9.4,-1.7) to[out=0,in=90] pic[pos=0.7,sloped]{arrow} (9.8,-2) to[out=-90,in=0] pic[pos=0.3,sloped]{arrow} (9.4,-2.3);
\node[right] at (10.7,-0.8) {$ \CW= 4 \,$ Planar Triangles};
\node[right] at (11.2,-1.6) {$+ n \, K_1 K_1 + \nt \, R R$};
\node at (4.5,0.4) {$b$};
\node at (7.5,0.4) {$\bt$};
\node at (2.6,-1) {$K_1$};
\node at (9.4,-1) {$R$};
\node at (6.8,-0.9) {$c$};
\node at (5.2,-0.9) {$\ct$};
\node at (4.3,-0.9) {$d$};
\node at (7.7,-0.8) {$\dt$};
\node at (4,-2.3) {$B_2$};
\node at (8,-2.3) {$\Bt_2$};
\node at (1.9,-2) {$n$};
\node at (10.1,-2) {$\nt$};
\epic \ee
Then we confine the $SU$ node with \eqref{SUbuildingBlock}.
\be \label{SU2Nantisym'T2} \scalebox{0.9}{\bpic[node distance=2cm,gSUnode/.style={circle,red,draw,minimum size=8mm},gUSpnode/.style={circle,blue,draw,minimum size=8mm},fnode/.style={rectangle,draw,minimum size=8mm}]    
\node at (-4,3) {$\CT_2:$};
\node[fnode,red] (F1) at (0,0) {$1$};
\node[gUSpnode] (G1) at (-1.8,2) {$2N$};
\node[gUSpnode] (G2) at (1.8,2) {$2N$};
\node[fnode] (F2) at (-2.7,-0.2) {$3$};
\node[fnode,orange] (F3) at (2.7,-0.2) {$3$};
\node[fnode,violet] (F4) at (-1,-2.2) {$1$};
\node[fnode,blue] (F5) at (1,-2.2) {$1$};
\draw (G1) -- (F1);
\draw (G1) -- (G2);
\draw (G1) -- pic[pos=0.5,sloped]{arrow} (F2);
\draw (G2) -- (F1);
\draw (G2) -- pic[pos=0.7,sloped]{arrow} (F3);
\draw (F1) -- (F4);
\draw (F1) -- (F5);
\draw (F2) -- pic[sloped,very thick]{arrow=latex reversed} (F4);
\draw (F3) -- pic[sloped,very thick]{arrow=latex reversed} (F5);
\draw (F4) -- (F5);
\draw (-3.1,0.1) to[out=180,in=90] pic[pos=0.2,sloped]{arrow} (-3.5,-0.2) to[out=-90,in=180] pic[pos=0.8,sloped]{arrow} (-3.1,-0.5);
\draw (3.1,0.1) to[out=0,in=90] pic[pos=0.7,sloped]{arrow} (3.5,-0.2) to[out=-90,in=0] pic[pos=0.3,sloped]{arrow} (3.1,-0.5);
\node[right] at (4,2) {$\CW= n \, K_1 K_1 + \, \nt \, R R $};
\node[right] at (4.2,1) {$+ \, \e_{2N} \, \e^{2N} \, \left[K_2^{2N-1} K_1 R B_2 \Bt_2 \right.   $};
\node[right] at (4.5,0) {$\left. +\, K_2^{2N} \a_N \right] + \, B_0 B_2 K_1 X$};
\node[right] at (4.5,-1) {$+ \, \Bt_0 \Bt_2 R \Xt + X \Xt K_2$};
\node[right] at (4.5,-2) {$+ \, \Bt_0 B_0 \a_N $};
\node at (-2.8,0.9) {$K_1$};
\node at (0,2.4) {$K_2$};
\node at (2.8,0.9) {$R$};
\node at (2.2,-1.4) {$\Bt_2$};
\node at (0,-2.5) {$\a_N$};
\node at (-2.1,-1.4) {$B_2$};
\node at (-1.1,0.8) {$X$};
\node at (1.2,0.8) {$\Xt$};
\node at (0.9,-1) {$\Bt_0$};
\node at (-0.9,-1) {$B_0$};
\node at (-3.7,-0.6) {$n$};
\node at (3.7,-0.6) {$\nt$};
\epic} \ee 
Once again some terms in the superpotential in \eqref{SU2Nantisym'T2} come from the determinant of the $(2N+1)\times(2N+1)$ meson matrix $\P$
\be 
\det \P \equiv \det
\begin{pmatrix}
K_2 & (b \ct) \\
(c \bt) & \a_N 
\end{pmatrix}
= \e^{a_1 \dots a_{2N+1}} \, \e_{b_1 \dots b_{2N+1}} \, \P_{a_1}^{b_1} \cdots \P_{a_{2N+1}}^{b_{2N+1}} \nn
\ee
\be 
\longrightarrow \e^{2N} \, \e_{2N} \, \left[K_2^{2N-1} \, (b \ct) \, (c \bt) + K_2^{2N} \a_N \right]
\ee
$(b \ct)$ and $(c \bt)$ are massive fields and their expressions are obtained by the E.O.M of $d$ and $\dt$ from $\CT_{1'}$ 
\begin{align*}
\text{E.O.M: \, from \,} d:  (b \ct) &= K_1 B_2 \\
\text{from \,} \dt: (c \bt) &= R \Bt_2
\end{align*} 
Then we confine the left $Usp(2N)$ node with \eqref{UspbuildingBlock} to get
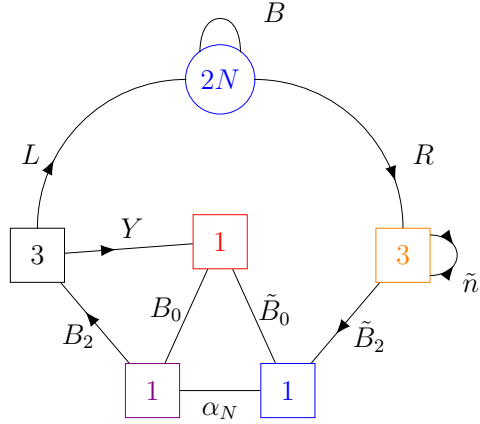
\be \label{SU2Nantisym'T3} \scalebox{0.9}{\bpic[node distance=2cm,gSUnode/.style={circle,red,draw,minimum size=8mm},gUSpnode/.style={circle,blue,draw,minimum size=8mm},fnode/.style={rectangle,draw,minimum size=8mm}]     
\node at (-3.5,4) {$\CT_3:$};
\node[fnode,red] (F1) at (0,0) {$1$};
\node[gUSpnode] (G1) at (0,2.4) {$2N$};
\node[fnode] (F2) at (-2.7,-0.2) {$3$};
\node[fnode,orange] (F3) at (2.7,-0.2) {$3$};
\node[fnode,violet] (F4) at (-1,-2.2) {$1$};
\node[fnode,blue] (F5) at (1,-2.2) {$1$};
\draw (F1) -- pic[pos=0.6,sloped]{arrow} (F2);
\draw (F1) -- (F4);
\draw (F1) -- (F5);
\draw (F2) -- pic[sloped,very thick]{arrow=latex reversed} (F4);
\draw (F3) -- pic[sloped,very thick]{arrow=latex reversed} (F5);
\draw (F4) -- (F5);
\draw (-2.7,0.2) to[out=90,in=180] pic[pos=0.3,sloped,very thick]{arrow} (-0.5,2.4);
\draw (2.7,0.2) to[out=90,in=0] pic[pos=0.2,sloped,very thick]{arrow} (0.5,2.4);
\draw (3.1,0.1) to[out=0,in=90] pic[pos=0.7,sloped]{arrow} (3.5,-0.2) to[out=-90,in=0] pic[pos=0.3,sloped]{arrow} (3.1,-0.5);
\draw (-0.3,2.8) to[out=90,in=180] (0,3.3) to[out=0,in=90] (0.3,2.8);
\node[right] at (4,2.5) {$\CW= \nt \, R R +\,  \a_N \, \tr\left( B^{N} \right)$};
\node[right] at (4.5,1.5) {$+ \, Y B_0 B_2 + \, \Bt_0 B_0 \a_N $};
\node[right] at (4.5,0.5) {$+ \, \e_{2N} \, \left(B^{N-1} L B_2 R \Bt_2 \right)$};
\node[right] at (4.5,-0.5) {$+ \, \e_{2N} \, \e^3 \, \left[B^{N-1} L^2 Y \right.$};
\node[right] at (4.5,-1.5) {$\left. + \, B^{N-2} L^3 R \Bt_2 \Bt_0 \right] $};
\node at (0.8,3.4) {$B$};
\node at (-2.8,1.3) {$L$};
\node at (3,1.3) {$R$};
\node at (2.2,-1.4) {$\Bt_2$};
\node at (0,-2.5) {$\a_N$};
\node at (-2.1,-1.4) {$B_2$};
\node at (0.8,-1) {$\Bt_0$};
\node at (-0.8,-1) {$B_0$};
\node at (3.7,-0.6) {$\nt$};
\node at (-1.3,0.2) {$Y$};
\epic} \ee 
We notice that the fields $n$ and $\Xt$ get a mass. To obtain \eqref{SU2Nantisym'T3} from \eqref{SU2Nantisym'T2} we have to compute the Pfaffan of the following $(2N+4)\times(2N+4)$ antisymmetric matrix
\be 
\Pf \phi \equiv \Pf
\begin{pmatrix}
B  & L & (K_2 X) \\
& (K_1 K_1)  & (K_1 X) \\
& & 0
\end{pmatrix}
\sim \e_{a_1 \dots a_{2N+4}} \, \phi^{a_1 a_2} \cdots \phi^{a_{2N+3} a_{2N+4}}
\ee
\begin{align*}
\text{E.O.M: \, from \,} &n : (K_1 K_1) = 0 \\
\text{from \,} &\Xt : (K_2 X) = [R \Bt_2] \Bt_0 
\end{align*}
\be 
\longrightarrow \, \e_{2N} \, \e^3 \, \left[B^{N-1} \, L^2 \, Y + B^{N-2} \, L^3  (K_2 X) \right] 
\ee
All in all we get \eqref{SU2Nantisym'T3}. The next step is to confine the $Usp(2N-2)$ node using the result \eqref{UspTDec}. Therefore we have to flip the whole tower of $\tr(B^k)$. Notice that the last one $\tr(B^{N})$ is already flipped by $\a_N$. We also flip $\Bt_0$ and $B_0$ because we do not want the flipper $ \a_N $ to appear elsewhere in the superpotential. By doing so we kill $3$ terms in the superpotential of \eqref{SU2Nantisym'T3} : $Y B_0 B_2, \, \Bt_0 B_0 \a_N, \, \e_{2N} \, \e^3 \, (B^{N-2} L^3 R \Bt_2 \Bt_0)$.

Using the mapping \eqref{mapSU2Nantisym'1} we see that in the original frame it amounts to flip the tower of $(A \, \At)^k, k=1,\dots,N-1$ and also $\e_{2N} \, A^N$, $\e_{2N} \, \At^N$.

\be \label{mapSU2Nantisym'1}
\scalebox{0.82}{$
\ba{c}\CT_1 \\
Q \, (A \, \At)^k \, \Qt \\
(A \, \At)^n \\
\At \, (A \, \At)^m \, Q^2 \\
A \, (A \, \At)^m \, \Qt^2 \\
\e_{2N} \, A^N  \\
\e_{2N} \, \At^N  \\
\e_{2N} \, (A^{N-1} \, Q^2) \\
\e_{2N} \, (\At^{N-1} \, \Qt^2) 
\ea
\Longleftrightarrow
\ba{c} \CT_{1'} \\
K_1 b \, (b b \, \bt \bt)^k \, R \bt) \\
(b b \, \bt \bt)^n \\
\bt \bt \, (b b \, \bt \bt)^m (K_1 b)^2 \\
b b \, (b b \, \bt \bt)^m (R \bt)^2 \\
\e_{2N} \, (b b)^N \\
\e_{2N} \, (\bt \bt)^N \\
B_2 \\
\Bt_2
\ea
\Longleftrightarrow 
\ba{c}\CT_2 \\
K_1 K_2 (K_2 K_2)^{k} R \\
(K_2 K_2)^n \\
K_1 (K_2 K_2)^{m+1} K_1 \\
R (K_2 K_2)^{m+1} R \\
B_0 \\
\Bt_0 \\
B_2 \\
\Bt_2
\ea
\Longleftrightarrow 
\ba{c}\CT_3 \\
L \, B^{k} \, R \\
B^n \\
L \, B^{m} \, L \\
R \, B^{m+1} \, R \\
B_0 \\
\Bt_0 \\
B_2 \\
\Bt_2 
\ea
\quad \quad
\ba{c}\\
k=0,\dots,N-1 \\
n=1,\dots,N-1 \\
m=0,\dots,N-2 \\
m=0,\dots,N-2 \\
\\
\\
\\
\\
    
\ea
$}
\ee
The summary of this flipping procedure is the following duality 
\be \label{SU2Nantisym'T1Flip} \scalebox{0.9}{\bpic[node distance=2cm,gSUnode/.style={circle,red,draw,minimum size=8mm},gUSpnode/.style={circle,blue,draw,minimum size=8mm},fnode/.style={rectangle,draw,minimum size=8mm}]   
\node at (0.5,1.8) {$\CT_{1,flip}:$};
\node[gSUnode] (G1) at (3,0) {$2N$};
\node[fnode,orange] (F1) at (4,-2) {$3$};
\node[fnode] (F2) at (2,-2) {$3$};
\draw (G1) -- pic[pos=0.6,sloped]{arrow} (F1);
\draw (G1) -- pic[pos=0.4,sloped]{arrow} (F2);
\draw (2.5,0.2) to[out=135,in=-90] pic[pos=0.1,sloped]{arrow} (2.3,0.6) to[out=90,in=180] (2.5,0.9) to[out=0,in=90] pic[pos=0.9,sloped]{arrow} (2.8,0.5);  
\draw (3.5,0.2) to[out=45,in=-90] pic[pos=1,sloped,very thick]{arrow} (3.7,0.6) to[out=90,in=0] (3.5,0.9) to[out=180,in=90] pic[pos=0.4,sloped,very thick]{arrow} (3.2,0.5);
\node[right] at (0.4,-3.3) {$ \CW= \sum_{i=1}^{N-1} \, \a_i \, \tr(A \, \At)^i$};
\node[right] at (0.4,-4.3) {$+ \b \left(\e_{2N} \, A^{N} \right) + \tilde \b \left(\e^{2N} \, \At^{N} \right)$};
\node at (2,1.2) {$A$};
\node at (4,1.2) {$\At$};
\node at (4,-0.9) {$\Qt$};
\node at (2,-0.9) {$Q$};
\node at (6,0) {$\llra$};
\node at (8,1.8) {$\CT_{3,flip}:$};
\node[fnode,red] (F3) at (10.6,-2.2) {$1$};
\node[gUSpnode] (G2) at (10.6,0.4) {$2N$};
\node[fnode] (F4) at (-2.7+10.6,-2.2) {$3$};
\node[fnode,orange] (F5) at (2.7+10.6,-2.2) {$3$};
\node[fnode,violet] (F6) at (9.6,-4.2) {$1$};
\node[fnode,blue] (F7) at (11.6,-4.2) {$1$};
\draw (F3) -- pic[pos=0.6,sloped]{arrow} (F4);
\draw (F4) -- pic[sloped,very thick]{arrow=latex reversed} (F6);
\draw (F5) -- pic[sloped,very thick]{arrow=latex reversed} (F7);
\draw (F6) -- (F7);
\draw (-2.7+10.6,-1.8) to[out=90,in=180] pic[pos=0.4,sloped]{arrow} (-0.9+10.9,0.4);
\draw (2.7+10.6,-1.8) to[out=90,in=0] pic[pos=0.3,sloped]{arrow} (0.9+10.3,0.4);
\draw (10.3,0.8) to[out=90,in=180] (10.6,1.3) to[out=0,in=90] (10.9,0.8);
\draw (13.7,-1.9) to[out=0,in=90] pic[pos=0.7,sloped]{arrow} (14.1,-2.2) to[out=-90,in=0] pic[pos=0.3,sloped]{arrow} (13.7,-2.5);
\node[right] at (7,-5.8) {$\CW= \nt \, R R + \, \e_{2N} \left(B^{N-1} L  R B_2 \Bt_2 \right)  $};
\node[right] at (7,-6.8) {$+ \, \e_{2N} \, \e^3 \, \left(B^{N-1} L^2 Y\right) + \, \sum_{i=2}^{N} \a_i \, \tr(B^i)$};
\node at (14.4,-2.2) {$\nt$};
\node at (9.3,-1.9) {$Y$};
\node at (11.2,1.4) {$B$};
\node at (8.7,-0.6) {$L$};
\node at (12.5,-0.7) {$R$};
\node at (12.9,-3.4) {$\Bt_2$};
\node at (8.2,-3.3) {$B_2$};
\node at (10.6,-4.6) {$\a_N$};
\epic} \ee
The final step is to confine the $Usp(2N)$ using the result from \eqref{UspTDecSplit}. The WZ is
\be \label{SU2Nantisym'T4Flip} \scalebox{0.9}{\bpic[node distance=2cm,gSUnode/.style={circle,red,draw,minimum size=8mm},gUSpnode/.style={circle,blue,draw,minimum size=8mm},fnode/.style={rectangle,draw,minimum size=8mm}]     
\node at (6.2,1.8) {$\CT_{4,flip}:$};
\node[fnode] (F1) at (9,0) {$3$};
\node[fnode,orange] (F2) at (12,0) {$3$};
\node[fnode,violet] (F3) at (9,-1.8) {$1$};
\node[fnode,blue] (F4) at (12,-1.8) {$1$};
\draw (F1) -- pic[sloped,very thick]{arrow=latex reversed} (F3);
\draw (F2) -- pic[sloped,very thick]{arrow=latex reversed} (F4);
\draw (9.4,-0.3) -- pic[pos=0.6,sloped]{arrow} (11.6,-0.3);
\node at (10.5,0.1) {$\vdots$};
\draw (9.4,0.3) -- pic[pos=0.6,sloped]{arrow} (11.6,0.3);
\draw (8.6,0.2) to[out=180,in=90] pic[pos=0.2,sloped,very thick]{arrow=latex reversed} (8.2,0) to[out=-90,in=180] pic[pos=0.8,sloped,very thick]{arrow=latex reversed} (8.6,-0.2);
\node at (7.9,0) {$\dots$};
\draw (8.6,0.4) to[out=180,in=90] pic[pos=0.2,sloped,very thick]{arrow=latex reversed} (7.5,0) to[out=-90,in=180] pic[pos=0.8,sloped,very thick]{arrow=latex reversed} (8.6,-0.4);
\draw (12.4,0.2) to[out=0,in=90] pic[pos=0.4,sloped,very thick]{arrow=latex reversed} (12.8,0) to[out=-90,in=0] pic[pos=0.6,sloped,very thick]{arrow=latex reversed} (12.4,-0.2);
\node at (13.2,0) {$\dots$};
\draw (12.4,0.4) to[out=0,in=90] pic[pos=0.6,sloped,very thick]{arrow=latex reversed} (13.5,0) to[out=-90,in=0] pic[pos=0.4,sloped,very thick]{arrow=latex reversed} (12.4,-0.4);
\node[right] at (4.3,-3.1) {$ \CW= \, \Bt_2 B_2 M_N +\,  \left.\sum_{i,j,k=1}^{N} \, \e_3 \, \e^3 \, \left(\Ht_i \, M_j \, H_k + M_i \, M_j \, M_k \right) \de_{i+j+k,2N+1}\right|_{H_N = \, \Ht_1 \, =0}$};
\node at (10.5,-0.6) {$M_1$};
\node at (10.5,0.6) {$M_N$};
\node at (7,0.7) {$H_1, \dots, H_{N-1}$};
\node at (13.7,0.7) {$\Ht_2, \dots, \Ht_N$};
\node at (12.5,-1) {$\Bt_2$};
\node at (8.5,-1) {$B_2$};
\epic} \ee 
The fields $\Ht_1$ and $H_N$ have been set to 0 by the E.O.M of $Y$ and $\nt$. It kills N terms in the sum. As a consistency check, for $N=2$ we recover the superpotential given in \cite{Csaki:1996zb} Sec.3.1.5. The mapping of the chiral ring in the flipping case is 
\be
\ba{c}\CT_{1,flip} \\
Q \, (A \, \At)^k \, \Qt \\
\At \, (A \, \At)^m \, Q^2 \\
A \, (A \, \At)^m \, \Qt^2 \\
\e_{2N} \, (A^{N-1} \, Q^2) \\
\e_{2N} \, (\At^{N-1} \, \Qt^2) 
\ea
\Longleftrightarrow
\ba{c} \CT_{3,flip} \\
L \, B^{k} \, R \\
L \, B^{m} \, L \\
R \, B^{m+1} \, R \\
B_2 \\
\Bt_2 
\ea
\Longleftrightarrow 
\ba{c}\CT_{4,flip} \\
M_{k+1} \\
H_{m+1} \\
\Ht_{m+2} \\
B_2 \\
\Bt_2  
\ea
\quad \quad
\ba{l}
k=0,\dots,N-1 \\
m=0,\dots,N-2 \\
m=0,\dots,N-2 \\
\\
\ea
\ee

\section{$SU(M)$ with {\tiny\ydiagram{1,1}} $+ M \, \overline{{\tiny\ydiagram{1}}} + 4 \, {\tiny\ydiagram{1}}$ series}\label{SUa}

\subsection{Odd rank: $M = 2N+1$} 
The gauge theory is $SU(2N+1)$ with one antisymmetric field, $2N+1$ antifundamentals and 4 fundamentals, with continuous global symmetry $SU(2N+1)_{\Qt} \times SU(4)_{Q} \times U(1)^2$.
\be \label{SU2N+11antisymT1} \scalebox{0.9}{\bpic[node distance=2cm,gSUnode/.style={circle,red,draw,minimum size=8mm},gUSpnode/.style={circle,blue,draw,minimum size=8mm},fnode/.style={rectangle,draw,minimum size=8mm}]  
\node at (4,1.5) {$\CT_1:$};
\node[gSUnode] (G1) at (6,0) {$2N+1$};
\node[fnode] (F1) at (5,-2.5) {$2N+1$};
\node[fnode] (F2) at (7.5,-2.5) {$4$};
\draw (G1) -- pic[pos=0.4,sloped,very thick]{arrow=latex reversed} (F1);
\draw (G1) -- pic[pos=0.5,sloped,very thick]{arrow=latex reversed} (F2);
\draw (6.5,0.7) to[out=90,in=0] pic[pos=0.1,sloped]{arrow} (6,1.4) to[out=180,in=90] pic[pos=0.7,sloped,very thick]{arrow=latex reversed} (5.5,0.7);
\node[right] at (9,0) {$ \CW= 0$};
\node at (6.7,1.4) {$A$};
\node at (5.1,-1) {$\Qt$};
\node at (7.1,-1) {$Q$};
\epic} \ee
The chiral ring generators are
\begin{itemize}
\item $Q \Qt \sim Q_i^{\a} \Qt_\a^I$ (transforming in the $(\, \scalebox{0.5}{\ydiagram{1}} \, , \overline{\scalebox{0.5}{\ydiagram{1}}} \,)$ of $SU(2N+1)_{\Qt} \times SU(4)_{Q}$)
\item $A\Qt^2 \sim A^{\a \b} \Qt^{[I}_\a \Qt^{J]}_\b$ $ \sim (\, \scalebox{0.5}{\ydiagram{1,1}} \, ,1)$ 
\item $A^N Q \sim \e_{\a_1\dots\a_{2N+1}}A^{\a_1\a_2} \dots A^{\a_{2N-1}\a_{2N}} Q_{i}^{\a_{2N+1}}$ $\sim (1, \overline{\scalebox{0.5}{\ydiagram{1}}} \,)$ 
\item $A^{N-1} Q^3 \sim \e_{\a_1\dots\a_{2N+1}} \e^{ijkl} \, A^{\a_1\a_2} \dots A^{\a_{2N-3}\a_{2N-2}} Q_{i}^{\a_{2N-1}} Q_{j}^{\a_{2N}} Q_{k}^{\a_{2N+1}}$ $\sim (1, \scalebox{0.5}{\ydiagram{1}} \,)$ 
\item $\Qt^{2N+1} \sim \e^{\a_1\dots\a_{2N+1}} \e_{I_1\dots I_{2N+1}} \, \Qt_{\a_1}^{I_1} \dots \Qt_{\a_{2N+1}}^{I_{2N+1}}$ $\sim (1,1)$ 
\end{itemize}
We deconfine the antisymmetric field using \eqref{SU2N+1Deconfinement}. We get
\be \label{SU2N+11antisymT1'} \scalebox{0.9}{\bpic[node distance=2cm,gSUnode/.style={circle,red,draw,minimum size=8mm},gUSpnode/.style={circle,blue,draw,minimum size=8mm},fnode/.style={rectangle,draw,minimum size=8mm}]  
\node at (1,1.5) {$\CT_{1'}:$};
\node[gSUnode] (G1) at (6,0) {$2N+1$};
\node[gUSpnode] (G2) at (3,0) {$2N-2$};
\node[fnode] (F1) at (5,-2.5) {$2N+1$};
\node[fnode] (F2) at (7.5,-2.5) {$4$};
\node[fnode,violet] (F3) at (3,-2.5) {$1$};
\draw (G1) -- pic[pos=0.4,sloped]{arrow} (G2);
\draw (G1) -- pic[pos=0.4,sloped,very thick]{arrow=latex reversed} (F1);
\draw (G1) -- pic[pos=0.5,sloped,very thick]{arrow=latex reversed} (F2);
\draw (G1) -- pic[pos=0.5,sloped,very thick]{arrow=latex reversed} (F3.north east);
\draw (G2) -- (F3);
\node[right] at (8,-1) {$ \CW=$ 1 Planar Triangle};
\node at (5.9,-1.2) {$\Qt$};
\node at (7.1,-1.2) {$Q$};
\node at (4.5,0.4) {$b$};
\node at (2.8,-1.4) {$l$};
\node at (4.2,-1) {$\ct$};
\epic} \ee
We then confine the $SU$ gauge node using \eqref{SUbuildingBlock}. The field $l$ becomes massive. Then the determinant of the meson matrix is easily computed. After rescaling of the fields to put a $+1$ in front of each term we obtain
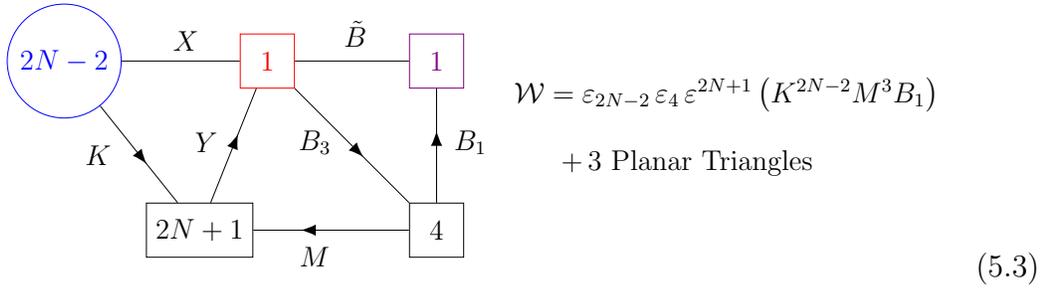
\be \label{SU2N+11antisymT2} \scalebox{0.9}{\bpic[node distance=2cm,gSUnode/.style={circle,red,draw,minimum size=8mm},gUSpnode/.style={circle,blue,draw,minimum size=8mm},fnode/.style={rectangle,draw,minimum size=8mm}]   
\node at (1,1.5) {$\CT_2:$};
\node[gUSpnode] (G1) at (3,0) {$2N-2$};
\node[fnode,red] (F1) at (6,0) {$1$};
\node[fnode] (F2) at (5,-2.5) {$2N+1$};
\node[fnode] (F3) at (8.5,-2.5) {$4$};
\node[fnode,violet] (F4) at (8.5,0) {$1$};
\draw (G1) -- (F1);
\draw (G1) -- pic[pos=0.6,sloped]{arrow} (F2);
\draw (F1) -- pic[pos=0.4,sloped]{arrow} (F2);
\draw (F1) -- pic[pos=0.6,sloped]{arrow} (F3);
\draw (F1) -- (F4);
\draw (F2) -- pic[pos=0.4,sloped,very thick]{arrow=latex reversed} (F3);
\draw (F4) -- pic[pos=0.5,sloped,very thick]{arrow=latex reversed} (F3);
\node[right] at (9.5,-0.5) {$ \CW= \e_{2N-2} \, \e_4 \, \e^{2N+1}  \left(K^{2N-2} M^3 B_1\right)$};
\node[right] at (10.2,-1.5) {$+\, 3$ Planar Triangles};
\node at (4.8,0.3) {$X$};
\node at (7.3,0.4) {$\Bt$};
\node at (6.7,-1.2) {$B_3$};
\node at (5.1,-1.2) {$Y$};
\node at (3.5,-1.4) {$K$};
\node at (9,-1.2) {$B_1$};
\node at (6.7,-2.9) {$M$};
\epic} \ee
We then S-confines the $Usp(2N-2)$ gauge node using \eqref{UspbuildingBlock}. The result of the integration of the massive field $Y$, the computation of the Pfaffian superpotential and the rescaling of the fields is the following WZ model
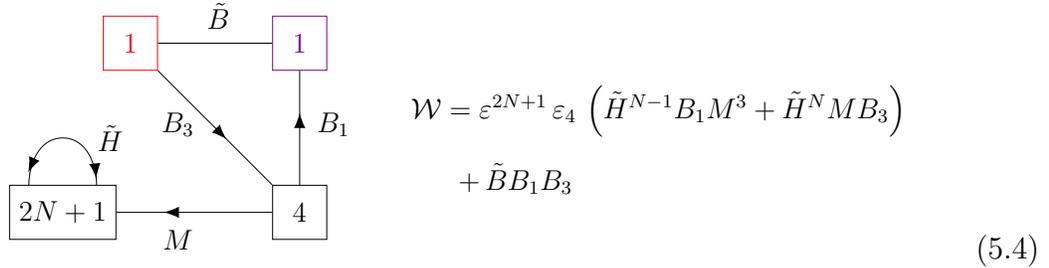
\be \label{SU2N+11antisymT3} \scalebox{0.9}{\bpic[node distance=2cm,gSUnode/.style={circle,red,draw,minimum size=8mm},gUSpnode/.style={circle,blue,draw,minimum size=8mm},fnode/.style={rectangle,draw,minimum size=8mm}]  
\node at (3.5,1) {$\CT_3:$};
\node[fnode,red] (F1) at (6,0) {$1$};
\node[fnode] (F2) at (5,-2.5) {$2N+1$};
\node[fnode] (F3) at (8.5,-2.5) {$4$};
\node[fnode,violet] (F4) at (8.5,0) {$1$};
\draw (F1) -- pic[pos=0.6,sloped]{arrow} (F3);
\draw (F1) -- (F4);
\draw (F2) -- pic[pos=0.4,sloped,very thick]{arrow=latex reversed} (F3);
\draw (F4) -- pic[pos=0.5,sloped,very thick]{arrow=latex reversed} (F3);
\draw (5.5,-2.1) to[out=90,in=0] pic[pos=0.1,sloped]{arrow} (5,-1.4) to[out=180,in=90] pic[pos=0.6,sloped,very thick]{arrow=latex reversed} (4.5,-2.1);
\node[right] at (10,-1) {$\CW= \e^{2N+1} \, \e_4 \, \left(\Ht^{N-1} B_1 M^3 + \Ht^N M B_3 \right)$};
\node[right] at (10.7,-2) {$+ \, \Bt B_1 B_3$};
\node at (7.3,0.4) {$\Bt$};
\node at (6.7,-1.2) {$B_3$};
\node at (9,-1.2) {$B_1$};
\node at (6.7,-2.9) {$M$};
\node at (5.7,-1.4) {$\Ht$};
\epic} \ee

\noindent We recover the result of Section 3.1.3 of \cite{Csaki:1996zb}. The mapping of the chiral ring generators across the different frames is given in \eqref{mapSU2N+11antisym}.
\be \label{mapSU2N+11antisym} 
\ba{c}\CT_1 \\
Q \, \Qt  \\
A \, \Qt^2  \\
\e_{2N+1} \, \Qt^{2N+1} \\
\e_{2N+1} \, (A^N \, Q)  \\
\e_{2N+1} \, (A^{N-1} \, Q^3) 
\ea
\quad \Longleftrightarrow \quad
\ba{c} \CT_{1'} \\
Q \, \Qt \\
b \, b \, \Qt^2 \\
\e_{2N+1} \, \Qt^{2N+1} \\
\ct \, Q \\
\e_{2N+1} \, (b^{2N-2} \, Q^3)  
\ea
\quad \Longleftrightarrow \quad
\ba{c}\CT_2 \\
M \\
K^2 \\
\Bt \\
B_1 \\
B_3
\ea
\quad \Longleftrightarrow \quad
\ba{c} \CT_{3} \\
M \\
\Ht \\
\Bt \\
B_1 \\
B_3
\ea
\ee

\subsection{Even rank: $M = 2N$} \label{SU(2N)1Antisym}
Now we study the even rank version of the previous theory. The continuous global symmetry is $SU(2N)_{\Qt} \times SU(4)_Q \times U(1)^2$.
\be \label{SU2N1antisymT1} \scalebox{0.9}{\bpic[node distance=2cm,gSUnode/.style={circle,red,draw,minimum size=8mm},gUSpnode/.style={circle,blue,draw,minimum size=8mm},fnode/.style={rectangle,draw,minimum size=8mm}]  
\node at (-2.5,1.5) {$\CT_1:$};
\node[gSUnode] (G1) at (0,0) {$2N$};
\node[fnode] (F1) at (-1,-2) {$2N$};
\node[fnode] (F2) at (1,-2) {$4$};
\draw (G1) -- pic[pos=0.4,sloped,very thick]{arrow=latex reversed} (F1.north);
\draw (G1) -- pic[pos=0.5,sloped,very thick]{arrow=latex reversed} (F2.north);
\draw (0.3,0.4) to[out=90,in=0] pic[pos=0.1,sloped]{arrow} (0,0.8) to[out=180,in=90] pic[pos=0.5,sloped,very thick]{arrow=latex reversed} (-0.3,0.4);
\node[right] at (-0.8,-3) {$ \CW= 0$};
\node at (0.6,0.9) {$A$};
\node at (-1,-0.7) {$\Qt$};
\node at (1,-0.7) {$Q$};
\node at (2.5,0) {$\equiv$};
\node[gSUnode] (G2) at (6,0) {$2N$};
\node[fnode] (F3) at (4.5,-2) {$2N$};
\node[fnode] (F4) at (6.5,-2) {$3$};
\node[fnode,violet] (F5) at (7.5,-2) {$1$};
\draw (G2) -- pic[pos=0.4,sloped,very thick]{arrow=latex reversed} (F3);
\draw (G2) -- pic[pos=0.5,sloped,very thick]{arrow=latex reversed} (F4);
\draw (G2) -- pic[pos=0.5,sloped,very thick]{arrow=latex reversed} (F5);
\draw (6.3,0.4) to[out=90,in=0] pic[pos=0.1,sloped]{arrow} (6,0.8) to[out=180,in=90] pic[pos=0.5,sloped,very thick]{arrow=latex reversed} (5.7,0.4);
\node[right] at (5.2,-3) {$ \CW= 0$};
\node at (6.6,0.9) {$A$};
\node at (5,-0.7) {$\Qt$};
\node at (6,-0.9) {$q$};
\node at (7.1,-0.9) {$F$};
\epic} \ee
We deconfine the antisymmetric using \eqref{SU2NDeconfinement} which implies the breaking of the $SU(4)_Q$ into $SU(3)_q \times U(1)_F$. The chiral ring generators in the split form are 
\begin{itemize}
\item $q \Qt \sim q_i^{\a} \Qt_\a^I$ (transforming in the $(\, \scalebox{0.5}{\ydiagram{1}} \, , \overline{\scalebox{0.5}{\ydiagram{1}}} \,)$ of $SU(2N)_{\Qt} \times SU(3)_q$)
\item $F \Qt \sim F^{\a} \Qt_\a^I$  $\sim (\, \scalebox{0.5}{\ydiagram{1}} \, , 1)$
\item $A\Qt^2 \sim A^{\a \b} \Qt^{[I}_\a \Qt^{J]}_\b$ $\sim (\, \scalebox{0.5}{\ydiagram{1,1}} \, , 1)$ 
\item $A^N \sim \e_{\a_1\dots\a_{2N}}A^{\a_1\a_2} \dots A^{\a_{2N-1}\a_{2N}}$ $\sim (1,1)$ 
\item $A^{N-1} q^2 \sim \e_{\a_1\dots\a_{2N}}A^{\a_1\a_2} \dots A^{\a_{2N-3}\a_{2N-2}} q_{[i}^{\a_{2N-1}} q_{j]}^{\a_{2N}}$ $\sim (1, \overline{\scalebox{0.5}{\ydiagram{1,1}}} \,)$ 
\item $A^{N-1} q F \sim \e_{\a_1\dots\a_{2N}}A^{\a_1\a_2} \dots A^{\a_{2N-3}\a_{2N-2}} q_{i}^{\a_{2N-1}} F^{\a_{2N}}$ $\sim (1, \overline{\scalebox{0.5}{\ydiagram{1}}} \,)$ 
\item $A^{N-2} q^3 F \sim \e_{\a_1\dots\a_{2N+1}} \e^{ijk} \, A^{\a_1\a_2} \dots A^{\a_{2N-5}\a_{2N-4}} q_{i}^{\a_{2N-3}} q_{j}^{\a_{2N-2}} q_{k}^{\a_{2N-1}} F^{\a_{2N}}$ $\sim (1,1)$ 
\item $\Qt^{2N} \sim \e^{\a_1\dots\a_{2N}} \e_{I_1\dots I_{2N}} \, \Qt_{\a_1}^{I_1} \dots \Qt_{\a_{2N}}^{I_{2N}}$ $\sim (1,1)$ 
\end{itemize}	
\be \label{SU2N1antisymT1'} \bpic[node distance=2cm,gSUnode/.style={circle,red,draw,minimum size=8mm},gUSpnode/.style={circle,blue,draw,minimum size=8mm},fnode/.style={rectangle,draw,minimum size=8mm}]  
\node at (1,1.5) {$\CT_{1'}:$};
\node[gSUnode] (G1) at (6,0) {$2N$};
\node[gUSpnode] (G2) at (3,0) {$2N-2$};
\node[fnode] (F1) at (6,-2.5) {$2N$};
\node[fnode] (F2) at (7.5,-2.5) {$3$};
\node[fnode,violet] (F3) at (3,-2.5) {$1$};
\node[fnode,blue] (F4) at (4.5,-2.5) {$1$};
\draw (G1) -- pic[pos=0.4,sloped]{arrow} (G2);
\draw (G1) -- pic[pos=0.5,sloped]{arrow} (F1);
\draw (G1) -- pic[pos=0.5,sloped,very thick]{arrow=latex reversed} (F2.north);
\draw (G1) -- pic[pos=0.4,sloped,very thick]{arrow=latex reversed} (F4.north);
\draw (G2) -- (F3);
\draw (G2) -- (F4);
\draw (F3) -- (F4);
\node[right] at (8.5,-1) {$ \CW=$ 2 Planar Triangles};
\node at (6.3,-1.2) {$\Qt$};
\node at (7.2,-1.2) {$q$};
\node at (4.7,0.4) {$b$};
\node at (2.8,-1.4) {$l$};
\node at (4.8,-1) {$\ct$};
\node at (4.2,-1.4) {$d$};
\node at (3.7,-2.8) {$B_0$};
\epic \ee
The next step is once again to confine the $SU$ gauge group using \eqref{SUbuildingBlock}.
\be \label{SU2N1antisymT2} \scalebox{0.9}{\bpic[node distance=2cm,gSUnode/.style={circle,red,draw,minimum size=8mm},gUSpnode/.style={circle,blue,draw,minimum size=8mm},fnode/.style={rectangle,draw,minimum size=8mm}]   
\node at (1.5,3.5) {$\CT_2:$};
\node[gUSpnode] (G1) at (3,0) {$2N-2$};
\node[fnode,red] (F1) at (6,0) {$1$};
\node[fnode] (F2) at (6,-2.5) {$2N$};
\node[fnode] (F3) at (8.5,0) {$3$};
\node[fnode,violet] (F4) at (4,2.5) {$1$};
\node[fnode,blue] (F5) at (6,2.5) {$1$};
\draw (G1) -- (F1);
\draw (G1) -- pic[pos=0.6,sloped]{arrow} (F2.north west);
\draw (G1) -- (F4);
\draw (F1) -- pic[pos=0.6,sloped,very thick]{arrow=latex reversed} (F2);
\draw (F1) -- pic[pos=0.6,sloped]{arrow} (F3);
\draw (F1) -- (F5);
\draw (F2.north east) -- pic[pos=0.4,sloped,very thick]{arrow=latex reversed} (F3.south west);
\draw (F5.east) -- pic[pos=0.6,sloped,very thick]{arrow=latex reversed} (F3.north);
\draw (F4) -- (F5);
\node[right] at (9.5,1) {$ \CW= \e_{2N-2} \,  \e^{2N} \, \e_3 \left(K^{2N-2} M_{(1)}^2 B_2^{(2)}\right. $};
\node[right] at (10,0) {$+ \left. K^{2N-3} M_{(1)}^3 l B_0  \right) + \, B_0 \Bt X l$};
\node[right] at (10,-1) {+ 3 Planar Triangles};
\node at (4.8,-0.3) {$X$};
\node at (5.7,1.2) {$\Bt$};
\node at (7.3,0.4) {$B_2^{(1)}$};
\node at (6.4,-1.2) {$Y$};
\node at (4.3,-1.6) {$K$};
\node at (7.8,1.7) {$B_2^{(2)}$};
\node at (7.6,-1.6) {$M_{(1)}$};
\node at (3.3,1.4) {$l$};
\node at (5,2.8) {$B_0$};
\epic} \ee
The last step is to confine the $Usp(2N-2)$. We get the following WZ theory
\be \label{SU2N1antisymT3} \scalebox{0.9}{\bpic[node distance=2cm,gSUnode/.style={circle,red,draw,minimum size=8mm},gUSpnode/.style={circle,blue,draw,minimum size=8mm},fnode/.style={rectangle,draw,minimum size=8mm}]
\node at (2.5,3) {$\CT_3:$};
\node[fnode,red] (F1) at (6,0) {$1$};
\node[fnode] (F2) at (6,-2.5) {$2N$};
\node[fnode] (F3) at (8.5,0) {$3$};
\node[fnode,violet] (F4) at (3.5,0) {$1$};
\node[fnode,blue] (F5) at (6,2.5) {$1$};
\draw (F1) -- (F4);
\draw (F1) -- pic[pos=0.6,sloped]{arrow} (F3);
\draw (F1) -- (F5);
\draw (F2.north east) -- pic[pos=0.4,sloped,very thick]{arrow=latex reversed} (F3.south west);
\draw (F4.south east) -- pic[pos=0.6,sloped]{arrow} (F2.north west);
\draw (F5.south east) -- pic[pos=0.6,sloped,very thick]{arrow=latex reversed} (F3.north west);
\draw (F4) -- (F5);
\draw (6.3,-2.9) to[out=-90,in=0] pic[pos=0.1,sloped]{arrow} (6,-3.5) to[out=-180,in=-90] pic[pos=0.6,sloped,very thick]{arrow=latex reversed} (5.7,-2.9);  
\node[right] at (9.5,0.5) {$ \CW= \e^{2N} \, \e_3 \, \left(\Ht^N B_4 + \Ht^{N-1} M_1^2 B_2^{(2)} \right. $};
\node[right] at (10,-0.5) {$\left. + \Ht^{N-1} M_1 M_2 B_2^{(1)}  + \, \Ht^{N-2} M_1^3 M_2 B_0  \right)$};
\node[right] at (10,-1.5) {$+ \, \Bt B_0 B_4 + \Bt B_2^{(2)} B_2^{(1)}$};
\node at (5.7,1.2) {$\Bt$};
\node at (7.3,0.4) {$B_2^{(1)}$};
\node at (7.8,1.5) {$B_2^{(2)}$};
\node at (7.6,-1.6) {$M_1$};
\node at (4.3,-1.6) {$M_2$};
\node at (4.4,1.5) {$B_0$};
\node at (4.8,-0.3) {$B_4$};
\node at (6.7,-3.4) {$\Ht$};
\epic} \ee
The final superpotential can be repackaged in a manifest $SU(4)$ invariant way as 
\be \label{SU2N1antisymT4}
\CW = \e^{2N} \, \e_4 \left(\Ht^N B_4 + \Ht^{N-1} M^2 B_2 + \Ht^{N-2} M^4 B_0 + \Bt B_0 B_4 + \Bt B_2^2\right) \,,\ee
where $Q\Qt \lra M=\{M_1,M_2\}$ is a fundamental of $SU(4)_Q$ and $A^{N-1} Q^2 \lra B_2=\{\e_3 B_2^{(1)},B_2^{(2)}\}$ is an antisymmetric of $SU(4)_Q$. We recover the result of Section 3.1.2 of \cite{Csaki:1996zb}. The mapping of the chiral ring generators across the frames is given in \eqref{mapSU2N1antisym}.
\be \label{mapSU2N1antisym} 
\ba{c}\CT_1 \\
q \, \Qt \\
F \, \Qt \\
A \, \Qt^2 \\
\e_{2N} \, \Qt^{2N} \\
\e_{2N} \, A^N \\
\e_{2N} \, (A^{N-1} \, q^2) \\
\e_{2N} \, (A^{N-1} \, q \, F) \\
\e_{2N} \, (A^{N-2} \, q^3 \, F) 
\ea
\quad \Longleftrightarrow \quad
\ba{c} \CT_{1'}\\
q \, \Qt \\
l \, b \, \Qt \\
b \, b \, \Qt^2 \\
\e_{2N} \, \Qt^{2N} \\
B_0 \\
\e_{2N} \, (b^{2N-2} \, q^2) \\
\ct \, q \\
\e_{2N} \, (b^{2N-4} \, l \, b \, q^3)  
\ea
\quad \Longleftrightarrow \quad
\ba{c}\CT_2\\
M_1 \\
l \, K \\
K^2 \\
\Bt\\
B_0 \\
B_2^{(1)} \\
B_2^{(2)} \\
l \, X
\ea
\quad \Longleftrightarrow \quad
\ba{c} \CT_{3}\\
M_1 \\
M_2 \\
\Ht \\
\Bt \\
B_0 \\
B_2^{(1)} \\
B_2^{(2)}\\
B_4
\ea
\ee

\section{Four 'sporadic' cases: $SU(5), SU(6), SU(7)$} \label{sporadic}
In this section we show how the strategy of deconfining antisymmetric fields allow proving the remaining four S-confining cases in the list of \cite{Csaki:1996sm, Csaki:1996zb}. In the two $SU(5)$ cases, there are extra subtelties when computing the superpotential because of the presence of degenerate operators, see also \cite{BAJEOTBENVENUTI2}.

\subsection{$SU(7)$ with 2 {\tiny \ydiagram{1,1}} + 6 $\overline{{\tiny\ydiagram{1}}}$}
The gauge theory is $SU(7)$ with two antisymmetric fields and $6$ antifundamentals, with continuous global symmetry $SU(2)_A \times SU(6)_{\Qt} \times U(1)$.
\be \hspace{-1.5cm} \label{SU7T1} \bpic[node distance=2cm,gSUnode/.style={circle,red,draw,minimum size=8mm},gUSpnode/.style={circle,blue,draw,minimum size=8mm},fnode/.style={rectangle,draw,minimum size=8mm}]  
\node at (4,0.9) {$\CT_1:$};
\node[gSUnode] (G2) at (6,0) {$7$};
\node[fnode] (F3) at (6,-2) {$6$};
\draw (G2) -- pic[sloped]{arrow} (F3);
\draw (5.6,0.2) to[out=110,in=90] pic[pos=0.3,sloped]{arrow} (5.1,0) to[out=270,in=250] pic[pos=0.7,sloped]{arrow} (5.6,-0.2);
\draw (6.4,0.2) to[out=70,in=90] pic[pos=0.5,sloped,very thick]{arrow=latex reversed} (6.9,0) to[out=270,in=290] pic[pos=0.5,sloped,very thick]{arrow=latex reversed} (6.4,-0.2);
\node[right] at (8,-0.6) {$ \CW= 0$};
\node at (4.7,0) {$A_{+1}$};
\node at (7.3,0) {$A_{-1}$};
\node at (6.5,-1) {$\Qt$};
\epic \ee
The chiral ring generators are 
\begin{itemize}
\item $A\Qt^2 \sim A_i^{\a_1 \a_2} \Qt^{[I}_{\a_1} \Qt^{J]}_{\a_2}$\footnote{$\a_i=1,\ldots,7$ are gauge indices. $i,j,k,l\dots=1,2$ are $SU(2)_A$ flavor indices. $I,J,\ldots=1,\ldots,6$ are $SU(6)_{\Qt}$ flavor indices.} (transforming in the $(\, \scalebox{0.5}{\ydiagram{1}} \, ,\scalebox{0.5}{\ydiagram{1,1}} \,)$ of $SU(2)_A \times SU(6)_{\Qt}$).
\item $\e_7 \, (A^4 \Qt) \sim \e_{\a_1 \dots \a_7} \, \e^{kl} \, A_i^{\a_1\a_2}A_j^{\a_3\a_4}A_k^{\a_5\a_6}A_l^{\a_7\b} \Qt_{\beta}^{I} \sim (\, \scalebox{0.5}{\ydiagram{2}} \, ,\scalebox{0.5}{\ydiagram{1}} \,)$
\end{itemize}
We will deconfine each antisymmetric field using \eqref{SU2N+1Deconfinement}. Notice that this step breaks the global $SU(2)_A$ symmetry into a $U(1)_A$. In \eqref{SU7T1}, the subscript of the two antisymmetric fields, $A_{\pm 1}$ correspond to the $U(1)_A$ charge (corresponding to the weight of the representation of $SU(2)_A$).  So let's give the chiral ring in this split form
\begin{itemize}
\item $A_{+1}\Qt^2 \sim A_{+1}^{\a_1 \a_2} \Qt^{[I}_{\a_1} \Qt^{J]}_{\a_2}$ $ \sim \scalebox{0.5}{\ydiagram{1,1}} \,$ of $SU(6)_{\Qt}$
\item $A_{-1}\Qt^2 \sim A_{-1}^{\a_1 \a_2} \Qt^{[I}_{\a_1} \Qt^{J]}_{\a_2}$ $ \sim \scalebox{0.5}{\ydiagram{1,1}} $
\item $\e_7 \, (A_{+1} A_{+1} A_{[+1} A_{-1]} \Qt) \sim \e_{\a_1 \dots \a_7} \, A_{+1}^{\a_1\a_2} A_{+1}^{\a_3\a_4} \left(A_{+1}^{\a_5\a_6} A_{-1}^{\a_7\b} - A_{-1}^{\a_5\a_6} A_{+1}^{\a_7\b}\right) \Qt_{\beta}^{I}$ $ \sim \scalebox{0.5}{\ydiagram{1}} $
\item $\e_7 \, (A_{(+1} A_{-1)} A_{[+1} A_{-1]} \Qt) \sim \e_{\a_1 \dots \a_7} \, A_{(+1}^{\a_1\a_2} A_{-1)}^{\a_3\a_4} \, A_{[+1}^{\a_5\a_6} A_{-1]}^{\a_7\b} \, \Qt_{\beta}^{I}$ $ \sim \scalebox{0.5}{\ydiagram{1}} $
\item $\e_7 \, (A_{-1} A_{-1} A_{[+1} A_{-1]} \Qt) \sim \e_{\a_1 \dots \a_7} \, A_{-1}^{\a_1\a_2} A_{-1}^{\a_3\a_4} \, A_{[+1}^{\a_5\a_6} A_{-1]}^{\a_7\b} \, \Qt_{\beta}^{I}$ $ \sim \scalebox{0.5}{\ydiagram{1}} $
\end{itemize}
\be \label{SU7T1'} \scalebox{0.9}{\bpic[node distance=2cm,gSUnode/.style={circle,red,draw,minimum size=8mm},gUSpnode/.style={circle,blue,draw,minimum size=8mm},fnode/.style={rectangle,draw,minimum size=8mm}]    
\node at (-4,2.3) {$\CT_{1'}:$};
\node[gSUnode] (G1) at (0,0) {$7$};
\node[gUSpnode] (G2) at (-2.5,-1) {$4$};
\node[gUSpnode] (G3) at (2.5,-1) {$4$};
\node[fnode] (F1) at (0,-2) {$6$};
\node[fnode,violet] (F2) at (-2.5,1) {$1$};
\node[fnode,blue] (F3) at (2.5,1) {$1$};
\draw (G2) -- pic[pos=0.7,sloped]{arrow} (G1);
\draw (G1) -- pic[sloped,very thick]{arrow=latex reversed} (G3);
\draw (G1) -- pic[sloped]{arrow} (F1);
\draw (G1) -- pic[pos=0.4,sloped,very thick]{arrow=latex reversed} (F2);
\draw (G1) -- pic[sloped]{arrow} (F3);
\draw (G2) -- (F2);
\draw (G3) -- (F3); 
\node[right] at (4,0) {$\CW= 2 \,$ Planar Triangles};
\node at (-2.9,0) {$l_{-\frac{7}{2}}$};
\node at (2.9,0) {$l_{+\frac{7}{2}}$};
\node at (-1.2,0.8) {$\ct_{+3}$};
\node at (1.2,0.8) {$\ct_{-3}$};
\node at (-1.2,-0.9) {$b_{+\frac{1}{2}}$};
\node at (1.4,-0.9) {$b_{-\frac{1}{2}}$};
\node at (0.4,-1) {$\Qt$};
\epic} \ee 
We then confine the $SU$ gauge node with \eqref{SUbuildingBlock}. The fields $l_{-\frac{7}{2}}$ and $l_{+\frac{7}{2}}$ get a mass. After integrating them out, computing $\det(\text{mesons})$ of degree $8$ and rescale the fields we get
\be \label{SU7T2} \scalebox{0.9}{\bpic[node distance=2cm,gSUnode/.style={circle,red,draw,minimum size=8mm},gUSpnode/.style={circle,blue,draw,minimum size=8mm},fnode/.style={rectangle,draw,minimum size=8mm}]   
\node at (-4,2.3) {$\CT_2:$};
\node[fnode,red] (F1) at (0,0) {$1$};
\node[gUSpnode] (G1) at (-2.7,-0.5) {$4$};
\node[gUSpnode] (G2) at (2.7,-0.5) {$4$};
\node[fnode] (F2) at (0,-2) {$6$};
\node[fnode,blue] (F3) at (-2,1.5) {$1$};
\node[fnode,violet] (F4) at (2,1.5) {$1$};
\draw (G1) -- (F1);
\draw (G1) -- pic[pos=0.6,sloped]{arrow} (F2);
\draw (G1) -- (F3);
\draw (G2) -- (F1);
\draw (G2) -- pic[sloped,very thick]{arrow=latex reversed} (F2);
\draw (G2) -- (F4);
\draw (F1) -- pic[sloped,very thick]{arrow=latex reversed} (F2);
\draw (F1) -- (F3);
\draw (F1) -- (F4);  
\node[right] at (4,0) {$\CW= \e_4 \, \e_4 \, \e^6 \, \left(K_{+\frac{1}{2}}^3 \, K_{-\frac{1}{2}}^3 \, C_{-\frac{5}{2}} \, C_{+\frac{5}{2}} \right)$};
\node[right] at (4,-1) {+ 4 Planar Triangles};
\node at (-2.8,0.6) {$C_{-\frac{5}{2}}$};
\node at (2.8,0.6) {$C_{+\frac{5}{2}}$};
\node at (1.6,-1.7) {$K_{-\frac{1}{2}}$};
\node at (-1.5,-1.6) {$K_{+\frac{1}{2}}$};
\node at (-1.4,0.1) {$D_{-\frac{1}{2}}$};
\node at (1.4,0.1) {$D_{+\frac{1}{2}}$};
\node at (-0.7,1) {$B_{+3}$};
\node at (0.7,1) {$B_{-3}$};
\node at (0.4,-1) {$V_0$};
\epic} \ee 
Notice that $B_{+3}$, $B_{-3}$ and $V_0$ are gauge singlets but are zero in the chiral ring, this is a quantum effect that can be seen dualizing the $Usp$ nodes.

We can now S-confine the left $Usp(4)$ with \eqref{UspbuildingBlock}. After integrating out massive fields, computing the Pfaffian superpotential and rescale the fields to put a $+1$ in front of each terms we get 
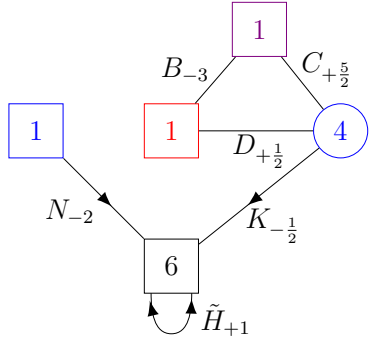
\be \label{SU7T3} \scalebox{0.9}{ \bpic[node distance=2cm,gSUnode/.style={circle,red,draw,minimum size=8mm},gUSpnode/.style={circle,blue,draw,minimum size=8mm},fnode/.style={rectangle,draw,minimum size=8mm}]    
\node at (-3,2.3) {$\CT_3:$};
\node[fnode,red] (F1) at (0,0) {$1$};
\node[gUSpnode] (G1) at (2.5,0) {$4$};
\node[fnode] (F2) at (0,-2) {$6$};
\node[fnode,blue] (F3) at (-2,0) {$1$};
\node[fnode,violet] (F4) at (1.3,1.5) {$1$};
\draw (G1) -- (F1);
\draw (G1) -- pic[sloped,very thick]{arrow=latex reversed} (F2);
\draw (G1) -- (F4);
\draw (F2) -- pic[pos=0.4,sloped]{arrow} (F3);
\draw (F1) -- (F4);
\draw (0.3,-2.4) to[out=-90,in=0] pic[pos=0.1,sloped]{arrow} (0,-3) to[out=-180,in=-90] pic[pos=0.6,sloped,very thick]{arrow=latex reversed} (-0.3,-2.4);  
\node[right] at (4,0) {$\CW= \e_4 \, \e^6 \, \left(\Ht_{+1} N_{-2} K_{-\frac{1}{2}}^3 C_{+\frac{5}{2}}\right)$};
\node[right] at (4,-1) {$+ \, \e^6\, \left(\Ht_{+1}^2 N_{-2} K_{-\frac{1}{2}} D_{+\frac{1}{2}} \right) + B_{-3} D_{+\frac{1}{2}} C_{+\frac{5}{2}} $};
\node at (2.3,0.9) {$C_{+\frac{5}{2}}$};
\node at (0.2,0.9) {$B_{-3}$};
\node at (1.3,-0.3) {$D_{+\frac{1}{2}}$};
\node at (1.5,-1.4) {$K_{-\frac{1}{2}}$};
\node at (-1.5,-1.2) {$N_{-2}$};
\node at (0.8,-2.8) {$\Ht_{+1}$};
\epic} \ee 
Finally, we repeat the same operation with the last gauge group and get the WZ model with quartic superpotential
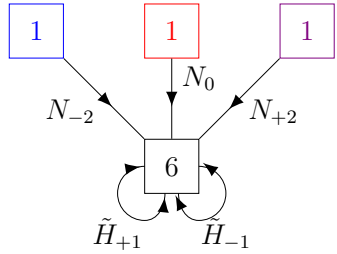
\be \label{SU7T4} \scalebox{0.9}{\bpic[node distance=2cm,gSUnode/.style={circle,red,draw,minimum size=8mm},gUSpnode/.style={circle,blue,draw,minimum size=8mm},fnode/.style={rectangle,draw,minimum size=8mm}] 
\node at (-3,1.3) {$\CT_4:$};
\node[fnode,red] (F1) at (0,0) {$1$};
\node[fnode] (F2) at (0,-2) {$6$};
\node[fnode,blue] (F3) at (-2,0) {$1$};
\node[fnode,violet] (F4) at (2,0) {$1$};
\draw (F1) -- pic[pos=0.6,sloped]{arrow} (F2);
\draw (F2) -- pic[pos=0.4,sloped]{arrow} (F3);
\draw (F2) -- pic[pos=0.5,sloped,very thick]{arrow=latex reversed} (F4);
\draw (-0.4,-2) to[out=180,in=90] pic[pos=0.1,sloped]{arrow} (-0.8,-2.4) to[out=270,in=180] (-0.4,-2.8) to[out=0,in=-90] pic[pos=0.9,sloped]{arrow} (-0.1,-2.4);  
\draw (0.4,-2) to[out=0,in=90] pic[pos=0.4,sloped,very thick]{arrow=latex reversed} (0.8,-2.4) to[out=270,in=0] (0.4,-2.8) to[out=180,in=270] pic[pos=0.6,sloped,very thick]{arrow=latex reversed} (0.1,-2.4);  
\node[right] at (3.6,-1) {$\CW= \e^6 \, \big(\Ht_{+1} \Ht_{-1} N_{+2} N_{-2}$};
\node[right] at (3.6,-2) {$+ \, \Ht_{+1}^2 N_{0} N_{-2}  + \, \Ht_{-1}^2 N_{+2} N_{0} \big)$};
\node at (-1.5,-1.2) {$N_{-2}$};
\node at (0.4,-0.7) {$N_{0}$};
\node at (1.5,-1.2) {$N_{+2}$};
\node at (-0.8,-3) {$\Ht_{+1}$};
\node at (0.8,-3) {$\Ht_{-1}$};
\epic} \ee 
The final superpotential can be repackaged in a manifest $SU(2)_A$ invariant way as 
\be \CW = \Ht^2 N^2\,,\ee
where $A\Qt^2 \lra \Ht=\{\Ht_{+1},\Ht_{-1}\}$ is a $SU(2)_A$ doublet and $A^4\Qt \lra N=\{N_{+2},N_{0},N_{-2}\}$ is a 2-symmetric tensor of $SU(2)_A$ (also called the triplet). We recover the result of Section 3.1.10 of \cite{Csaki:1996zb}. The mapping of the chiral ring generators is
\be \label{mapSU71}
\scalebox{0.87}{$
\ba{c}\CT_1 \\
A_{+1} \, \Qt^2 \\
A_{-1} \, \Qt^2 \\
\e_7 \, (A_{+1} A_{+1} A_{[+1} A_{-1]} \Qt) \\
\e_7 \, (A_{(+1} A_{-1)} A_{[+1} A_{-1]} \Qt) \\
\e_7 \, (A_{-1} A_{-1} A_{[+1} A_{-1]} \Qt) 
\ea
\Leftrightarrow
\ba{c} \CT_{1'}\\
b_{+\frac{1}{2}} b_{+\frac{1}{2}} \, \Qt^2 \\
b_{-\frac{1}{2}} b_{-\frac{1}{2}} \, \Qt^2 \\
b_{-\frac{1}{2}} \ct_{+3} \, b_{-\frac{1}{2}} \Qt \\
\e_7 \, (b_{+\frac{1}{2}}^3 \, b_{-\frac{1}{2}}^3 \, b_{[+\frac{1}{2}} \, b_{-\frac{1}{2}]} \Qt) \\
b_{+\frac{1}{2}} \ct_{-3} \, b_{+\frac{1}{2}} \Qt
\ea
\Leftrightarrow 
\ba{c}\CT_2\\
K_{+\frac{1}{2}} K_{+\frac{1}{2}} \\
K_{-\frac{1}{2}} K_{-\frac{1}{2}} \\
C_{+\frac{5}{2}} K_{-\frac{1}{2}} \\
D_{-\frac{1}{2}} K_{+\frac{1}{2}} = D_{+\frac{1}{2}} K_{-\frac{1}{2}} \\
C_{-\frac{5}{2}} K_{+\frac{1}{2}}
\ea
\Leftrightarrow 
\ba{c}\CT_3 \\
\Ht_{+1} \\
K_{-\frac{1}{2}} K_{-\frac{1}{2}} \\
C_{+\frac{5}{2}} K_{-\frac{1}{2}} \\
D_{+\frac{1}{2}} K_{-\frac{1}{2}} \\
N_{-2}
\ea
\Leftrightarrow
\ba{c} \CT_{4}\\
\Ht_{+1} \\
\Ht_{-1}  \\
N_{+2} \\
N_{0} \\
N_{-2} 
\ea
$}
\ee

\subsection{$SU(6)$ with 2 {\tiny \ydiagram{1,1}} + 5 $\overline{{\tiny\ydiagram{1}}}$ + 1 ${\tiny \ydiagram{1}}$}
The second sporadic case is the $SU(6)$ gauge theory with two antisymmetric fields, $5$ antifundamentals and $1$ fundamental, with continuous global symmetry $SU(2)_A \times SU(5)_{\Qt} \times U(1)^2$.
\be \label{SU6T1} \bpic[node distance=2cm,gSUnode/.style={circle,red,draw,minimum size=8mm},gUSpnode/.style={circle,blue,draw,minimum size=8mm},fnode/.style={rectangle,draw,minimum size=8mm}]  
\node at (3.6,1) {$\CT_1:$};
\node[gSUnode] (G1) at (6,0) {$6$};
\node[fnode] (F1) at (7,-2) {$5$};
\node[fnode,violet] (F2) at (5,-2) {$1$};
\draw (G1) -- pic[sloped]{arrow} (F1);
\draw (G1) -- pic[pos=0.3,sloped]{arrow} (F2);
\draw (5.6,0.2) to[out=110,in=90] pic[pos=0.3,sloped]{arrow} (5.1,0) to[out=270,in=250] pic[pos=0.7,sloped]{arrow} (5.6,-0.2);
\draw (6.4,0.2) to[out=70,in=90] pic[pos=0.5,sloped,very thick]{arrow=latex reversed} (6.9,0) to[out=270,in=290] pic[pos=0.5,sloped,very thick]{arrow=latex reversed} (6.4,-0.2);
\node[right] at (8,-0.4) {$ \CW= 0$};
\node at (4.7,0) {$A_{+1}$};
\node at (7.3,0) {$A_{-1}$};
\node at (7,-1) {$\Qt$};
\node at (5,-1) {$Q$};
\epic \ee
The chiral ring generators are 
\begin{itemize}
\item $Q \, \Qt \sim Q^{\a} \Qt^{I}_{\a}$ transforming in the $(1, \scalebox{0.5}{\ydiagram{1}} \,)$ of $SU(2)_A \times SU(5)_{\Qt}$
\item $A\Qt^2 \sim A_i^{\a_1 \a_2} \Qt^{[I}_{\a_1} \Qt^{J]}_{\a_2}$ $ \sim (\, \scalebox{0.5}{\ydiagram{1}} \, ,\scalebox{0.5}{\ydiagram{1,1}} \,)$
\item $\e_6 \, A^3 \sim \e_{\a_1 \dots \a_6} \, A_{(i}^{\a_1 \a_2} A_{j}^{\a_3 \a_4} A_{k)}^{\a_5 \a_6}$ $ \sim (\, \scalebox{0.5}{\ydiagram{3}} \, ,1)$
\item $\e_6 \, A^3 Q \Qt \sim \e_{\a_1 \dots \a_6} \, \e^{i_2 i_3} \, A_{i_1}^{\a_1 \a_2} A_{i_2}^{\a_3 \a_4} A_{i_3}^{\a_5 \b} Q^{\a_6} \Qt_{\b}^I$ $ \sim (\, \scalebox{0.5}{\ydiagram{1}} \, , \scalebox{0.5}{\ydiagram{1}} \,)$
\item $\e_6 \, (A^4 \Qt^2) \sim \e_{\a_1 \dots \a_6} \, \e^{ik} \, \e^{jl} \, A_i^{\a_1\a_2}A_j^{\a_3\a_4}A_k^{\a_5\b_1}A_l^{\a_6\b_2} \Qt_{\beta_1}^{[I} \Qt_{\beta_2}^{J]}$ $ \sim (1,\scalebox{0.5}{\ydiagram{1,1}} \,)$
\end{itemize}
We will deconfine $A_1$, using \eqref{SU2NDeconfinement} and then our result in Section~\ref{SU(2N)1Antisym}. This step breaks the $SU(2)_A$ symmetry into $U(1)_A$. In \eqref{SU6T1}, the subscript of the two antisymmetric fields, $A_{\pm 1}$ correspond to the $U(1)_A$ charge. So let's give the chiral ring in this split form
\begin{itemize}
\item \scalebox{0.95}{$A_{+1} \Qt^2 \sim A_{+1}^{\a_1 \a_2} \Qt^{[I}_{\a_1} \Qt^{J]}_{\a_2}$ $ \sim \scalebox{0.5}{\ydiagram{1,1}} \,$ of $SU(5)_{\Qt}$}
\item \scalebox{0.95}{$A_{-1} \Qt^2 \sim A_{-1}^{\a_1 \a_2} \Qt^{[I}_{\a_1} \Qt^{J]}_{\a_2}$ $ \sim \scalebox{0.5}{\ydiagram{1,1}} $}
\item \scalebox{0.95}{$\e_6 \, A_{+1}^3 \equiv \e_6 \, (A^3)_{111} \sim \e_{\a_1 \dots \a_6} \, A_{+1}^{\a_1\a_2} A_{+1}^{\a_3 \a_4} A_{+1}^{\a_5 \a_6} \sim 1$}
\item \scalebox{0.95}{$\e_6 \, (A_{+1}^2 A_{-1}) \equiv \e_6 \, (A^3)_{112} \sim \e_{\a_1 \dots \a_6} \, A_{+1}^{\a_1\a_2} A_{+1}^{\a_3 \a_4} A_{-1}^{\a_5 \a_6} \sim 1$}
\item \scalebox{0.95}{$\e_6 \, (A_{+1} A_{-1}^2) \equiv \e_6 \, (A^3)_{122} \sim \e_{\a_1 \dots \a_6} \, A_{+1}^{\a_1\a_2} A_{-1}^{\a_3 \a_4} A_{-1}^{\a_5 \a_6} \sim 1$}
\item \scalebox{0.95}{$\e_6 \, A_{-1}^3 \equiv \e_6 \, (A^3)_{222} \sim \e_{\a_1 \dots \a_6} \, A_{-1}^{\a_1\a_2} A_{-1}^{\a_3 \a_4} A_{-1}^{\a_5 \a_6} \sim 1$}
\item \scalebox{0.95}{$\e_6 \, (A_{+1} A_{[+1} A_{-1]} Q \Qt) \sim \e_{\a_1 \dots \a_6} \, A_{+1}^{\a_1\a_2} \left(A_{+1}^{\a_3 \a_4} A_{-1}^{\a_5 \b} - A_{-1}^{\a_3 \a_4} A_{+1}^{\a_5 \b}\right) Q^{\a_6} \Qt_{\beta}^{I}$ $ \sim \scalebox{0.5}{\ydiagram{1}} $}
\item \scalebox{0.95}{$\e_6 \, (A_{-1} A_{[+1} A_{-1]} Q \Qt) \sim \e_{\a_1 \dots \a_6} \, A_{-1}^{\a_1\a_2} \left(A_{+1}^{\a_3 \a_4} A_{-1}^{\a_5 \b} - A_{-1}^{\a_3 \a_4} A_{+1}^{\a_5 \b}\right) Q^{\a_6} \Qt_{\beta}^{I}$ $ \sim \scalebox{0.5}{\ydiagram{1}} $}
\item \scalebox{0.95}{$\e_6 \, (A_{[+1} A_{-1]} A_{[+1} A_{-1]} \Qt^2) \sim \e_{\a_1 \dots \a_6} \, \left(A_{[+1}^{\a_1 \a_2} A_{-1]}^{\a_5 \b_1} \right) \left(A_{[+1}^{\a_3 \a_4} A_{-1]}^{\a_6 \b_2} \right) \Qt_{\beta_1}^{[I} \Qt_{\beta_2}^{J]} \sim 1 $}
\end{itemize}
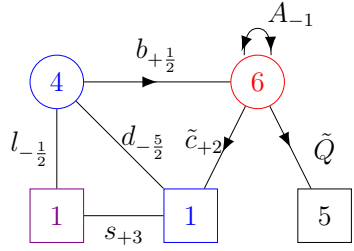
\be \label{SU6T1'} \scalebox{0.89}{\bpic[node distance=2cm,gSUnode/.style={circle,red,draw,minimum size=8mm},gUSpnode/.style={circle,blue,draw,minimum size=8mm},fnode/.style={rectangle,draw,minimum size=8mm}]  
\node at (1,1.5) {$\CT_{1'}:$};
\node[gSUnode] (G1) at (6,0) {$6$};
\node[gUSpnode] (G2) at (3,0) {$4$};
\node[fnode] (F1) at (7,-2) {$5$};
\node[fnode,blue] (F2) at (5,-2) {$1$};
\node[fnode,violet] (F3) at (3,-2) {$1$};
\draw (G1) -- pic[sloped]{arrow} (G2);
\draw (G1) -- pic[sloped]{arrow} (F1);
\draw (G1) -- pic[pos=0.4,sloped,very thick]{arrow= latex reversed} (F2);
\draw (G2) -- (F3);
\draw (G2) -- (F2);
\draw (F2) -- (F3);
\draw (6.2,0.4) to[out=90,in=0] pic[pos=0.1,sloped]{arrow} (6,0.8) to[out=180,in=90] pic[pos=0.5,sloped,very thick]{arrow=latex reversed} (5.8,0.4);
\node[right] at (8.5,-1) {$ \CW= 2 \,$ Planar Triangles};
\node at (6.5,1) {$A_{-1}$};
\node at (7,-1) {$\Qt$};
\node at (4.5,0.4) {$b_{+\frac{1}{2}}$};
\node at (2.6,-1) {$l_{-\frac{1}{2}}$};
\node at (5.2,-0.9) {$\ct_{+2}$};
\node at (4.3,-0.9) {$d_{-\frac{5}{2}}$};
\node at (4,-2.3) {$s_{+3}$};
\epic} \ee
Now we want to use the result of the section \eqref{SU(2N)1Antisym}, specialized in the case N=3. Before using the superpotential \eqref{SU2N1antisymT4}, we should split the $6$ antifundamentals into $5+1$. More precisely, $\Ht$, the antisymmetric of the global $SU(6)_{\Qt}$ there, split into $\Ht_{-1}$, an antisymmetric of $SU(5)_{\Qt}$ and $D_{+1}$ a fundamental. Similarly, $M$, the fundamental of $SU(6)$ there, split into a fundamental, $N_{+\frac{1}{2}}$ and a singlet (under the global $SU(5)_{\Qt}$ symmetry) $(b\ct)_{+\frac{5}{2}}$. We should also split $B_2$, the antisymmetric of $SU(4)_{Q}$ into a traceless antisymmetric tensor of $Usp(4)$ $B_{-1}$ and a singlet $s_{-1}$. Finally, we rename the three singlets $B_4, B_0$ and $\Bt$ there as $s_{+1},s_{-3}$ and $\b_{+2}$. After this splitting, the use of \eqref{SU2N1antisymT4} give
\be \label{SU6TIntermediate} \scalebox{0.89}{\bpic[node distance=2cm,gSUnode/.style={circle,red,draw,minimum size=8mm},gUSpnode/.style={circle,blue,draw,minimum size=8mm},fnode/.style={rectangle,draw,minimum size=8mm}]  
\node[gUSpnode] (G1) at (3,0) {$4$};
\node[fnode] (F1) at (6,0) {$5$};
\node[fnode,blue] (F2) at (5,-2) {$1$};
\node[fnode,violet] (F3) at (3,-2) {$1$};
\draw (G1) -- pic[pos=0.6,sloped]{arrow} (F1);
\draw (G1) -- (F3);
\draw (F2) -- (F3);
\draw (F2) -- pic[pos=0.6,sloped]{arrow} (F1);
\draw (6.3,0.4) to[out=90,in=0] pic[pos=0.1,sloped]{arrow} (6,0.8) to[out=180,in=90] pic[pos=0.6,sloped,very thick]{arrow=latex reversed} (5.7,0.4);
\draw (3.3,0.3) to[out=90,in=0] (3,0.7) to[out=180,in=90] (2.7,0.3);
\draw (3.4,-0.2) -- (4.8,-1.6);
\draw (3.2,-0.4) -- (4.6,-1.8);
\node[right] at (8,0) {$ \CW= \e_5 \, \big( \Ht_{-1}^2 D_{+1} s_{+1} + N_{+\frac{1}{2}}^2 \Ht_{-1} D_{+1} s_{-1} + B_{-1} N_{+\frac{1}{2}}^2 \Ht_{-1} D_{+1} $};
\node[right] at (8.6,-1) {$+ N_{+\frac{1}{2}} (b\ct)_{+\frac{5}{2}} \Ht_{-1}^2 s_{-1} + B_{-1} N_{+\frac{1}{2}} (b\ct)_{+\frac{5}{2}} \Ht_{-1}^2 + N_{+\frac{1}{2}}^4 D_{+1} s_{-3}$};
\node[right] at (8.6,-2) {$+ N_{+\frac{1}{2}}^3 (b\ct)_{+\frac{5}{2}} \Ht_{-1} s_{-3} \big) + s_{+1} s_{-3} \b_{+2} + \b_{+2} s_{-1}^2 + \b_{+2} B_{-1}^2$};
\node[right] at (8.6,-3) {$+ l_{-\frac{1}{2}} d_{-\frac{5}{2}} s_{+3} + d_{-\frac{5}{2}} (b\ct)_{+\frac{5}{2}}$};
\node at (6.5,1) {$\Ht_{-1}$};
\node at (3.3,1) {$B_{-1}$};
\node at (4.5,0.4) {$N_{+\frac{1}{2}}$};
\node at (2.6,-1) {$l_{-\frac{1}{2}}$};
\node at (6.1,-1) {$D_{+1}$};
\node at (4.6,-0.9) {$(b\ct)_{+\frac{5}{2}}$};
\node at (3.6,-1.2) {$d_{-\frac{5}{2}}$};
\node at (4,-2.3) {$s_{+3}$};
\node at (5,-3) {+ singlets $ (s_{+1}, s_{-1}, s_{-3} \, \& \, \b_{+2})$};
\epic} \ee
After integrating out the massive fields $d_{-\frac{5}{2}}, (b\ct)_{+\frac{5}{2}}$ and rescaling the fields we obtain 

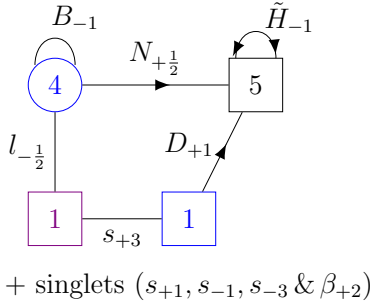
\be \label{SU6T2} \scalebox{0.89}{\bpic[node distance=2cm,gSUnode/.style={circle,red,draw,minimum size=8mm},gUSpnode/.style={circle,blue,draw,minimum size=8mm},fnode/.style={rectangle,draw,minimum size=8mm}]  
\node at (1.5,1.5) {$\CT_2:$};
\node[gUSpnode] (G1) at (3,0) {$4$};
\node[fnode] (F1) at (6,0) {$5$};
\node[fnode,blue] (F2) at (5,-2) {$1$};
\node[fnode,violet] (F3) at (3,-2) {$1$};
\draw (G1) -- pic[pos=0.6,sloped]{arrow} (F1);
\draw (G1) -- (F3);
\draw (F2) -- (F3);
\draw (F2) -- pic[pos=0.6,sloped]{arrow} (F1);
\draw (6.3,0.4) to[out=90,in=0] pic[pos=0.1,sloped]{arrow} (6,0.8) to[out=180,in=90] pic[pos=0.6,sloped,very thick]{arrow=latex reversed} (5.7,0.4);
\draw (3.3,0.3) to[out=90,in=0] (3,0.7) to[out=180,in=90] (2.7,0.3);
\node[right] at (7.6,0) {$ \CW= \e_5 \, \big( \Ht_{-1}^2 D_{+1} s_{+1} + N_{+\frac{1}{2}}^2 \Ht_{-1} D_{+1} s_{-1} + B_{-1} N_{+\frac{1}{2}}^2 \Ht_{-1} D_{+1}$};
\node[right] at (8.2,-1) {$+ N_{+\frac{1}{2}} l_{-\frac{1}{2}} \Ht_{-1}^2 s_{+3} s_{-1} + B_{-1} N_{+\frac{1}{2}} l_{-\frac{1}{2}} \Ht_{-1}^2 s_{+3} + N_{+\frac{1}{2}}^4 D_{+1} s_{-3}$};
\node[right] at (8.2,-2) {$+ N_{+\frac{1}{2}}^3 l_{-\frac{1}{2}} \Ht_{-1} s_{+3} s_{-3} \big) + s_{+1} s_{-3} \b_{+2} + \b_{+2} s_{-1}^2 + \b_{+2} B_{-1}^2$};
\node at (6.5,1) {$\Ht_{-1}$};
\node at (3.3,1) {$B_{-1}$};
\node at (4.5,0.4) {$N_{+\frac{1}{2}}$};
\node at (2.6,-1) {$l_{-\frac{1}{2}}$};
\node at (5,-0.9) {$D_{+1}$};
\node at (4,-2.3) {$s_{+3}$};
\node at (5,-3) {+ singlets $ (s_{+1}, s_{-1}, s_{-3} \, \& \, \b_{+2})$};
\epic} \ee
Now we want to confine the $Usp(4)$ gauge group with the antisymmetric and $6$ flavors. Unfortunately, we cannot immediately use our result about $Usp(2N)$ of Section \ref{Usp2N} for the following reason. The flipper $\b_{+2}$ appears in \eqref{SU6T2} in three different terms but we know the final superpotential only when we flip the whole tower of the traces of the antisymmetric as in \eqref{UspT0}. In this case since the rank is small, it's easy to apply our strategy of the Section \ref{Usp2N} starting with $\cW=0$ to get
\be \label{Usp4W=0} \scalebox{0.9}{\bpic[node distance=2cm,gSUnode/.style={circle,red,draw,minimum size=8mm},gUSpnode/.style={circle,blue,draw,minimum size=8mm},fnode/.style={rectangle,draw,minimum size=8mm}]  
\node[gUSpnode,minimum size=1cm] (G1) at (0,0) {$4$};
\node[fnode] (F1) at (2.5,0) {$5$};
\node[fnode,violet] (F2) at (0,-1.8) {$1$};
\draw (G1) -- pic[pos=0.7,sloped]{arrow} (F1) node[midway,below] {$N_{+\frac{1}{2}}$};
\draw (G1) -- (F2) node[midway,left] {$l_{-\frac{1}{2}}$};
\draw (0.3,0.4) to[out=90,in=0]  (0,0.8) to[out=180,in=90] (-0.3,0.4);
\node[right] at (-0.5,-3) {$ \CW= 0$};
\node at (0.6,0.9) {$B_{-1}$};
\node at (4.5,0) {$\equiv$};
\node[fnode,minimum size=1cm] (F3) at (7,0) {$5$};
\node[fnode,orange] (F4) at (6.1,-2) {$1$};
\node[fnode,violet] (F5) at (7.9,-2) {$1$};
\draw (F3) -- pic[pos=0.3,sloped]{arrow} (F4) node[midway,left] {$D_{-1}$};
\draw (F3) -- pic[pos=0.4,sloped,very thick]{arrow=latex reversed} (F5) node[midway,right] {$M_0$};
\draw (7.2,0.5) to[out=90,in=0] pic[pos=0.1,sloped]{arrow} (7,0.9) to[out=180,in=90] pic[pos=0.6,sloped,very thick]{arrow=latex reversed} (6.8,0.5);
\draw (6.6,0.5) to[out=90,in=-180] pic[pos=0.4,sloped,very thick]{arrow=latex reversed} (7,1.2) to[out=0,in=90] pic[pos=0.8,sloped]{arrow} (7.4,0.5);
\node[right] at (3.2,-3) {$\CW = \e_5 \, \left(T_{-2} \Ht_{+1}^2 M_0 + \P_0 \Ht_{+1} D_{-1} + M_0 \P_0^2\right)$};
\node at (8.2,1) {$\P_0, \Ht_{+1}$};
\epic} \ee  
The mapping is $N_{+\frac{1}{2}} N_{+\frac{1}{2}} \llra \Ht_1, N_{+\frac{1}{2}} l_{-\frac{1}{2}} \llra M_0, N_{+\frac{1}{2}}^2 B_{-1} \llra \P_0, N_{+\frac{1}{2}} l_{-\frac{1}{2}} B_{-1} \llra D_{-1} $ and $B_{-1}^2 \llra T_{-2}$.

We now use this result into \eqref{SU6T2}. We see that the singlets $\b_{+2}$ and $T_{-2}$ become massive. After integrating them out and rescaling fields, we get the final result 

\be \label{SU6T3} \scalebox{0.9}{\bpic[node distance=2cm,gSUnode/.style={circle,red,draw,minimum size=8mm},gUSpnode/.style={circle,blue,draw,minimum size=8mm},fnode/.style={rectangle,draw,minimum size=8mm}]    
\node at (-3,1) {$\CT_3:$};
\node[fnode,blue] (F1) at (-1,0) {$1$};
\node[fnode] (F2) at (0,-2) {$5$};
\node[fnode,orange] (F3) at (-2,0) {$1$};
\node[fnode,violet] (F4) at (1.5,0) {$1$};
\draw (F1) -- pic[pos=0.6,sloped]{arrow} (F2);
\draw (F2) -- pic[pos=0.4,sloped]{arrow} (F3);
\draw (F2) -- pic[pos=0.5,sloped,very thick]{arrow=latex reversed} (F4);
\draw (-0.4,-2) to[out=180,in=90] pic[pos=0.1,sloped]{arrow} (-0.8,-2.4) to[out=270,in=180] (-0.4,-2.8) to[out=0,in=-90] pic[pos=0.9,sloped]{arrow} (-0.1,-2.4);  
\draw (0.4,-2) to[out=0,in=90] pic[pos=0.4,sloped,very thick]{arrow=latex reversed} (0.8,-2.4) to[out=270,in=0] (0.4,-2.8) to[out=180,in=270] pic[pos=0.6,sloped,very thick]{arrow=latex reversed} (0.1,-2.4);  
\node[right] at (2.6,-0.5) {$\CW= \e_5 \big( \Ht_{-1}^2 D_{+1} s_{+1} + \Ht_{+1} \Ht_{-1} D_{+1} s_{-1} + \Ht_{-1}^2 D_{-1} s_{+3}$};
\node[right] at (2.6,-1.5) {$+ \Ht_{+1}^2 D_{+1} s_{-3} + \P_0 \Ht_{-1} D_{+1} + \P_0 \Ht_{+1} D_{-1} + M_0 \Ht_{+1} \Ht_{-1} s_{+3} s_{-3}$};
\node[right] at (2.6,-2.5) {$+ M_0 \Ht_{-1}^2 s_{+3} s_{-1} + M_0 \Ht_{+1}^2 s_{+1} s_{-3} + M_0 \Ht_{+1}^2 s_{-1}^2 + M_0 \P_0^2 \big)$};
\node at (1.2,-1.2) {$M_0$};
\node at (-1.5,-1.2) {$D_{+1}$};
\node at (-0.1,-0.8) {$D_{-1}$};
\node at (-0.8,-3.1) {$\Ht_{+1},\Ht_{-1}$};
\node at (1,-2.8) {$\P_0$};
\node at (0,-3.8) {+ singlets $ (s_{+3}, s_{+1}, s_{-1} \, \& \,  s_{-3})$};
\epic} \ee 
The final superpotential can be repackaged in a manifest $SU(2)_{A}$ invariant way as 
\be \CW = \Ht^2 D S + \P_0 \Ht D + M_0 \Ht^2 S^2 + M_0 \P_0^2 \,,\ee
where $A^3 \lra S=\{s_{+3}, s_{+1}, s_{-1}, s_{-3}\}$ is a completely symmetric 3-tensor of $SU(2)_A$, $A\Qt^2 \lra \Ht=\{\Ht_{+1},\Ht_{-1}\}$, $A^3 Q \Qt \lra D=\{D_{+1},D_{-1}\}$ are $SU(2)_A$ doublets and $Q \Qt \lra M_0$, $A^4 \Qt^2 \lra \P_0$ are $SU(2)_A$ singlets. We recover the result of Section 3.1.9 of \cite{Csaki:1996zb}. The final mapping of the chiral ring generators is 
\be \label{mapSU61}
\ba{c}\CT_1 \\
Q \, \Qt \\
A_{+1} \, \Qt^2 \\
A_{-1} \, \Qt^2 \\
\e_6 \, A_{+1}^3 \\
\e_6 \, (A_{+1}^2 \, A_{-1}) \\
\e_6 \, (A_{-1}^2 \, A_{+1}) \\
\e_6 \, A_{-1}^3 \\
\e_6 \, (A_{+1} \, A_{[+1} \, A_{-1]} \, Q \, \Qt) \\
\e_6 \, (A_{-1} \, A_{[+1} \, A_{-1]} \, Q \, \Qt)  \\
\e_6 \, (A_{[+1} \, A_{-1]} \, A_{[+1} \, A_{-1]} \, \Qt^2) 
\ea
\Longleftrightarrow
\ba{c} \CT_{1'} \\
l_{-\frac{1}{2}} \, b_{+\frac{1}{2}} \, \Qt \\
b_{+\frac{1}{2}} b_{+\frac{1}{2}} \, \Qt^2  \\
A_{-1} \, \Qt^2 \\
s_{+3} \\
\e_6 \, (b_{+\frac{1}{2}}^4 \, A_{-1}) \\
\e_6 \, (b_{+\frac{1}{2}}^2 \, A_{-1}^2) \\
\e_6 \, (A_{-1}^3) \\
A_{-1} \, \ct_{+2} \, \Qt \\
\e_6 \, (A_{-1}^2 \, b_{+\frac{1}{2}}^3 \, l_{-\frac{1}{2}} \, \Qt)  \\
\e_6 \, (b_{+\frac{1}{2}}^4 \, A_{-1}^2 \, \Qt^2) 
\ea
\Longleftrightarrow 
\ba{c}\CT_2\\
l_{-\frac{1}{2}} \, N_{+\frac{1}{2}} \\
N_{+\frac{1}{2}} \, N_{+\frac{1}{2}} \\
\Ht_{-1} \\
s_{+3} \\
s_{+1} \\
s_{-1} \\
s_{-3} \\
D_{+1} \\
N_{+\frac{1}{2}} \, l_{-\frac{1}{2}} \, B_{-1} \\
N_{+\frac{1}{2}}^2 \, B_{-1}   
\ea
\Longleftrightarrow
\ba{c} \CT_{3}\\
M_0 \\
\Ht_1 \\
\Ht_{-1}  \\
s_{+3} \\
s_{+1} \\
s_{-1} \\
s_{-3} \\
D_{+1} \\
D_{-1} \\
\P_0
\ea
\ee
  
\subsection{$SU(5)$ with 2 {\tiny \ydiagram{1,1}} + 4 $\overline{{\tiny\ydiagram{1}}}$ + 2 ${\tiny \ydiagram{1}}$} \label{DegenerateOpPhenomenon}
The third case is the $SU(5)$ gauge theory with $2$ antisymmetric, $4$ antifundamental and $2$ fundamental fields with continuous global symmetry $SU(2)_A \times SU(4)_{\Qt} \times SU(2)_{Q} \times U(1)^2$.
\be \label{SU52antisymT1} \bpic[node distance=2cm,gSUnode/.style={circle,red,draw,minimum size=8mm},gUSpnode/.style={circle,blue,draw,minimum size=8mm},fnode/.style={rectangle,draw,minimum size=8mm}]  
\node at (3.6,1) {$\CT_1:$};
\node[gSUnode] (G1) at (6,0) {$5$};
\node[fnode] (F1) at (7,-2) {$4$};
\node[fnode] (F2) at (5,-2) {$2$};
\draw (G1) -- pic[sloped]{arrow} (F1);
\draw (G1) -- pic[pos=0.3,sloped]{arrow} (F2);
\draw (5.6,0.2) to[out=110,in=90] pic[pos=0.3,sloped]{arrow} (5.1,0) to[out=270,in=250] pic[pos=0.7,sloped]{arrow} (5.6,-0.2);
\draw (6.4,0.2) to[out=70,in=90] pic[pos=0.5,sloped,very thick]{arrow=latex reversed} (6.9,0) to[out=270,in=290] pic[pos=0.5,sloped,very thick]{arrow=latex reversed} (6.4,-0.2);
\node[right] at (8,-0.4) {$ \CW= 0$};
\node at (4.7,0) {$A_{+1}$};
\node at (7.3,0) {$A_{-1}$};
\node at (7,-1) {$\Qt$};
\node at (5,-1) {$Q$};
\epic \ee
The chiral ring generators are
\begin{itemize}
\item $Q \, \Qt \sim Q_i^{\a} \Qt^{I}_{\a}$ transforming in the $(1, \scalebox{0.5}{\ydiagram{1}} \,, \scalebox{0.5}{\ydiagram{1}} \, )$ of $SU(2)_{A} \times SU(4)_{\Qt} \times SU(2)_{Q}$
\item $A\Qt^2 \sim A_a^{\a_1 \a_2} \, \Qt^{[I}_{\a_1} \Qt^{J]}_{\a_2}$ $ \sim (\, \scalebox{0.5}{\ydiagram{1}} \, ,\scalebox{0.5}{\ydiagram{1,1}} \,,1)$
\item $\e_5 \, (A^2 Q) \sim \e_{\a_1 \dots \a_5} A_{(a}^{\a_1 \a_2} A_{b)}^{\a_3 \a_4} \, Q_{i}^{\a_5}$ $ \sim (\, \scalebox{0.5}{\ydiagram{2}} \, ,1, \, \scalebox{0.5}{\ydiagram{1}} \,)$
\item $\e_5 \, (A^3 \, \Qt) \sim \e_{\a_1 \dots \a_5} \, \e^{b_1 b_2} \, A_{a}^{\a_1 \a_2} A_{b_1}^{\a_3 \a_4} A_{b_2}^{\a_5 \b} \Qt_{\b}^{I}$ $ \sim (\, \scalebox{0.5}{\ydiagram{1}} \, , \scalebox{0.5}{\ydiagram{1}} \,, 1)$
\item $\e_5 \, (A^2 \, Q^2 \, \Qt) \sim \e_{\a_1 \dots \a_5} \, \e^{b_1 b_2} \, \e^{i_1,i_2} \, A_{b_1}^{\a_1 \a_2} A_{b_2}^{\a_3 \b} Q_{i_1}^{\a_4} Q_{i_2}^{\a_5} \Qt_{\b}^{I}$ $ \sim (1, \scalebox{0.5}{\ydiagram{1}} \,, 1)$
\end{itemize}
We now deconfine the two antisymmetric using \eqref{SU2N+1Deconfinement} and so we break $SU(2)_A$ into $U(1)_A$. In \eqref{SU52antisymT1}, the subscript of the two antisymmetric fields, $A_{\pm 1}$ correspond to the $U(1)_A$ charge (corresponding to the weight of the representation of $SU(2)_A$). After the splitting, the chiral ring generators become

\begin{itemize}
\item $A_{+1}\Qt^2 \sim A_{+1}^{\a_1 \a_2} \, \Qt^{[I}_{\a_1} \Qt^{J]}_{\a_2}$ $ \sim (\, \scalebox{0.5}{\ydiagram{1,1}} \,,1)$ of $SU(4)_{\Qt} \times SU(2)_{Q}$
\item $A_{-1}\Qt^2 \sim A_{-1}^{\a_1 \a_2} \, \Qt^{[I}_{\a_1} \Qt^{J]}_{\a_2}$ $ \sim (\, \scalebox{0.5}{\ydiagram{1,1}} \,,1)$
\item $\e_5 \, (A_{+1}^2 Q) \sim \e_{\a_1 \dots \a_5} A_{+1}^{\a_1 \a_2} A_{+1}^{\a_3 \a_4} \, Q_{i}^{\a_5}$ $ \sim (1, \, \scalebox{0.5}{\ydiagram{1}} \,)$
\item $\e_5 \, (A_{+1} A_{-1} Q) \sim \e_{\a_1 \dots \a_5} A_{(+1}^{\a_1 \a_2} A_{-1)}^{\a_3 \a_4} \, Q_{i}^{\a_5}$ $ \sim (1, \, \scalebox{0.5}{\ydiagram{1}} \,)$
\item $\e_5 \, (A_{-1}^2 Q) \sim \e_{\a_1 \dots \a_5} A_{-1}^{\a_1 \a_2} A_{-1}^{\a_3 \a_4} \, Q_{i}^{\a_5}$ $ \sim (1, \, \scalebox{0.5}{\ydiagram{1}} \,)$
\item $\e_5 \, (A_{+1} A_{[+1} A_{-1]} \, \Qt) \sim \e_{\a_1 \dots \a_5} \, A_{+1}^{\a_1 \a_2} A_{[+1}^{\a_3 \a_4} A_{-1]}^{\a_5 \b} \Qt_{\b}^{I}$ $ \sim (\, \scalebox{0.5}{\ydiagram{1}} \,, 1)$
\item $\e_5 \, (A_{-1} A_{[+1} A_{-1]} \, \Qt) \sim \e_{\a_1 \dots \a_5} \, A_{-1}^{\a_1 \a_2} A_{[+1}^{\a_3 \a_4} A_{-1]}^{\a_5 \b} \Qt_{\b}^{I}$ $ \sim (\, \scalebox{0.5}{\ydiagram{1}} \,, 1)$
\item $\e_5 \, (A_{[+1} A_{-1]} \, Q^2 \, \Qt) \sim \e_{\a_1 \dots \a_5} \, \e^{i_1,i_2} \, A_{[+1}^{\a_1 \a_2} A_{-1]}^{\a_3 \b} Q_{i_1}^{\a_4} Q_{i_2}^{\a_5} \Qt_{\b}^{I}$ $ \sim (\, \scalebox{0.5}{\ydiagram{1}} \,, 1)$
\end{itemize}
\be \label{SU52antisymT1'} \scalebox{0.9}{\bpic[node distance=2cm,gSUnode/.style={circle,red,draw,minimum size=8mm},gUSpnode/.style={circle,blue,draw,minimum size=8mm},fnode/.style={rectangle,draw,minimum size=8mm}]   
\node at (2,3) {$\CT_{1'}:$};
\node[gSUnode] (G1) at (6,0) {$5$};
\node[gUSpnode] (G2) at (3,0) {$2$};
\node[gUSpnode] (G3) at (9,0) {$2$};
\node[fnode] (F1) at (7,-2) {$4$};
\node[fnode] (F2) at (5,-2) {$2$};
\node[fnode,violet] (F3) at (4,2) {$1$};
\node[fnode,orange] (F4) at (8,2) {$1$};
\draw (G1) -- pic[sloped]{arrow} (G2);
\draw (G1) -- pic[sloped,very thick]{arrow=latex reversed} (G3);
\draw (G1) -- pic[pos=0.6,sloped]{arrow} (F1);
\draw (G1) -- pic[pos=0.4,sloped]{arrow} (F2);
\draw (G1) -- pic[sloped,very thick]{arrow=latex reversed} (F3);
\draw (G1) -- pic[sloped]{arrow} (F4);
\draw (G2) -- (F3);
\draw (G3) -- (F4);
\node[right] at (10.5,0) {$ \CW= 2 \,$Planar Triangles};
\node at (5,-1) {$Q$};
\node at (7,-1) {$\Qt$};
\node at (4.5,-0.4) {$b_{+\frac{1}{2}}$};
\node at (7.5,-0.4) {$b_{-\frac{1}{2}}$};
\node at (5.4,1.1) {$\ct_{+2}$};
\node at (6.5,1.1) {$\ct_{-2}$};
\node at (3.1,1) {$l_{-\frac{5}{2}}$};
\node at (8.9,1) {$l_{+\frac{5}{2}}$};
\epic} \ee
Once again, the subscripts in \eqref{SU52antisymT1'} correspond to the $U(1)_A$ charges of the fields. The next step is to confine the $SU(5)$ gauge group with \eqref{SUbuildingBlock}. Notice that $l_{-\frac{5}{2}}$ and $l_{+\frac{5}{2}}$ will become massive. After integrating them out, computing the $\det(mesons)$ of degree 6 and rescaling the fields we obtain
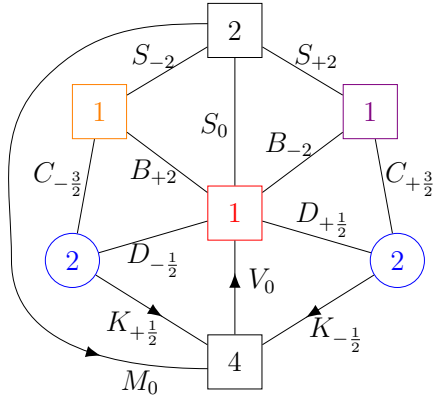
\be \label{SU52antisymT2} \scalebox{0.9}{\bpic[node distance=2cm,gSUnode/.style={circle,red,draw,minimum size=8mm},gUSpnode/.style={circle,blue,draw,minimum size=8mm},fnode/.style={rectangle,draw,minimum size=8mm}]     
\node at (-3.6,3.8) {$\CT_2:$};
\node[fnode,red] (F1) at (0,0) {$1$};
\node[gUSpnode] (G1) at (-2.4,-0.7) {$2$};
\node[gUSpnode] (G2) at (2.4,-0.7) {$2$};
\node[fnode] (F2) at (0,-2.2) {$4$};
\node[fnode,orange] (F3) at (-2,1.5) {$1$};
\node[fnode,violet] (F4) at (2,1.5) {$1$};
\node[fnode] (F5) at (0,2.7) {$2$};
\draw (G1) -- (F1);
\draw (G1) -- pic[pos=0.6,sloped]{arrow} (F2);
\draw (G1) -- (F3);
\draw (G2) -- (F1);
\draw (G2) -- pic[sloped,very thick]{arrow=latex reversed} (F2);
\draw (G2) -- (F4);
\draw (F5) -- (F3);
\draw (F5) -- (F4);
\draw (F1) -- pic[sloped,very thick]{arrow=latex reversed} (F2);
\draw (F1) -- (F3);
\draw (F1) -- (F4);
\draw (F1) -- (F5);
\draw (-0.4,2.8) to[out=180,in=45] (-2.8,2) to[out=-135,in=90] (-3.3,-0.7) to[out=-90,in=180] pic[pos=0.6,sloped]{arrow}  (-0.4,-2.3);
\node[right] at (4,1.5) {$\CW = \, \e_4 \, \e^2 \, \e^2 \, \e^2 \, \left[M_0^2 C_{-\frac{3}{2}} K_{+\frac{1}{2}} C_{+\frac{3}{2}} K_{-\frac{1}{2}} \right.$};
\node[right] at (4,0.5) {$+ \, K_{+\frac{1}{2}}^2 K_{-\frac{1}{2}}^2 S_{-2} S_{+2} + \, K_{+\frac{1}{2}}^2 C_{+\frac{3}{2}} K_{-\frac{1}{2}} S_{-2} M_0$};
\node[right] at (4,-0.5) {$\left. + \, K_{-\frac{1}{2}}^2 C_{-\frac{3}{2}} K_{+\frac{1}{2}} S_{+2} M_0 \right] + \, S_{0} V_0 M_0$};
\node[right] at (4,-1.5) {+ 6 Planar Triangles};
\node at (-2.6,0.5) {$C_{-\frac{3}{2}}$};
\node at (2.6,0.5) {$C_{+\frac{3}{2}}$};
\node at (-1.2,2.3) {$S_{-2}$};
\node at (1.2,2.3) {$S_{+2}$};
\node at (-0.3,1.3) {$S_{0}$};
\node at (1.5,-1.8) {$K_{-\frac{1}{2}}$};
\node at (-1.5,-1.7) {$K_{+\frac{1}{2}}$};
\node at (-1.2,-0.7) {$D_{-\frac{1}{2}}$};
\node at (1.3,-0.1) {$D_{+\frac{1}{2}}$};
\node at (-1.2,0.6) {$B_{+2}$};
\node at (0.8,1) {$B_{-2}$};
\node at (0.4,-1) {$V_0$};
\node at (-1.4,-2.5) {$M_0$};
\epic} \ee 
Then we use \eqref{UspbuildingBlock} for the left $Usp(2)$. The fields $B_{+2}$ and $V_0$ get a mass. We have to integrate them out and compute the Pfaffian superpotential. Let us write more explicitely the Pfaffian term because it will be useful later. The mesons involving in the Pfaffian are $n_{-1} \equiv [C_{-\frac{3}{2}} K_{+\frac{1}{2}}], \, \Ht_{+1} \equiv [K_{+\frac{1}{2}} K_{+\frac{1}{2}}], \, [D_{-\frac{1}{2}} K_{+\frac{1}{2}}]$ and $[C_{-\frac{3}{2}} D_{-\frac{1}{2}}]$. We didn't give a name to the last two mesons because they are massive and will be integrate out by the E.O.M of $V_0$ and $B_{+2}$. The Pfaffian is then given by
\be 
\Pf
\begin{pmatrix}
\Ht_{+1} & [D_{-\frac{1}{2}} K_{+\frac{1}{2}}] & n_{-1} \\
& 0 & [C_{-\frac{3}{2}} D_{-\frac{1}{2}}] \\
& & 0
\end{pmatrix}
\sim \e_{4} \, \left[\Ht_{+1}^2 [C_{-\frac{3}{2}} D_{-\frac{1}{2}}] + \Ht_{+1} n_{-1} [D_{-\frac{1}{2}} K_{+\frac{1}{2}}] \right]
\ee
\begin{align*}
\text{E.O.M: \, from \,} & V_0: [D_{-\frac{1}{2}} K_{+\frac{1}{2}}] = K_{-\frac{1}{2}} D_{+\frac{1}{2}} + S_{0} M_0 \\
\text{from \,} & B_{+2}: [C_{-\frac{3}{2}} D_{-\frac{1}{2}}] = S_{-2} S_0
\end{align*}
Where we rescaled the fields. Therefore the Pfaffian gives the following contribution 
\be 
\Pf( ) = \e_4 \, \left[\Ht_{+1}^2 S_{-2} S_0 + \Ht_{+1} n_{-1} \left( K_{-\frac{1}{2}} D_{+\frac{1}{2}} + S_{0} M_0 \right) \right]
\ee
The theory after this $Usp(2)_{left}$ confinement is

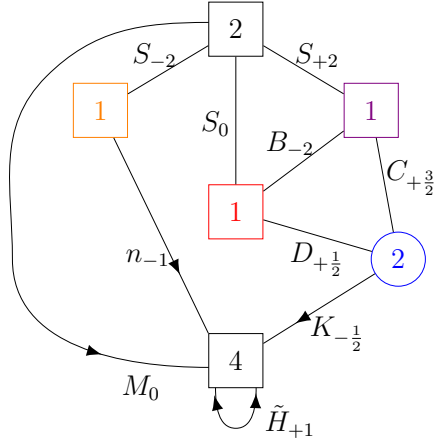
\be \label{SU52antisymT3} \scalebox{0.9}{\bpic[node distance=2cm,gSUnode/.style={circle,red,draw,minimum size=8mm},gUSpnode/.style={circle,blue,draw,minimum size=8mm},fnode/.style={rectangle,draw,minimum size=8mm}] 
\node at (-3.6,3.6) {$\CT_3:$};
\node[fnode,red] (F1) at (0,0) {$1$};
\node[gUSpnode] (G2) at (2.4,-0.7) {$2$};
\node[fnode] (F2) at (0,-2.2) {$4$};
\node[fnode,orange] (F3) at (-2,1.5) {$1$};
\node[fnode,violet] (F4) at (2,1.5) {$1$};
\node[fnode] (F5) at (0,2.7) {$2$};
\draw (G2) -- (F1);
\draw (G2) -- pic[pos=0.6,sloped,very thick]{arrow=latex reversed} (F2);
\draw (G2) -- (F4);
\draw (F5) -- (F3);
\draw (F5) -- (F4);
\draw (F1) -- (F4);
\draw (F1) -- (F5);
\draw (F2.north west) -- pic[pos=0.3,sloped]{arrow} (F3);
\draw (-0.4,2.8) to[out=180,in=45] (-2.8,2) to[out=-135,in=90] (-3.3,-0.7) to[out=-90,in=180] pic[pos=0.6,sloped]{arrow}  (-0.4,-2.3);
\draw (0.3,-2.6) to[out=-90,in=0] pic[pos=0.1,sloped]{arrow} (0,-3.2) to[out=-180,in=-90] pic[pos=0.6,sloped,very thick]{arrow=latex reversed} (-0.3,-2.6);  
\node[right] at (4,2) {$\CW= \, \e_4 \, \e^2 \, \e^2 \, \left[M_0^2 n_{-1} C_{+\frac{3}{2}} K_{-\frac{1}{2}} \right.$};
\node[right] at (4,1) {$+ \, \Ht_{+1} K_{-\frac{1}{2}}^2 S_{+2} S_{-2} + \, \Ht_{+1} C_{+\frac{3}{2}} K_{-\frac{1}{2}} S_{-2} M_0$};
\node[right] at (4,0) {$\left. + \, K_{-\frac{1}{2}}^2 n_{-1} S_{+2} M_0 \right] + \, \e_4 \, \left[ \Ht_{+1}^2 S_{-2} S_{0} \right. $};
\node[right] at (4,-1) {$\left. + \, \Ht_{+1} n_{-1} \left(K_{-\frac{1}{2}} D_{+\frac{1}{2}} + S_{0} M_0 \right) \right] $};
\node[right] at (4,-2) {$+ \, S_0 B_{-2} S_{+2} + \, B_{-2} D_{+\frac{1}{2}} C_{+\frac{3}{2}} $};
\node at (-1.2,2.3) {$S_{-2}$};
\node at (1.2,2.3) {$S_{+2}$};
\node at (-0.3,1.3) {$S_{0}$};
\node at (2.6,0.5) {$C_{+\frac{3}{2}}$};
\node at (1.5,-1.8) {$K_{-\frac{1}{2}}$};
\node at (0.8,1) {$B_{-2}$};
\node at (1.2,-0.7) {$D_{+\frac{1}{2}}$};
\node at (-1.3,-0.7) {$n_{-1}$};
\node at (-1.4,-2.6) {$M_0$};
\node at (0.8,-3.1) {$\Ht_{+1}$};
\epic} \ee 
The last step is to confine the other $Usp(2)$. The field $B_{-2}$ gets a mass. As in the last step, let us write the Pfaffian term. The mesons involve are: $\Ht_{-1} \equiv [K_{-\frac{1}{2}} K_{-\frac{1}{2}}], \, f_0 \equiv [K_{-\frac{1}{2}} D_{+\frac{1}{2}}], \, n_{+1} \equiv [C_{+\frac{3}{2}} K_{-\frac{1}{2}}]$ and $[C_{+\frac{3}{2}} D_{+\frac{1}{2}}]$. The last one is massive and will be integrate out with the E.O.M of $B_{-2}$. The Pfaffian is
\be 
\Pf
\begin{pmatrix}
\Ht_{-1} & f_0 & n_{+1} \\
& 0 & [C_{+\frac{3}{2}} D_{+\frac{1}{2}}] \\
& & 0
\end{pmatrix}
\sim \e_{4} \, \left[\Ht_{-1}^2 [C_{+\frac{3}{2}} D_{+\frac{1}{2}}] + \Ht_{-1} n_{+1} f_0 \right]
\ee
\begin{align*}
\text{E.O.M: \, from \,} & B_{-2}: [C_{+\frac{3}{2}} D_{+\frac{1}{2}}] = S_{+2} S_0
\end{align*}
Where we rescaled the fields. Therefore the Pfaffian gives the following contribution 
\be \label{PfaffianSU5T3}
\Pf( ) = \e_4 \, \left[\Ht_{-1}^2 S_{+2} S_0 + \Ht_{-1} n_{+1} f_0 \right]
\ee 
We get the following WZ model
\be \label{SU52antisymT4} \scalebox{0.9}{\bpic[node distance=2cm,gSUnode/.style={circle,red,draw,minimum size=8mm},gUSpnode/.style={circle,blue,draw,minimum size=8mm},fnode/.style={rectangle,draw,minimum size=8mm}]     
\node at (-3.3,2) {$\CT_4:$};
\node[fnode,red] (F1) at (0,0) {$1$};
\node[fnode] (F2) at (0,-2) {$4$};
\node[fnode,orange] (F3) at (-2,0) {$1$};
\node[fnode,violet] (F4) at (2,0) {$1$};
\node[fnode] (F5) at (0,2) {$2$};
\draw (F1) -- pic[pos=0.7,sloped]{arrow} (F2);
\draw (F2) -- pic[pos=0.4,sloped]{arrow} (F3);
\draw (F2) -- pic[pos=0.5,sloped,very thick]{arrow=latex reversed} (F4);
\draw (F1) -- (F5);
\draw (F3) -- (F5);
\draw (F4) -- (F5);
\draw (-0.4,2) to[out=180,in=45] (-2.4,1.4) to[out=-135,in=90] (-2.9,-0.2) to[out=-90,in=180] pic[pos=0.6,sloped]{arrow}  (-0.4,-1.8);
\draw (-0.4,-2) to[out=180,in=90] pic[pos=0.1,sloped]{arrow} (-0.8,-2.4) to[out=270,in=180] (-0.4,-2.8) to[out=0,in=-90] pic[pos=0.9,sloped]{arrow} (-0.1,-2.4);  
\draw (0.4,-2) to[out=0,in=90] pic[pos=0.4,sloped,very thick]{arrow=latex reversed} (0.8,-2.4) to[out=270,in=0] (0.4,-2.8) to[out=180,in=270] pic[pos=0.6,sloped,very thick]{arrow=latex reversed} (0.1,-2.4);  
\node[right] at (4,1) {$\CW= \e_4 \, \big( M_0^2 n_{-1} n_{+1} + \, \Ht_{+1} \Ht_{-1} S_{+2} S_{-2} $};
\node[right] at (4,0) {$+ \, \Ht_{+1}^2 S_{-2} S_{0} + \, \Ht_{-1}^2 S_{+2} S_{0} + \, \Ht_{+1} n_{1} S_{-2} M_0 $};
\node[right] at (4,-1) {$+ \, \Ht_{-1} n_{-1} S_{+2} M_0 + \, \Ht_{+1} n_{-1} \left(f_0 + S_0 M_0 \right)$};
\node[right] at (4,-2) {$+ \, \Ht_{-1} n_{+1} f_0 \, + \,$ \textcolor{red}{$\Ht_{-1} n_{+1} S_{0} M_0$} $\big)$};
\node at (-0.8,-3) {$\Ht_{+1}$};
\node at (0.8,-3) {$\Ht_{-1}$};
\node at (-2.1,1.2) {$M_0$};
\node at (-0.8,0.8) {$S_{-2}$};
\node at (0.3,0.8) {$S_{0}$};
\node at (1.7,0.8) {$S_{+2}$};
\node at (1.4,-1.2) {$n_{+1}$};
\node at (-1.4,-1.2) {$n_{-1}$};
\node at (0.3,-1) {$f_0$};
\epic} \ee 
The final superpotential can be repackaged in a manifest $SU(2)_{A}$ invariant way as 
\be\label{SU52T5} \CW = M_0^2 N^2 + \Ht^2 S^2 + \Ht N f_0 + \Ht N S M_0 \,,\ee
 where $A^2 Q \lra S=\{S_{+2}, S_{-2}, S_{0}\}$ is a symmetric 2-tensor of $SU(2)_A$, $A^3 \Qt \lra N=\{n_{-1}, n_{+1}\}$, $A \Qt^2 \lra \Ht=\{\Ht_{+1},\Ht_{-1}\}$ are $SU(2)_A$ doublets and $Q \Qt \lra M_0$, $A^2 Q^2 \Qt \lra f_0$ are $SU(2)_A$ singlets. We recover the result of Section 3.1.8 of \cite{Csaki:1996zb}.

Before moving on, we should comment on the \textcolor{red}{red} term in \eqref{SU52antisymT4}: $\Ht_{-1} n_{+1} S_{0} M_0$. Indeed, if we combine the superpotential in \eqref{SU52antisymT3} with the Pfaffian term \eqref{PfaffianSU5T3} we get the superpotential in \eqref{SU52antisymT4} \emph{without} this \textcolor{red}{red} term. So why did we add it and where does it come from? First, we remark that without this term it would not be possible to repackage the superpotential in a manifestly $SU(2)_A$ invariant way as in \eqref{SU52T5}. In addition, this term is invariant under all the global symmetries including the $U(1)^2 \times U(1)_A$. There is another argument that suggest the presence of this term. Suppose that after the frame $\CT_2$, we decide to confine in the reverse order meaning that we first confine the $Usp(2)_{right}$ and then the $Usp(2)_{left}$. With this order, the term $\Ht_{-1} n_{+1} S_{0} M_0$ is present in the frame $\CT_4$ and it would be the term $\Ht_{+1} n_{-1} S_{0} M_0$ missing. The last observation is that it would have been possible to add an extra term that respect all the global symmetries in all the previous frame ($\CT_1$ to $\CT_3$) and which lead to the \textcolor{red}{red} term in $\CT_4$. However, we believe that this extra term is forbidden in the frames $\CT_1, \CT_2, \CT_3$ because of chiral ring stability. Therefore, chiral ring stability will force us to wait up to the last frame before adding this extra term allowed by the global symmetries.

This subtle point seems to come from the fact that we have degenerate operators in $\CT_2$: $[D_{-\frac{1}{2}} K_{+\frac{1}{2}}], \, [K_{-\frac{1}{2}} D_{+\frac{1}{2}}]$ and $S_0 M_0$. It suggests that in presence of degenerate operators, applying a duality locally (to a particular node inside a quiver) would miss some informations \footnote{In a forthcoming paper \citep{BAJEOTBENVENUTI2}, we will see another instance of issues related to degenerate operators and the failure of applying dualities locally.}. 

One prescription that lead to the correct final superpotential is the following: When going from $\CT_3$ to $\CT_4$, during the computation of the Pfaffian superpotential \eqref{PfaffianSU5T3}, we should not use $f_0$ but the combination $f_0 + S_{0} M_0$. One can understand this prescription in the following way: In the frames $\CT_1$ to $\CT_2$ there is a $\bZ_2$ symmetry, corresponding to the Weyl reflection inside $SU(2)_A$, which maps a field with $U(1)_A$ charge $x$ to the field with charge $-x$. When we go to the frame $\CT_3$, this symmetry is not explicit anymore. Imposing the restoration of this symmetry in $\CT_4$ is enough to give the correct superpotential. This is the role of this prescription. 
 
The mapping of the chiral ring generators is 
\be \label{mapSU52antisym}
\scalebox{0.9}{$
\ba{c}\CT_1 \\
Q \, \Qt \\
A_{+1} \, \Qt^2  \\
A_{-1} \, \Qt^2  \\
\e_5 \, (A_{+1}^2 \, Q) \\
\e_5 \, (A_{-1}^2 \, Q) \\
\e_5 \, (A_{+1} \, A_{-1} \, Q) \\
\e_5 \, (A_{-1} \, A_{[+1} \, A_{-1]} \, \Qt) \\
\e_5 \, (A_{+1} \, A_{[+1} \, A_{-1]} \, \Qt)  \\
\e_5 \, (A_{[+1} \, A_{-1]} \, Q^2 \, \Qt )
\ea
\Leftrightarrow
\ba{c} \CT_{1'}\\
Q \, \Qt \\
b_{+\frac{1}{2}} b_{+\frac{1}{2}} \, \Qt^2  \\
b_{-\frac{1}{2}} b_{-\frac{1}{2}} \, \Qt^2  \\
\ct_{+2} \, Q  \\
\ct_{-2} \, Q  \\
\e_5 \, (b_{+\frac{1}{2}} \, b_{+\frac{1}{2}} \, b_{-\frac{1}{2}} \, b_{-\frac{1}{2}} \, Q) \\
b_{+\frac{1}{2}} \, \ct_{-2} \, b_{+\frac{1}{2}}  \, \Qt \\
b_{-\frac{1}{2}} \, \ct_{+2} \, b_{-\frac{1}{2}} \, \Qt \\
\e_5 \,(b_{+\frac{1}{2}} \, b_{-\frac{1}{2}} \, b_{[+\frac{1}{2}} \, b_{-\frac{1}{2}]} \, Q^2 \, \Qt)
\ea
\Leftrightarrow 
\ba{c}\CT_2\\
M_0 \\
K_{+\frac{1}{2}}^2 \\
K_{-\frac{1}{2}}^2 \\
S_{+2} \\
S_{-2} \\
S_{0} \\
C_{-\frac{3}{2}} \, K_{+\frac{1}{2}} \\
C_{+\frac{3}{2}} \, K_{-\frac{1}{2}} \\
D_{+\frac{1}{2}} \, K_{-\frac{1}{2}}  
\ea
\Leftrightarrow
\ba{c}\CT_3 \\
M_0 \\
\Ht_{+1} \\
K_{-\frac{1}{2}}^2 \\
S_{+2} \\
S_{-2} \\
S_{0} \\
n_{-1} \\
C_{+\frac{3}{2}} \, K_{-\frac{1}{2}} \\
D_{+\frac{1}{2}} \, K_{-\frac{1}{2}}
\ea
\Leftrightarrow
\ba{c} \CT_{4}\\
M_0 \\
\Ht_{+1} \\
\Ht_{-1} \\
S_{+2} \\
S_{-2} \\
S_{0} \\
n_{-1} \\
n_{-1} \\
f_0
\ea
$}
\ee
We stress that the last line in the mapping \eqref{mapSU52antisym} is ambiguous in the intermediate frames. Indeed, in the frames $\CT_2$ and $\CT_3$ there are multiple holomorphic operators with the same quantum numbers under all the global symmetries which should map to a single chiral ring generator.

\subsection{$SU(5)$ with 3 {\tiny \ydiagram{1,1}} + 3 $\overline{{\tiny\ydiagram{1}}}$}
Our last case is the $SU(5)$ gauge theory with $3$ antisymmetric and $3$ antifundamental fields with continuous global symmetry $SU(3)_A \times SU(3)_{\Qt} \times U(1)$.
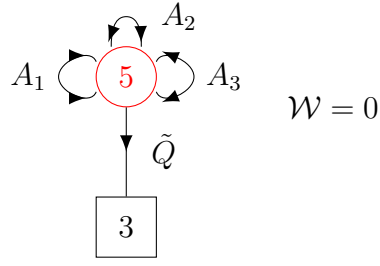
\be \label{SU53antisymT1} \bpic[node distance=2cm,gSUnode/.style={circle,red,draw,minimum size=8mm},gUSpnode/.style={circle,blue,draw,minimum size=8mm},fnode/.style={rectangle,draw,minimum size=8mm}]   
\node at (3.8,1.4) {$\CT_1:$};
\node[gSUnode] (G2) at (6,0) {$5$};
\node[fnode] (F3) at (6,-2) {$3$};
\draw (G2) -- pic[sloped]{arrow} (F3);
\draw (5.6,0.2) to[out=110,in=90] pic[pos=0.3,sloped]{arrow} (5.1,0) to[out=270,in=250] pic[pos=0.7,sloped]{arrow} (5.6,-0.2);
\draw (6.2,0.4) to[out=90,in=0] pic[pos=0.1,sloped]{arrow} (6,0.8) to[out=180,in=90] pic[pos=0.5,sloped,very thick]{arrow=latex reversed} (5.8,0.4);
\draw (6.4,0.2) to[out=70,in=90] pic[pos=0.5,sloped,very thick]{arrow=latex reversed} (6.9,0) to[out=270,in=290] pic[pos=0.5,sloped,very thick]{arrow=latex reversed} (6.4,-0.2);
\node[right] at (8,-0.4) {$ \CW= 0$};
\node at (4.7,0) {$A_1$};
\node at (6.7,0.8) {$A_2$};
\node at (7.3,0) {$A_3$};
\node at (6.5,-1) {$\Qt$};
\epic \ee
The chiral ring generators are
\begin{itemize}
\item $A \Qt^2 \sim A_a^{\a_1 \a_2} \, \Qt^{[I}_{\a_1} \Qt^{J]}_{\a_2}$ transforming in the $(\, \scalebox{0.5}{\ydiagram{1}} \, ,\scalebox{0.5}{\ydiagram{1,1}} \,)$ of $SU(3)_{A} \times SU(3)_{\Qt}$
\item $\e_5 \, \e_5 \, A^5 \sim \e_{\a_1 \dots \a_5} \, \e_{\b_1 \dots \b_5} \, \e^{b_1 b_2 b_3} \, A_{(a_1}^{\a_1 \b_1} A_{a_2)}^{\a_2 \b_2} A_{b_1}^{\a_3 \b_3} A_{b_2}^{\a_4 \b_4} A_{b_3}^{\a_5 \b_5}$ $ \sim (\, \scalebox{0.5}{\ydiagram{2}} \,, 1)$
\item $\e_5 \, (A^3 \, \Qt) \sim \e_{\a_1 \dots \a_5} \, \e^{c \, b_1 b_2} \, A_{a}^{\a_1 \a_2} A_{b_1}^{\a_3 \a_4} A_{b_2}^{\a_5 \b} \Qt_{\b}^{I}$ $ \sim (\, \scalebox{0.5}{\ydiagram{2,1}} \,, \scalebox{0.5}{\ydiagram{1}} \,)$
\end{itemize}
Now we deconfine the three antisymmetric, breaking $SU(3)_A$ to $U(1)_A^2$. Contrary to the others \say{sporadic} cases, the subscripts here do not correspond to the $U(1)_A$ charges. After the splitting the chiral ring generators are given by
\begin{itemize}
\item $A_1 \Qt^2 \sim A_1^{\a_1 \a_2} \, \Qt^{[I}_{\a_1} \Qt^{J]}_{\a_2}$ transforming in the $\scalebox{0.5}{\ydiagram{1,1}} $ of $SU(3)_{\Qt}$
\item $A_2 \Qt^2 \sim A_2^{\a_1 \a_2} \, \Qt^{[I}_{\a_1} \Qt^{J]}_{\a_2} \sim \scalebox{0.5}{\ydiagram{1,1}}$
\item $A_3 \Qt^2 \sim A_3^{\a_1 \a_2} \, \Qt^{[I}_{\a_1} \Qt^{J]}_{\a_2} \sim \scalebox{0.5}{\ydiagram{1,1}}$
\item $\e_5 \, \e_5 \, (A_1 A_1 A_{[1} A_2 A_{3]}) \equiv (A^5)_{11} \sim \e_{\a_1 \dots \a_5} \, \e_{\b_1 \dots \b_5} \, A_{1}^{\a_1 \b_1} A_{1}^{\a_2 \b_2} A_{[1}^{\a_3 \b_3} A_{2}^{\a_4 \b_4} A_{3]}^{\a_5 \b_5}$ $ \sim 1$
\item $\e_5 \, \e_5 \, (A_{(1} A_{2)} A_{[1} A_2 A_{3]}) \equiv (A^5)_{12} \sim \e_{\a_1 \dots \a_5} \, \e_{\b_1 \dots \b_5} \, A_{(1}^{\a_1 \b_1} A_{2)}^{\a_2 \b_2} A_{[1}^{\a_3 \b_3} A_{2}^{\a_4 \b_4} A_{3]}^{\a_5 \b_5}$ $ \sim 1$
\item $\e_5 \, \e_5 \, (A_{(1} A_{3)} A_{[1} A_2 A_{3]}) \equiv (A^5)_{13} \sim \e_{\a_1 \dots \a_5} \, \e_{\b_1 \dots \b_5} \, A_{(1}^{\a_1 \b_1} A_{3)}^{\a_2 \b_2} A_{[1}^{\a_3 \b_3} A_{2}^{\a_4 \b_4} A_{3]}^{\a_5 \b_5}$ $ \sim 1$
\item $\e_5 \, \e_5 \, (A_{2} A_{2} A_{[1} A_2 A_{3]}) \equiv (A^5)_{22} \sim \e_{\a_1 \dots \a_5} \, \e_{\b_1 \dots \b_5} \, A_{2}^{\a_1 \b_1} A_{2}^{\a_2 \b_2} A_{[1}^{\a_3 \b_3} A_{2}^{\a_4 \b_4} A_{3]}^{\a_5 \b_5}$ $ \sim 1$
\item $\e_5 \, \e_5 \, (A_{(2} A_{3)} A_{[1} A_2 A_{3]}) \equiv (A^5)_{23} \sim \e_{\a_1 \dots \a_5} \, \e_{\b_1 \dots \b_5} \, A_{(2}^{\a_1 \b_1} A_{3)}^{\a_2 \b_2} A_{[1}^{\a_3 \b_3} A_{2}^{\a_4 \b_4} A_{3]}^{\a_5 \b_5}$ $ \sim 1$
\item $\e_5 \, \e_5 \, (A_{3} A_{3} A_{[1} A_2 A_{3]}) \equiv (A^5)_{33} \sim \e_{\a_1 \dots \a_5} \, \e_{\b_1 \dots \b_5} \, A_{3}^{\a_1 \b_1} A_{3}^{\a_2 \b_2} A_{[1}^{\a_3 \b_3} A_{2}^{\a_4 \b_4} A_{3]}^{\a_5 \b_5}$ $ \sim 1$
\item $\e_5 \, (A_1 A_{[2} A_{3]} \, \Qt) \equiv (A^3 \, \Qt)_{11} \sim \e_{\a_1 \dots \a_5} \, A_{1}^{\a_1 \a_2} A_{[2}^{\a_3 \a_4} A_{3]}^{\a_5 \b} \Qt_{\b}^{I}$ $ \sim \scalebox{0.5}{\ydiagram{1}}$
\item $\e_5 \, (A_1 A_{[3} A_{1]} \, \Qt) \equiv (A^3 \, \Qt)_{12} \sim \e_{\a_1 \dots \a_5} \, A_{1}^{\a_1 \a_2} A_{[3}^{\a_3 \a_4} A_{1]}^{\a_5 \b} \Qt_{\b}^{I}$ $ \sim \scalebox{0.5}{\ydiagram{1}}$
\item $\e_5 \, (A_1 A_{[1} A_{2]} \, \Qt) \equiv (A^3 \, \Qt)_{13} \sim \e_{\a_1 \dots \a_5} \, A_{1}^{\a_1 \a_2} A_{[1}^{\a_3 \a_4} A_{2]}^{\a_5 \b} \Qt_{\b}^{I}$ $ \sim \scalebox{0.5}{\ydiagram{1}}$
\item $\e_5 \, (A_2 A_{[2} A_{3]} \, \Qt) \equiv (A^3 \, \Qt)_{21} \sim \e_{\a_1 \dots \a_5} \, A_{2}^{\a_1 \a_2} A_{[2}^{\a_3 \a_4} A_{3]}^{\a_5 \b} \Qt_{\b}^{I}$ $ \sim \scalebox{0.5}{\ydiagram{1}}$
\item $\e_5 \, (A_2 A_{[3} A_{1]} \, \Qt) \equiv (A^3 \, \Qt)_{22} \sim \e_{\a_1 \dots \a_5} \, A_{2}^{\a_1 \a_2} A_{[3}^{\a_3 \a_4} A_{1]}^{\a_5 \b} \Qt_{\b}^{I}$ $ \sim \scalebox{0.5}{\ydiagram{1}}$
\item $\e_5 \, (A_2 A_{[1} A_{2]} \, \Qt) \equiv (A^3 \, \Qt)_{23} \sim \e_{\a_1 \dots \a_5} \, A_{2}^{\a_1 \a_2} A_{[1}^{\a_3 \a_4} A_{2]}^{\a_5 \b} \Qt_{\b}^{I}$ $ \sim \scalebox{0.5}{\ydiagram{1}}$
\item $\e_5 \, (A_3 A_{[2} A_{3]} \, \Qt) \equiv (A^3 \, \Qt)_{31} \sim \e_{\a_1 \dots \a_5} \, A_{3}^{\a_1 \a_2} A_{[2}^{\a_3 \a_4} A_{3]}^{\a_5 \b} \Qt_{\b}^{I}$ $ \sim \scalebox{0.5}{\ydiagram{1}}$
\item $\e_5 \, (A_3 A_{[3} A_{1]} \, \Qt) \equiv (A^3 \, \Qt)_{32} \sim \e_{\a_1 \dots \a_5} \, A_{3}^{\a_1 \a_2} A_{[3}^{\a_3 \a_4} A_{1]}^{\a_5 \b} \Qt_{\b}^{I}$ $ \sim \scalebox{0.5}{\ydiagram{1}}$
\end{itemize}
We didn't include $\e_5 \, (A_3 A_{[1} A_{2]} \, \Qt) \equiv (A^3 \, \Qt)_{33} \sim \e_{\a_1 \dots \a_5} \, A_{3}^{\a_1 \a_2} A_{[1}^{\a_3 \a_4} A_{2]}^{\a_5 \b} \Qt_{\b}^{I}$ because the trace of the $SU(3)_A$ adjoint matrix should vanish which imposes the relation
\be \label{vanishingTrace}
(A^3 \, \Qt)_{11}+(A^3 \, \Qt)_{22}+(A^3 \, \Qt)_{33} = 0 \nn
\ee
\be \label{SU53antisymT1'} \scalebox{0.9}{\bpic[node distance=2cm,gSUnode/.style={circle,red,draw,minimum size=8mm},gUSpnode/.style={circle,blue,draw,minimum size=8mm},fnode/.style={rectangle,draw,minimum size=8mm}]    
\node at (-3.5,3.5) {$\CT_{1'}:$};
\node[gSUnode] (G1) at (0,0) {$5$};
\node[gUSpnode] (G2) at (-2.5,-1) {$2$};
\node[gUSpnode] (G3) at (-1,2.5) {$2$};
\node[gUSpnode] (G4) at (2.5,-1) {$2$};
\node[fnode] (F1) at (0,-2) {$3$};
\node[fnode,violet] (F2) at (-2.5,1) {$1$};
\node[fnode,orange] (F3) at (1.6,2.5) {$1$};
\node[fnode,blue] (F4) at (2.5,1) {$1$};
\draw (G2) -- pic[pos=0.6,sloped]{arrow} (G1);
\draw (G3) -- pic[pos=0.6,sloped]{arrow} (G1);
\draw (G4) -- pic[sloped,very thick]{arrow=latex reversed} (G1);
\draw (F1) -- pic[sloped,very thick]{arrow=latex reversed} (G1);
\draw (F2) -- pic[sloped,very thick]{arrow=latex reversed} (G1);
\draw (F3) -- pic[sloped]{arrow} (G1);
\draw (F4) -- pic[sloped]{arrow} (G1);
\draw (G2) -- (F2);
\draw (G3) -- (F3); 
\draw (G4) -- (F4);
\node[right] at (4,0) {$\CW= 3 \,$ Planar Triangles};
\node at (-2.8,0) {$l_1$};
\node at (0.3,2.8) {$l_2$};
\node at (2.8,0) {$l_3$};
\node at (-1.3,0.2) {$\ct_1$};
\node at (0.6,1.5) {$\ct_2$};
\node at (1.5,0.2) {$\ct_3$};
\node at (-1.2,-0.9) {$b_1$};
\node at (-1,1.5) {$b_2$};
\node at (1.4,-0.9) {$b_3$};
\node at (0.4,-1) {$\Qt$};
\epic} \ee 
Then we confine the $SU(5)$ using \eqref{SUbuildingBlock}. Once again we integrate out the massive fields ($l_1, l_2$ and $l_3$), we compute the determinant of the meson matrix and rescale the fields. We get 
\be \label{SU53antisymT2} \scalebox{0.9}{\bpic[node distance=2cm,gSUnode/.style={circle,red,draw,minimum size=8mm},gUSpnode/.style={circle,blue,draw,minimum size=8mm},fnode/.style={rectangle,draw,minimum size=8mm}]   
\node at (-3.7,3.5) {$\CT_2:$};
\node[fnode,red] (F1) at (0,0) {$1$};
\node[gUSpnode] (G1) at (-2.4,-0.7) {$2$};
\node[gUSpnode] (G2) at (2.4,-0.7) {$2$};
\node[gUSpnode] (G3) at (0,2.7) {$2$};
\node[fnode,blue] (F2) at (0,-2.2) {$1$};
\node[fnode,orange] (F3) at (-2,1.5) {$1$};
\node[fnode,violet] (F4) at (2,1.5) {$1$};
\node[fnode] (F5) at (0,-3.8) {$3$};
\draw (G1) -- (F1);
\draw (G1) -- (F2);
\draw (G1) -- (F3);
\draw (G2) -- (F1);
\draw (G2) -- (F2);
\draw (G2) -- (F4);
\draw (G3) -- (F1);
\draw (G3) -- (F3);
\draw (G3) -- (F4);
\draw (F1) -- (F2);
\draw (F1) -- (F3);
\draw (F1) -- (F4);
\draw (-2.4,-1.1) to[out=-90,in=135] (-1.6,-3) to[out=-45,in=180] pic[pos=0.5,sloped]{arrow} (-0.4,-3.6);
\draw (2.4,-1.1) to[out=-90,in=45] (1.6,-3) to[out=-135,in=0] pic[pos=0.3,sloped,very thick]{arrow=latex reversed} (0.4,-3.6);  
\draw (-0.4,2.7) to[out=180,in=45] (-2.8,2) to[out=-135,in=90] (-3.8,-0.7) to[out=-90,in=135] (-2.8,-3.4) to[out=-45,in=180] pic[pos=0.6,sloped]{arrow} (-0.4,-4);
\draw (0.4,-4) to[out=0,in=-145] pic[pos=0.5,sloped]{arrow} (2.8,-3.5) node {$+$};
\draw (0.4,0.1) to[out=20,in=180] (1.5,0.3) node {$+$};
\node[right] at (3.8,2) {$\CW= \e_3 \, \e^2 \, \e^2 \, \e^2 \, \left[K_5 M_2 K_2 K_3 M_1^2 \right.$};
\node[right] at (3.4,1) {$+ \,  K_2 M_3 K_4 K_5 M_1^2 + \, K_1 M_1 K_4 K_5 M_3^2$};
\node[right] at (3.4,0) {$+ \, K_6 M_1 K_2 K_3 M_2^2 + \, K_3 M_3 K_1 K_6 M_2^2$};
\node[right] at (3.4,-1) {$+ \, K_4 M_2 K_1 K_6 M_3^2 + \, K_1 M_1 K_5 M_2 K_3 M_3$};
\node[right] at (3.4,-2) {$\left. + \, K_6 M_1 K_4 M_2 K_2 M_3 \right] + B_7 B_1 M_1 $};
\node[right] at (3.4,-3) {$+ B_7 B_3 M_3 + B_7 B_5 M_2 $+ 6 Planar Triangles};
\node at (-2.6,0.5) {$K_1$};
\node at (-1.2,2.3) {$K_2$};
\node at (1.2,2.3) {$K_3$};
\node at (2.6,0.5) {$K_4$};
\node at (1.5,-1.6) {$K_5$};
\node at (-1.5,-1.6) {$K_6$};
\node at (-1.2,-0.7) {$B_1$};
\node at (-1.2,0.6) {$B_2$};
\node at (-0.3,1.3) {$B_3$};
\node at (0.8,1) {$B_4$};
\node at (1.2,-0.7) {$B_5$};
\node at (-0.3,-1) {$B_6$};
\node at (3.2,-3.6) {$B_7$};
\node at (-1.4,-2.6) {$M_1$};
\node at (2.4,-2.5) {$M_2$};
\node at (-2.5,-3.1) {$M_3$};
\epic} \ee 
The next step is to confine the $Usp(2)_{up}$, $B_2, B_4$ and $B_7$ get a mass. The result is 
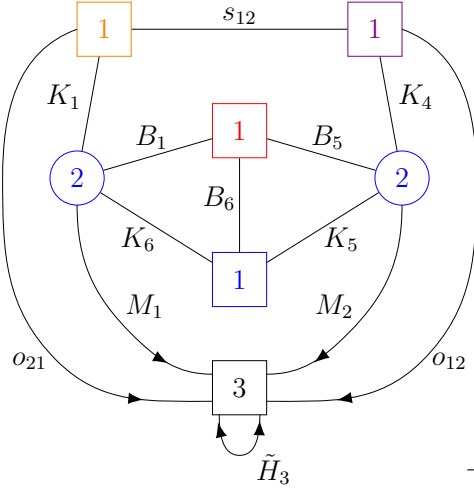
\be \label{SU53antisymT3} \scalebox{0.9}{\bpic[node distance=2cm,gSUnode/.style={circle,red,draw,minimum size=8mm},gUSpnode/.style={circle,blue,draw,minimum size=8mm},fnode/.style={rectangle,draw,minimum size=8mm}]    
\node at (-3.8,2.6) {$\CT_3:$};
\node[fnode,red] (F1) at (0,0) {$1$};
\node[gUSpnode] (G1) at (-2.4,-0.7) {$2$};
\node[gUSpnode] (G2) at (2.4,-0.7) {$2$};
\node[fnode,blue] (F2) at (0,-2.2) {$1$};
\node[fnode,orange] (F3) at (-2,1.5) {$1$};
\node[fnode,violet] (F4) at (2,1.5) {$1$};
\node[fnode] (F5) at (0,-3.8) {$3$};
\draw (G1) -- (F1);
\draw (G1) -- (F2);
\draw (G1) -- (F3);
\draw (G2) -- (F1);
\draw (G2) -- (F2);
\draw (G2) -- (F4);
\draw (F1) -- (F2);
\draw (F3) -- (F4);
\draw (-2.4,-1.1) to[out=-90,in=135] (-1.6,-3) to[out=-45,in=180] pic[pos=0.5,sloped]{arrow} (-0.4,-3.6);
\draw (2.4,-1.1) to[out=-90,in=45] (1.6,-3) to[out=-135,in=0] pic[pos=0.3,sloped,very thick]{arrow=latex reversed} (0.4,-3.6);
\draw (-2.4,1.5) to[out=200,in=90] (-3.5,-0.7) to[out=-90,in=135] (-2.7,-3.4) to[out=-45,in=180] pic[pos=0.6,sloped]{arrow} (-0.4,-4);
\draw (2.4,1.5) to[out=-20,in=90] (3.5,-0.7) to[out=-90,in=45] (2.7,-3.4) to[out=-135,in=0] pic[pos=0.5,sloped,very thick]{arrow=latex reversed} (0.4,-4);
\draw (0.3,-4.2) to[out=-90,in=0] pic[pos=0.1,sloped]{arrow} (0,-4.8) to[out=-180,in=-90] pic[pos=0.6,sloped,very thick]{arrow=latex reversed} (-0.3,-4.2);  
\node[right] at (4.2,2) {$\CW= \e_3 \, \e^2 \, \e^2 \, \big[K_5 M_2 s_{12} M_1^2$};
\node[right] at (4.2,1) {$+ o_{21} K_4 K_5  M_1^2 + \, K_1 M_1 K_4 K_5 \Ht_3$};
\node[right] at (4.2,0) {$+ \, K_6 M_1 s_{12} M_2^2 + \, o_{12} K_1 K_6 M_2^2$};
\node[right] at (4.2,-1) {$+ \, K_4 M_2 K_1 K_6 \Ht_3 + \, K_1 M_1 K_5 M_2 o_{12}$};
\node[right] at (4.2,-2) {$+ \, K_6 M_1 K_4 M_2 o_{21} \big] + \, B_1 B_6 K_6 $};
\node[right] at (4.2,-3) {$+ \, B_5 B_6 K_5 + \e_3 \big[o_{21} o_{12} B_1 M_1 $};
\node[right] at (4.2,-4) {$+ \, o_{21} o_{12} B_5 M_2 + \, o_{21} K_4 B_5 \Ht_3$};
\node[right] at (3.2,-5) {$+ \, o_{12} K_1 B_1 \Ht_3 + \, B_1 M_1 s_{12} \Ht_3 + \, B_5 M_2 s_{12} \Ht_3\big] $};
\node at (-2.6,0.5) {$K_1$};
\node at (2.6,0.5) {$K_4$};
\node at (1.5,-1.6) {$K_5$};
\node at (-1.5,-1.6) {$K_6$};
\node at (-1.3,-0.1) {$B_1$};
\node at (1.3,-0.1) {$B_5$};
\node at (-0.3,-1) {$B_6$};
\node at (-1.4,-2.6) {$M_1$};
\node at (1.4,-2.6) {$M_2$};
\node at (0,1.7) {$s_{12}$};
\node at (-3.1,-3.4) {$o_{21}$};
\node at (3.1,-3.4) {$o_{12}$};
\node at (0.5,-5) {$\Ht_3$};
\epic} \ee 
In the next two steps, we confine $Usp(2)_{left}$ and $Usp(2)_{right}$. It is really similar to the previous step. We obtain
\be \label{SU53antisymT4} \scalebox{0.9}{\bpic[node distance=2cm,gSUnode/.style={circle,red,draw,minimum size=8mm},gUSpnode/.style={circle,blue,draw,minimum size=8mm},fnode/.style={rectangle,draw,minimum size=8mm}]    
\node at (-3.4,2.5) {$\CT_4:$};
\node[fnode,red] (F1) at (0,0) {$1$};
\node[gUSpnode] (G2) at (2.4,-0.7) {$2$};
\node[fnode,blue] (F2) at (0,-2.2) {$1$};
\node[fnode,orange] (F3) at (-2,1.5) {$1$};
\node[fnode,violet] (F4) at (2,1.5) {$1$};
\node[fnode] (F5) at (0,-4.5) {$3$};
\draw (G2) -- (F1);
\draw (G2) -- (F2);
\draw (G2) -- (F4);
\draw (F3) -- (F4);
\draw (F3.south east) -- (F1.north west);
\draw (F2) -- pic[pos=0.6,sloped]{arrow} (F5);
\draw (F3.south west) to[out=-120,in=90] (-2.8,-1) to[out=-90,in=135] (-2.2,-3.6) to[out=-45,in=180] pic[pos=0.5,sloped]{arrow} (-0.4,-4.5);
\draw (F3.south) to[out=-90,in=90] (-2,-1) to[out=-90,in=180] (-0.4,-2.2);
\draw (-0.4,0) to[out=180,in=70] (-1.2,-0.5) node {$+$};
\draw (-0.4,-4.3) to[out=-180,in=-70] pic[pos=0.5,sloped]{arrow} (-1.5,-3) node{$+$};
\draw (2.4,-1.1) to[out=-90,in=45] (1.6,-3.7) to[out=-135,in=0] pic[pos=0.3,sloped,very thick]{arrow=latex reversed} (0.4,-4.3);
\draw (-2.4,1.5) to[out=200,in=90] (-3.5,-0.7) to[out=-90,in=135] (-2.7,-4.1) to[out=-45,in=180] pic[pos=0.6,sloped]{arrow} (-0.4,-4.7);
\draw (2.4,1.5) to[out=-20,in=90] (3.5,-0.7) to[out=-90,in=45] (2.7,-4.1) to[out=-135,in=0] pic[pos=0.5,sloped,very thick]{arrow=latex reversed} (0.4,-4.7);
\draw (0.3,-4.9) to[out=-90,in=0] pic[pos=0.1,sloped]{arrow} (0,-5.5) to[out=-180,in=-90] pic[pos=0.6,sloped,very thick]{arrow=latex reversed} (-0.3,-4.9);  
\node[right] at (4.2,1.5) {$\CW=  \e_3 \, \e^2 \, \big[K_5 M_2 s_{12} \Ht_1$};
\node[right] at (4.2,0.5) {$+ \, o_{21} K_4 K_5 \Ht_1 + \, o_{23} K_4 K_5 \Ht_3$};
\node[right] at (4.2,-0.5) {$+ \, o_{32} s_{12} M_2^2 + \, o_{12} s_{23} M_2^2 + \, K_4 M_2 s_{23} \Ht_3$};
\node[right] at (4.2,-1.5) {$+ o_{23} K_5 M_2 o_{12} + o_{32} K_4 M_2 o_{21} \big] $};
\node[right] at (4.2,-2.5) {$+ \e_3 \big[ o_{21} o_{12} o_{11} + o_{21} o_{12} B_5 M_2 $};
\node[right] at (4.2,-3.5) {$+ o_{21} K_4 B_5 \Ht_3 + o_{12} s_{22} \Ht_3 + o_{11} s_{12} \Ht_3 $};
\node[right] at (4.2,-4.5) {$+ B_5 M_2 s_{12} \Ht_3 + o_{23} o_{32} o_{11} + o_{23} o_{32} B_5 M_2 $};
\node[right] at (2.2,-5.5) {$+ o_{23} K_5 B_5 \Ht_1 + o_{32} s_{22} \Ht_1 + o_{11} s_{23} \Ht_1 + B_5 M_2 s_{23} \Ht_1 \big]$};
\node at (2.6,0.5) {$K_4$};
\node at (1.5,-1.6) {$K_5$};
\node at (1.3,-0.1) {$B_5$};
\node at (1.7,-2.8) {$M_2$};
\node at (0,1.7) {$s_{12}$};
\node at (-1.3,-1.7) {$s_{23}$};
\node at (-1.1,0.6) {$s_{22}$};
\node at (-3.3,-3.7) {$o_{21}$};
\node at (3.3,-3.7) {$o_{12}$};
\node at (-2.4,-2.7) {$o_{23}$};
\node at (-0.4,-3.4) {$o_{32}$};
\node at (-1.2,-2.8) {$o_{11}$};
\node at (1,-5.5) {$\Ht_1, \Ht_3$};
\epic} \ee 
\be \label{SU53antisymT5} \scalebox{0.9}{\bpic[node distance=2cm,gSUnode/.style={circle,red,draw,minimum size=8mm},gUSpnode/.style={circle,blue,draw,minimum size=8mm},fnode/.style={rectangle,draw,minimum size=8mm}]    
\node at (-3.4,2.6) {$\CT_5:$};
\node[fnode,red] (F1) at (0,0) {$1$};
\node[fnode,blue] (F2) at (0,-2.2) {$1$};
\node[fnode,orange] (F3) at (-2,1.5) {$1$};
\node[fnode,violet] (F4) at (2,1.5) {$1$};
\node[fnode] (F5) at (0,-4.5) {$3$};
\draw (F1) -- (F2);
\draw (F3) -- (F4);
\draw (F3.south east) -- (F1.north west);
\draw (F4.south west) -- (F1.north east);
\draw (-0.2,-2.6) -- pic[pos=0.6,sloped]{arrow} (-0.2,-4.1);
\draw (0.2,-2.6) -- pic[pos=0.6,sloped]{arrow} (0.2,-4.1);
\draw (F3.south west) to[out=-120,in=90] (-2.8,-1) to[out=-90,in=135] (-2.2,-3.6) to[out=-45,in=180] pic[pos=0.5,sloped]{arrow} (-0.4,-4.5);
\draw (F4.south east) to[out=-60,in=90] (2.8,-1) to[out=-90,in=45] (2.2,-3.6) to[out=-135,in=0] pic[pos=0.5,sloped,very thick]{arrow=latex reversed} (0.4,-4.5);
\draw (F3.south) to[out=-90,in=90] (-2,-1) to[out=-90,in=180] (-0.4,-2.2);
\draw (F4.south) to[out=-90,in=90] (2,-1) to[out=-90,in=0] (0.4,-2.2);
\draw (-0.4,0) to[out=180,in=70] (-1.2,-0.5) node {$+$};
\draw (0.4,0) to[out=0,in=110] (1.2,-0.5) node {$+$};
\draw (-0.4,-4.3) to[out=-180,in=-70] pic[pos=0.5,sloped]{arrow} (-1.5,-3) node{$+$};
\draw (0.4,-4.3) to[out=0,in=-110] pic[pos=0.5,sloped,very thick]{arrow=latex reversed} (1.5,-3) node{$+$};
\draw (-2.4,1.5) to[out=200,in=90] (-3.5,-0.7) to[out=-90,in=135] (-2.7,-4.1) to[out=-45,in=180] pic[pos=0.6,sloped]{arrow} (-0.4,-4.7);
\draw (2.4,1.5) to[out=-20,in=90] (3.5,-0.7) to[out=-90,in=45] (2.7,-4.1) to[out=-135,in=0] pic[pos=0.5,sloped,very thick]{arrow=latex reversed} (0.4,-4.7);
\draw (0.3,-4.9) to[out=-90,in=0] pic[pos=0.1,sloped]{arrow} (0,-5.5) to[out=-180,in=-90] pic[pos=0.6,sloped,very thick]{arrow=latex reversed} (-0.3,-4.9);  
\node[right] at (4,1) {$\CW= \e_3 \, \big[o_{31} s_{12} \Ht_1 + o_{21} s_{13} \Ht_1 + o_{23} s_{33} \Ht_1$};
\node[right] at (5,0) {$+ o_{32} s_{22} \Ht_1 + o_{11} s_{23} \Ht_1 + o_{22} s_{23} \Ht_1$};
\node[right] at (5,-1) {$+ o_{12} s_{23} \Ht_2 + o_{32} s_{12} \Ht_2 + o_{13} s_{33} \Ht_2$};
\node[right] at (5,-2) {$+ o_{31} s_{11} \Ht_2 + o_{22} s_{13} \Ht_2 + o_{11} s_{13} \Ht_2$};
\node[right] at (5,-3) {$+ o_{13} s_{23} \Ht_3 + o_{21} s_{11} \Ht_3 + o_{12} s_{22} \Ht_3$};
\node[right] at (5,-4) {$+ o_{23} s_{13} \Ht_3 + o_{11} s_{12} \Ht_3 + o_{22} s_{12} \Ht_3$};
\node[right] at (3.2,-5) {$+ o_{23} o_{31} o_{12} + o_{32} o_{13} o_{21} + o_{21} o_{12} o_{11} + o_{21} o_{12} o_{22}$};
\node[right] at (3.2,-6) {$+ o_{23} o_{32} o_{11} + o_{23} o_{32} o_{22} + o_{13} o_{31} o_{22} + o_{13} o_{31} o_{11} \big]$};
\node at (0,1.7) {$s_{12}$};
\node at (-1.3,-1.7) {$s_{23}$};
\node at (-1.1,0.6) {$s_{22}$};
\node at (1.3,-1.7) {$s_{13}$};
\node at (1.1,0.6) {$s_{11}$};
\node at (0.3,-1.1) {$s_{33}$};
\node at (-3.3,-3.7) {$o_{21}$};
\node at (3.3,-3.7) {$o_{12}$};
\node at (-2.4,-2.7) {$o_{23}$};
\node at (-0.6,-3.4) {$o_{32}$};
\node at (-1.2,-2.8) {$o_{11}$};
\node at (2.4,-2.7) {$o_{13}$};
\node at (0.6,-3.4) {$o_{31}$};
\node at (1.2,-2.8) {$o_{22}$};
\node at (1.3,-5.5) {$\Ht_1, \Ht_2, \Ht_3$};
\epic} \ee 
The final superpotential can be repackaged in a manifest $SU(2)_{A}$ invariant way as 
\be \label{WSU53antisym} \CW = O S \Ht + O^3 \,,\ee
where
\be \nn
\scalebox{0.95}{$
A^3 \Qt \lra O \equiv 
\begin{pmatrix}
o_{11} & o_{12} & o_{13} \\
o_{21} & o_{22} & o_{23} \\
o_{31} & o_{32} & -o_{11}-o_{22}
\end{pmatrix}, \,
A^5 \lra S \equiv
\begin{pmatrix}
s_{11} & s_{12} & s_{13} \\
s_{12} & s_{22} & s_{23} \\
s_{13} & s_{23} & s_{33}
\end{pmatrix} \, 
\text{and} \, A\Qt^2 \lra \Ht \equiv
\begin{pmatrix}
\Ht_1 \\
\Ht_2 \\
\Ht_3
\end{pmatrix}
$}
\ee 
$O$ transforms in the adjoint, $S$ as a symmetric 2-tensor and $\Ht$ in the fundamental of $SU(3)_A$. We recover the result of Section 3.1.7 of \cite{Csaki:1996zb}. The mapping of the chiral ring generators is 
\be \label{mapSU53antisym}
\scalebox{0.9}{$
\ba{c}\CT_1 \\
A_1 \, \Qt^2 \\
A_2 \, \Qt^2 \\
A_3 \, \Qt^2 \\
(A^5)_{11} \\
(A^5)_{12} \\
(A^5)_{13} \\
(A^5)_{22} \\
(A^5)_{23} \\
(A^5)_{33} \\
(A^3 \, \Qt)_{11} \\
(A^3 \, \Qt)_{12} \\
(A^3 \, \Qt)_{13} \\
(A^3 \, \Qt)_{21} \\
(A^3 \, \Qt)_{22} \\
(A^3 \, \Qt)_{23} \\
(A^3 \, \Qt)_{31} \\
(A^3 \, \Qt)_{32} \\
\ea
\Longleftrightarrow
\ba{c} \CT_{1'}\\
b_1 b_1 \, \Qt^2 \\
b_2 b_2 \, \Qt^2 \\
b_3 b_3 \, \Qt^2 \\
b_1^2 \, b_2^2 \, b_3^2 \, \ct_1 \\
b_3 b_3 \, \ct_1 \, \ct_2) \\
b_2 b_2 \, \ct_1 \, \ct_3) \\
b_1^2 \, b_2^2 \, b_3^2 \, \ct_2 \\
b_1 b_1 \, \ct_2 \, \ct_3) \\
b_1^2 \, b_2^2 \, b_3^2 \, \ct_3 \\
\e_5 \, (b_1 \, b_2^2 \, b_3^2) \, b_1 \, \Qt  \\
b_3 b_3 \, \ct_1 \, \Qt \\
b_2 b_2 \, \ct_1 \, \Qt \\
b_3 b_3 \, \ct_2 \, \Qt \\
\e_5 \, (b_2 \, b_1^2 \, b_3^2) \, b_2 \, \Qt \\
b_1 b_1 \, \ct_2 \, \Qt \\
b_2 b_2 \, \ct_3 \, \Qt \\
b_1 b_1 \, \ct_3 \, \Qt \\
\ea
\Longleftrightarrow 
\ba{c}\CT_2\\
M_1^2 \\
M_2^2 \\
M_3^2 \\
B_3 \, K_3 \\
K_2 \, K_3 \\
K_4 \, K_5 \\
B_1 \, K_1 \\
K_1 \, K_6 \\
B_5 \, K_5 \\
B_1 \, M_1 \\
K_3 \, M_3 \\
K_4 \, M_2 \\
K_2 \, M_3 \\
B_5 \, M_2 \\
K_1 \, M_1 \\
K_5 \, M_2 \\
K_6 \, M_1 \\
\ea
\Longleftrightarrow
\ba{c}\CT_3 \\
M_1^2 \\
M_2^2 \\
\Ht_3 \\
B_5 \, K_4 \\
s_{12} \\
K_4 \, K_5 \\
B_1 \, K_1 \\
K_1 \, K_6 \\
B_5 \, K_5 \\
B_1 \, M_1 \\
o_{12} \\
K_4 \, M_2 \\
o_{21} \\
B_5 \, M_2 \\
K_1 \, M_1 \\
K_5 \, M_2 \\
K_6 \, M_1 \\
\ea
\Longleftrightarrow
\ba{c} \CT_{4}\\
\Ht_1 \\
M_2^2 \\
\Ht_3 \\
B_5 \, K_4 \\
s_{12} \\
K_4 \, K_5 \\
s_{22} \\
s_{23} \\
B_5 \, K_5 \\
o_{11} \\
o_{12} \\
K_4 \, M_2 \\
o_{21} \\
B_5 \, M_2 \\
o_{23} \\
K_5 \, M_2 \\
o_{32} \\
\ea
\Longleftrightarrow
\ba{c} \CT_{5}\\
\Ht_1 \\
\Ht_2\\
\Ht_3 \\
s_{11} \\
s_{12} \\
s_{13} \\
s_{22} \\
s_{23} \\
s_{33} \\
o_{11} \\
o_{12} \\
o_{13} \\
o_{21} \\
o_{22} \\
o_{23} \\
o_{31} \\
o_{32} \\
\ea
$}
\ee
The same phenomenon with degenerate operators appears here, as in Section~\ref{DegenerateOpPhenomenon}. In this case they are the operators $B_1 M_1, B_5 M_2$ and $B_3 M_3$ in $\CT_2$. They are related by the F-term equation of $B_7 : B_1 M_1 + B_5 M_2 + B_3 M_3 =0$, which reproduces the traceless condition \eqref{vanishingTrace}. To reproduce the correct superpotential in the final frame $\CT_5$, we use our prescription of Section~\ref{DegenerateOpPhenomenon}. Concretly, when we go from $\CT_3$ to $\CT_4$ it is the combination $o_{11} + B_5 M_2$ that enters in the computation of the Pfaffian superpotential. Similarly from $\CT_4$ to $\CT_5$ it is the combination $o_{11} + o_{22}$ that should enter. Taking into account this subtle point, we can recover the final superpotential \eqref{WSU53antisym} with the correct $SU(3)_A$ global symmetry. 

\section{ Conclusions and outlook}

In this paper we have shown that all $4d$ $\cN=1$ S-confining gauge theories with a single gauge group, vanishing tree-level superpotential and rank-1 and/or rank-2 matter can be obtained from the basic Seiberg \eqref{SUbuildingBlock} and Intriligator-Pouliot \eqref{UspbuildingBlock} S-confining dualities \cite{Seiberg:1994bz, Intriligator:1995ne}. We have also obtained the confining superpotential in a closed form for all theories. We did this using new versions of the deconfinement technique of \cite{Berkooz:1995km}. 

Our result participates to the project of reducing the number of apparently independent dualities started in \cite{Berkooz:1995km}. It would be interesting to know how far we can go with this technique and in particular if dualities with matter only in the fundamental representation are strong enough to reproduce all the more complicated dualities.

There are many other directions to explore. An obvious one is trying to go beyond the classification of \cite{Csaki:1996zb} by considering more than one node quivers and/or non-vanishing tree-level superpotential \cite{Csaki:1998fm}. There are also S-confining theories involving non-quivers type of matter as rank-3 and $Spin$ gauge theories with chirals in the spinor representation \cite{Csaki:1996zb}. It would be really interesting if we can also obtain these theories from simpler dualities.

It would also be worth exploring beyond S-confining theories. For example, more general IR dualities involving rank-2 matter \cite{Kutasov:1995np, Intriligator:1995ff, Intriligator:1995ax, Brodie:1996xm} as well as self-dual theories \cite{Csaki:1996eu, Csaki:1997cu}. In a separate paper \cite{BAJEOTBENVENUTI2} we will report some progress on the last point.

\acknowledgments
We are grateful to Sara Pasquetti and Matteo Sacchi for various useful discussions.\\
Stephane Bajeot is partially supported by the INFN Research Project STEFI. Sergio Benvenuti is partially supported by the INFN Research Project GAST.

\newpage
\appendix

\section{Superpotential of Section \ref{Usp2N} in frame $\CT_k$} \label{SuperpotInTk}
In this appendix, we will obtain the form of the $Usp(2N)$ with one antisymmetric and $6$ fundamentals, studied in Section\ref{Usp2N}, after $k$ iterations of deconfinement/confinement. What is complicated is to keep track of the superpotential terms. We will do it in several steps. The key ingredient is to follow terms with the field $O_p$ because we know how they start to appear. For example, when we go from the frame $\CT_{0'}$ to the frame $\CT_1$, the confining Pfaffian superpotential has created 2 terms containing the field $O_1$: $\e_{2N-2} \, \e_5 \, \left[ A_1^{N-2} \, Q_1^2 M_1 O_1 + A_1^{N-3} \, Q_1^4 O_1 \right]$. But it is true more generally, when we go from the frame $\CT_{(p-1)'}$ to $\CT_{p}$ the confining Pfaffian superpotential creates 2 terms with the field $O_p$: $\e_{2N-2p} \, \e_5 \, \left[ A_p^{N-p-1} \, Q_p^2 M_p O_p + A_p^{N-p-2} \, Q_p^4 O_p \right]$.

So let's call $\Theta_{(p)}$ all the terms containing $O_p$. As a first step, we need to understand what is $\Theta_{(p)}$ in a frame $\CT_k$. Let's see how it is working for $\Theta_{(1)}$. In other words, we have to study the evolution of the two terms in $\CT_1$.

\noindent $\bullet \quad \e_{2N-2} \, \e_5 \, \left[ A_1^{N-2} \, Q_1^2 M_1 O_1 \right]$: It is easy to follow the evolution of this term. Indeed after 1 iteration it gives
\be 
\text{In \,} \, \CT_2: \e_{2N-4} \, \e_5 \, \left[ \cancel{A_2^{N-2} \, M_2 M_1 O_1} + A_2^{N-3} \, Q_2^2 M_1 O_1 \right] \nn
\ee
The first term is killed by the chiral ring stability argument near \eqref{ChiralRingStab}. To get the second, one power of the antisymmetric $A_1$ has been used with $Q_1^2$ to produce $Q_2^2$. We can repeat this k-times and we get 
\be 
\text{In \,} \, \CT_k: \e_{2N-2k} \, \e_5 \, \left[ A_k^{N-k-1} \, Q_k^2 M_1 O_1 \right] \nn
\ee
 
\noindent $\bullet \quad \e_{2N-2} \, \e_5 \, \left[ A_1^{N-3} \, Q_1^4 O_1 \right]$: this term is trickier to keep track. After 1 iteration it produces
\be 
\text{In \,} \, \CT_2 : \e_{2N-6} \, \e_5 \, \left[ \cancel{A_2^{N-3} \, M_2^2 O_1} + \underset{\star_1}{A_2^{N-4} \, Q_2^2 M_2 O_1} + \underset{\star_2}{A_2^{N-5} \, Q_2^4 O_1} \right] \nn
\ee
Once again the first term vanishes due to chiral ring stability. The term $\star_1$ has been obtained by using one power of $A_1$ in combination with $Q_1^2$ to produce $Q_2^2$ and $Q_1^2$ produces $M_2$. It is of the same kind as the one previously studied. Therefore $\star_1$ produces in $\CT_k$
\be 
\text{In \,} \, \CT_k: \e_{2N-2k-2} \, \e_5 \, \left[ A_k^{N-k-2} \, Q_k^2 M_2 O_1 \right] \nn
\ee 
The term $\star_2$ has been produced by using two powers of $A_1$ with $Q_1^4$ to get $Q_2^4$. It will once again produce 3 terms in $\CT_3$: 1 killed by chiral ring stability, 1 with $Q_3^2$ and one less power of the antisymmetric and 1 with $Q_3^4$ and two less power of the antisymmetric.

\noindent Therefore in $\CT_k$, $\star_2$ produces a sum of terms with $Q_k^2$ and only one term with $Q_k^4$. The first term in the sum is: $\e_{2N-2k-4} \, \e_5 \, \left[ A_k^{N-k-3} \, Q_k^2 M_3 O_1 \right]$ and the last one is: 

\noindent $\e_{2N-4k+2} \, \e_5 \, \left[ A_k^{N-2k} \, Q_k^2 M_k O_1 \right]$ and there are all the terms in between. The term with $Q_k^4$ is given by: $\e_{2N-4k+2} \, \e_5 \, \left[ A_k^{N-2k-1} \, Q_k^4 O_1 \right]$. So the term $\star_2$ produces
\be 
\text{In \,} \, \CT_k: \e_{2N-4k+2} \, \e_5 \, \left[ A_k^{N-2k-1} \, Q_k^4 O_1 \right] + \sum_{i=3}^{k} \e_{2N-2k-2i+2} \, \e_5 \, \left[ A_k^{N-k-i} \, Q_k^2 M_i O_1 \right] \nn
\ee 
The last result is valid until $N-2k-1 \ge 0$, which correspond to the power of the antisymmetric field in the term with $Q_k^4$. Since $k$ should be an integer, it should satisfy: $k \le \lfloor \frac{N-1}{2} \rfloor$.

\noindent To summarize: $\Theta_{(1)}$ in the frame $\CT_k$ with $1 \le k \le \lfloor \frac{N-1}{2} \rfloor$ is given by
\be 
\Theta_{(1)} = \e_{2N-4k+2} \, \e_5 \, \left[ A_k^{N-2k-1} \, Q_k^4 O_1 \right] + \sum_{i=1}^{k} \e_{2N-2k-2i+2} \, \e_5 \, \left[ A_k^{N-k-i} \, Q_k^2 M_i O_1 \right] \nn
\ee 
Applying the same reasoning, it is not complicated to get the expression for the generic case: $\Theta_{(p)}$ in the frame $\CT_k$ with $p \le k \le k_{max}^{(p)} \equiv \lfloor \frac{N+p-2}{2} \rfloor$ is
\be \label{Theta_p}
\Theta_{(p)} = \e_{2N-4k+2p} \, \e_5 \, \left[ A_k^{N-2k-2+p} \, Q_k^4 O_p \right] + \sum_{i=p}^{k} \e_{2N-2k-2i+2p} \, \e_5 \, \left[ A_k^{N-k-i-1+p} \, Q_k^2 M_i O_p \right]
\ee 
The value of $k_{max}^{(p)}$ is obtained by requiring that the power of $A_k$, in the term with $Q_k^4$, is positive. This finishes the first step.
\newline

The second step is to determine the evolution of $\Theta_{(p)}$ from $\CT_{k_{max}^{(p)}}$ to the end $\CT_{N-1}$ and in a general frame in between. It will be useful to distinguish when the combination that enters in $k_{max}^{(p)}$: $N+p-2$ is even or odd.

\noindent$\bullet \quad \e_{2N-4k+2p} \, \e_5 \, \left[ A_k^{N-2k-2+p} \, Q_k^4 O_p \right]$: 
\begin{enumerate}
\item Case: $N+p-2$ even: $k_{max}^{(p)} = \frac{N+p-2}{2}$
\begin{align}
\text{In \,} \, &\CT_{k_{max}^{(p)}}: \e_5 \, \left[ Q_{k_{max}^{(p)}}^4 O_p \right]  \nn \\
\text{In \,} \, &\CT_{k_{max}^{(p)} + 1 + t}: \e_5 \, \left[ M_{k_{max}^{(p)}+1}^2 O_p \right] \quad t \ge 0 \label{even1} 
\end{align}
\item Case: $N+p-2$ odd: $k_{max}^{(p)} = \frac{N+p-3}{2}$
\begin{align}
\text{In \,} \, &\CT_{k_{max}^{(p)}}: \e_2 \, \e_5 \, \left[ A_{k_{max}^{(p)}} Q_{k_{max}^{(p)}}^4 O_p \right] \nn \\
\text{In \,} \, &\CT_{k_{max}^{(p)} + 1}: \e_5 \, \left[ M_{k_{max}^{(p)}+1} Q_{k_{max}^{(p)}+1}^2 O_p \right] \label{odd1}  \\
\text{In \,} \, &\CT_{k_{max}^{(p)} + 2 + t}: \e_5 \, \left[ M_{k_{max}^{(p)}+1} M_{k_{max}^{(p)}+2} O_p \right] \quad t \ge 0 \label{odd2} 
\end{align}
\end{enumerate}

\noindent$\bullet \quad \sum_{i=p}^{k} \e_{2N-2k-2i+2p} \, \e_5 \, \left[ A_k^{N-k-i-1+p} \, Q_k^2 M_i O_p \right]$ : 
\begin{enumerate}
\item Case: $N+p-2$ even: $k_{max}^{(p)} = \frac{N+p-2}{2}$
\begin{align}
\text{In \,} \, &\CT_{k_{max}^{(p)}}: \sum_{i=p}^{k_{max}^{(p)}} \e_{2k_{max}^{(p)}+4-2i} \, \e_5 \, \left[ A_{k_{max}^{(p)}}^{k_{max}^{(p)}+1-i} \, Q_{k_{max}^{(p)}}^2 M_i O_p \right] \nn \\
\text{In \,} \, &\CT_{k_{max}^{(p)}+1}: \sum_{i=p}^{k_{max}^{(p)}} \e_{2k_{max}^{(p)}+2-2i} \, \e_5 \, \left[ A_{k_{max}^{(p)}+1}^{k_{max}^{(p)}-i} \, Q_{k_{max}^{(p)}+1}^2 M_i O_p \right] \nn \\
\text{In \,} \, &\CT_{k_{max}^{(p)}+2}: \e_5 \, \left[ M_{k_{max}^{(p)}} M_{k_{max}^{(p)}+2} O_p \right] + \sum_{i=p}^{k_{max}^{(p)}-1} \e_{2k_{max}^{(p)}-2i} \, \e_5 \, \left[ A_{k_{max}^{(p)}+2}^{k_{max}^{(p)}-1-i} \, Q_{k_{max}^{(p)}+2}^2 M_i O_p \right] \nn \\
&\vdots \nn \\
\text{In \,} \, &\CT_{k_{max}^{(p)} + 1 + t}: \sum_{i=1}^{t} \e_5 \, \left[ M_{k_{max}^{(p)}+1-i} M_{k_{max}^{(p)}+1+i} O_p \right] \hspace{1cm} t=0,\dots,k_{max}^{(p)}-p \nn \\
&\hspace{1.7cm}+ \sum_{i=p}^{k_{max}^{(p)}-t} \e_{2k_{max}^{(p)}+2-2i-2t} \, \e_5 \, \left[ A_{k_{max}^{(p)}+1+t}^{k_{max}^{(p)}-t-i} \, Q_{k_{max}^{(p)}+1+t}^2 M_i O_p \right] \label{even2}
\end{align}
Notice that $2k_{max}^{(p)}+1-p = N-1$. So we have the expression up to the end.
\item Case: $N+p-2$ odd: $k_{max}^{(p)} = \frac{N+p-3}{2}$
\begin{align}
\text{In \,} \, &\CT_{k_{max}^{(p)}}: \sum_{i=p}^{k_{max}^{(p)}} \e_{2k_{max}^{(p)}+6-2i} \, \e_5 \, \left[ A_{k_{max}^{(p)}}^{k_{max}^{(p)}+2-i} \, Q_{k_{max}^{(p)}}^2 M_i O_p \right] \nn \\
\text{In \,} \, &\CT_{k_{max}^{(p)}+1}: \sum_{i=p}^{k_{max}^{(p)}} \e_{2k_{max}^{(p)}+4-2i} \, \e_5 \, \left[ A_{k_{max}^{(p)}+1}^{k_{max}^{(p)}+1-i} \, Q_{k_{max}^{(p)}+1}^2 M_i O_p \right] \label{odd3}
\end{align}
\begin{align}
&\text{In \,} \, \CT_{k_{max}^{(p)}+2}: \sum_{i=p}^{k_{max}^{(p)}} \e_{2k_{max}^{(p)}+2-2i} \, \e_5 \, \left[ A_{k_{max}^{(p)}+2}^{k_{max}^{(p)}-i} \, Q_{k_{max}^{(p)}+2}^2 M_i O_p \right] \nn \\
&\text{In \,} \, \CT_{k_{max}^{(p)}+3}: \e_5 \, \left[ M_{k_{max}^{(p)}} M_{k_{max}^{(p)}+3} O_p \right] + \sum_{i=p}^{k_{max}^{(p)}-1} \e_{2k_{max}^{(p)}-2i} \, \e_5 \, \left[ A_{k_{max}^{(p)}+3}^{k_{max}^{(p)}-1-i} \, Q_{k_{max}^{(p)}+3}^2 M_i O_p \right] \nn \\
&\vdots \nn \\
&\text{In \,} \, \CT_{k_{max}^{(p)} + 2 + t}: \sum_{i=1}^{t} \e_5 \, \left[ M_{k_{max}^{(p)}+1-i} M_{k_{max}^{(p)}+2+i} O_p \right] \hspace{1cm} t=0,\dots,k_{max}^{(p)}-p \nn \\
&\hspace{1.7cm}+ \sum_{i=p}^{k_{max}^{(p)}-t} \e_{2k_{max}^{(p)}+2-2i-2t} \, \e_5 \, \left[ A_{k_{max}^{(p)}+2+t}^{k_{max}^{(p)}-t-i} \, Q_{k_{max}^{(p)}+2+t}^2 M_i O_p \right] \label{odd4} 
\end{align}
Once again $2k_{max}^{(p)}+2-p = N-1$. So we have the expression up to the end.
\end{enumerate}
This finishes the second step. The third and last step is to combine all the previous ingredients to write the superpotential in a generic frame. So let's fix k ($1 \le k \le N-1$) we want the superpotential in $\CT_k$. As we said it contains terms with $O_p$ with $1 \le p \le k$. Also, depending on $p$, there are two kinds of terms depending on whether $k_{max}^{(p)}$ is greater or smaller than $k$. The terms with $M^2 O_p \, \& \, Q_k^2$ as in \eqref{even2}-\eqref{odd4} that we call terms A and the terms with $Q_k^4 \, \& \, Q_k^2$ as in \eqref{Theta_p} that we call terms B. Let's call $p_\star$, the precise p that is at the junction between the terms A and B. It is given by
\be 
\frac{N+p_\star -2}{2} \overset{!}{=} k \quad \Rightarrow \quad p_\star = 2k + 2 -N \label{kmax}
\ee
Therefore obviously $k_{max}^{(p_\star)} = k$. Then for $p \ge p_\star$: $k_{max}^{(p)} = \lfloor \frac{N+p-2}{2} \rfloor \ge k$ and for $p < p_\star$: $k_{max}^{(p)} < k$ which is what we wanted. To recap, for $1 \le p \le p_\star-1$ we have terms A and for $p_\star \le p \le k$ we have terms B. So the superpotential takes the following form
\be 
W_k = H(-p_\star) \underbrace{\sum_{p=1}^{k} \Theta_{(p)}}_B + H(p_\star-\e) \left( \, \underbrace{\sum_{p=1}^{p_\star-1} \Theta_{(p)}}_A + \underbrace{\sum_{p=p_\star}^{k} \Theta_{(p)}}_B \right)
\ee
We have included the Heavyside step function: $H(x) = \begin{cases} 1 & \text{if $x\ge0$} \\
0 & \text{if $x<0$}
\end{cases} $ to treat also the case $p_\star \le 0$. $\e$ is a small positive quantity ($\ll 1$) to avoid double counting in the case $p_\star=0$. Now let's look in turn at terms A and B.
\newline

Terms B: By definition they satisfy $p \le k \le k_{max}^{(p)} \, \forall p \in \, $B$\, \equiv \, \forall p \ge p_\star$. In this case it is easy and $\Theta_{(p)}$ is given by \eqref{Theta_p}. 
\newline

Terms A: By definition we have $k > k_{max}^{(p)} \, \forall p \in \, $A$\, \equiv \, \forall p < p_\star$. It is more complicated in this case. The first thing to notice is that $N+p_\star-2 \, (=2k)$ is necessarily even. Equivalently we can say that $p_\star$ has the same parity has $N$. 

\noindent For $p=p_\star-1: k_{max}^{(p_\star -1)} = \lfloor \frac{N+p_\star-2-1}{2} \rfloor = \lfloor k + \frac{-1}{2} \rfloor = k-1$. So in the frame $\CT_k$ we are 1 frame above $k_{max}^{(p_\star -1)}$. In addition $N+(p_\star-1)-2$ is odd therefore we are in the situation of \eqref{odd1} and \eqref{odd3}. Conclusion $\Theta_{(p_\star-1)}$ is given by
\be 
\Theta_{(p_\star-1)} = \e_5 \, \left[ M_{k} Q_{k}^2 O_{p_\star-1} \right] + \sum_{i=p_\star-1}^{k-1} \e_{2k+2-2i} \, \e_5 \, \left[ A_{k}^{k-i} \, Q_{k}^2 M_i O_{p_\star-1} \right]
\ee   
Then we go on. For $p= p_\star-1-t: k_{max}^{(p_\star -1-t)} = k + \lfloor \frac{-1-t}{2} \rfloor$ and $t=1,\dots,p_\star-2$. In addition $N+ (p_\star-1-t)-2$ has the opposite parity of $t$. therefore to continue and use our \eqref{even2} and \eqref{odd4} we should separate between the odd and even value of $t$. That is, we rewrite the sum A as 
\be 
\sum_{p=1}^{p_\star-1} \Theta_{(p)} = \Theta_{(p_\star-1)} + \sideset{}{'}\sum_{t_{odd}=1}^{t_{odd}^{max}} \underbrace{\Theta_{(p_\star-1-t_{odd})}}_{\text{Even parity}} + \sideset{}{'}\sum_{t_{even}=2}^{t_{even}^{max}} \underbrace{\Theta_{(p_\star-1-t_{even})}}_{\text{Odd parity}}
\ee
Where the \say{$\sideset{}{'}\sum$} means $t_{min}, t_{min}+2, t_{min}+4, \dots$ It's not complicated to get the expression for $t_{odd}^{max}$ and $t_{even}^{max}$. They are given by $t_{odd}^{max} = 2\lceil \frac{p_\star}{2} \rceil -3$ and $t_{even}^{max} = 2\lfloor \frac{p_\star}{2} \rfloor -2$. 

\noindent Now $\Theta_{(p_\star-1-t_{odd})}$ in $\CT_k$ is given by \eqref{even1} and \eqref{even2}. To use these formulas we need to do the following translation: $k \overset{!}{=} k_{max}^{(p_\star -1-t_{odd})} + 1 +t = k - \frac{1+t_{odd}}{2}+1+t \Rightarrow t = \frac{1+t_{odd}}{2}-1$. Also $k_{max}^{p_\star-1-t_{odd}} - t = k+1-(1+t_{odd})=k-t_{odd}$. So 
\begin{align}
\Theta_{(p_\star-1-t_{odd})} &= \sum_{i=0}^{\frac{1+t_{odd}}{2}-1} \e_5 \, \left[ M_{k-\frac{1+t_{odd}}{2}+1-i} M_{k-\frac{1+t_{odd}}{2}+1+i} O_{p_\star-1-t_{odd}} \right] \nn \\
&\hspace{1.7cm}+ \sum_{i=p_\star-1-t_{odd}}^{k-t_{odd}} \e_{2k+2-2i-2t_{odd}} \, \e_5 \, \left[ A_{k}^{k-t_{odd}-i} \, Q_{k}^2 M_i O_{p_\star-1-t_{odd}} \right]
\end{align}
Similarly, $\Theta_{(p_\star-1-t_{even})}$ in $\CT_k$ is found using \eqref{odd2} and \eqref{odd4}. Once again to use the formula we need to impose : $k \overset{!}{=} k_{max}^{(p_\star -1-t_{even})} +2+t = k - \frac{t_{even}}{2}+1+t \Rightarrow t = \frac{t_{even}}{2}-1$. Also $k_{max}^{p_\star-1-t_{even}} - t=k-t_{even}$. So 
\begin{align}
\Theta_{(p_\star-1-t_{even})} &= \sum_{i=0}^{\frac{t_{even}}{2}-1} \e_5 \, \left[ M_{k-\frac{t_{even}}{2}-i} M_{k-\frac{t_{even}}{2}+1+i} O_{p_\star-1-t_{even}} \right] \nn \\
&\hspace{1cm}+ \sum_{i=p_\star-1-t_{even}}^{k-t_{even}} \e_{2k+2-2i-2t_{even}} \, \e_5 \, \left[ A_{k}^{k-t_{even}-i} \, Q_{k}^2 M_i O_{p_\star-1-t_{even}} \right]
\end{align}
Combining all results, the superpotential in $\CT_k$ is
\begin{align}
&W_k= H(-p_\star) \sum_{p=1}^{k} \Bigg( \sum_{i=p}^{k} \e_{2N-2k-2i+2p} \, \e_5 \, \left[ A_k^{N-k-i-1+p} \, Q_k^2 M_i O_p \right]  \nn \\
& + \e_{2N-4k+2p} \, \e_5 \, \left[ A_k^{N-2k-2+p} \, Q_k^4 O_p \right] \Bigg) + H(p_\star-\e) \sum_{p=p_\star}^{k} \Bigg( \e_{2N-4k+2p} \, \e_5 \, \left[ A_k^{N-2k-2+p} \, Q_k^4 O_p \right] \nn \\
& +\sum_{i=p}^{k} \e_{2N-2k-2i+2p} \, \e_5 \, \left[ A_k^{N-k-i-1+p} \, Q_k^2 M_i O_p \right] \Bigg) \nn \\
&+H(p_\star-1-\e)\left(\e_5 \, \left[ M_{k} Q_{k}^2 O_{p_\star-1} \right] + \sum_{i=p_\star-1}^{k-1} \e_{2k+2-2i} \, \e_5 \, \left[ A_{k}^{k-i} \, Q_{k}^2 M_i O_{p_\star-1} \right] \right) \nn \\ 
&+H(p_\star-2-\e)\sideset{}{'}\sum_{t_{odd}=1}^{2\lceil \frac{p_\star}{2} \rceil -3} \left(\sum_{i=0}^{\frac{1+t_{odd}}{2}-1} \e_5 \, \left[ M_{k-\frac{1+t_{odd}}{2}+1-i} M_{k-\frac{1+t_{odd}}{2}+1+i} O_{p_\star-1-t_{odd}} \right]\right. \nn \\
&\hspace{2.3cm} \left.+ \sum_{i=p_\star-1-t_{odd}}^{k-t_{odd}} \e_{2k+2-2i-2t_{odd}} \, \e_5 \, \left[ A_{k}^{k-t_{odd}-i} \, Q_{k}^2 M_i O_{p_\star-1-t_{odd}} \right]\right) \nn \\
&+H(p_\star-3-\e)\sideset{}{'}\sum_{t_{even}=2}^{2\lfloor \frac{p_\star}{2} \rfloor -2} \left(\sum_{i=0}^{\frac{t_{even}}{2}-1} \e_5 \, \left[ M_{k-\frac{t_{even}}{2}-i} M_{k-\frac{t_{even}}{2}+1+i} O_{p_\star-1-t_{even}} \right] \right. \nn \\
&\hspace{2.3cm} \left.+ \sum_{i=p_\star-1-t_{even}}^{k-t_{even}} \e_{2k+2-2i-2t_{even}} \, \e_5 \, \left[ A_{k}^{k-t_{even}-i} \, Q_{k}^2 M_i O_{p_\star-1-t_{even}} \right] \right) 
\end{align}
By renaming indices it is possible to write $W_k$ as

\begin{align}
&W_k= H(-p_\star) \sum_{p=1}^{k} \Bigg( \sum_{i=p}^{k} \e_{2N-2k-2i+2p} \, \e_5 \, \left[ A_k^{N-k-i-1+p} \, Q_k^2 M_i O_p \right]  \nn \\
& + \e_{2N-4k+2p} \, \e_5 \, \left[ A_k^{N-2k-2+p} \, Q_k^4 O_p \right] \Bigg) + H(p_\star-\e) \sum_{p=p_\star}^{k} \Bigg( \e_{2N-4k+2p} \, \e_5 \, \left[ A_k^{N-2k-2+p} \, Q_k^4 O_p \right] \nn \\
& +\sum_{i=p}^{k} \e_{2N-2k-2i+2p} \, \e_5 \, \left[ A_k^{N-k-i-1+p} \, Q_k^2 M_i O_p \right] \Bigg) \nn \\
&+H(p_\star-1-\e)\left(\e_5 \, \left[ M_{k} Q_{k}^2 O_{p_\star-1} \right] + \sum_{i=p_\star-1}^{k-1} \e_{2k+2-2i} \, \e_5 \, \left[ A_{k}^{k-i} \, Q_{k}^2 M_i O_{p_\star-1} \right] \right) \nn \\ 
&+H(p_\star-2-\e) \sideset{}{'}\sum_{l=N-2\lceil \frac{N}{2} \rceil +2}^{2k-N} \left(\sum_{j=\frac{N+l}{2}}^{N+l-k} \e_5 \, \underset{\textcolor{blue}{\star_1}}{\left[ M_{j} M_{N+l-j} O_{l} \right]}\right. \nn \\
&\hspace{4cm} \left.+ \sum_{i=l}^{N+l-k-1} \e_{2N+2l-2i-2k} \, \e_5 \, \underset{\textcolor{red}{\star_1}}{\left[ A_{k}^{N+l-i-k-1} \, Q_{k}^2 M_i O_{l} \right]} \right) \nn \\
&+H(p_\star-3-\e) \sideset{}{'}\sum_{m=N-2\lfloor \frac{N}{2} \rfloor +1}^{2k-N-1} \left(\sum_{j=\frac{N+m-1}{2}}^{N+m-k-1} \e_5 \, \underset{\textcolor{blue}{\star_2}}{\left[ M_{j} M_{N+m-j} O_{m} \right]} \right. \nn \\
&\hspace{4cm} \left.+ \sum_{i=m}^{N+m-k-1} \e_{2N+2m-2i-2k} \, \e_5 \, \underset{\textcolor{red}{\star_2}}{\left[ A_{k}^{N+m-k-i-1} \, Q_{k}^2 M_i O_{m} \right]} \right) 
\end{align}
Where we recall that \say{$\sideset{}{'}\sum$} means $l_{min}, l_{min}+2, l_{min}+4, \dots$ and similar for $m$.\\
We can do even better, $\textcolor{red}{\star_1} + \textcolor{red}{\star_2}$ can combine together to give
\be 
\sum_{p=1}^{2k-N} \, \sum_{i=p}^{N+p-k-1} \e_{2N+2p-2i-2k} \, \e_5 \, \left[ A_{k}^{N+p-k-i-1} \, Q_{k}^2 M_i O_{p} \right] \nn
\ee 
Less obviously, $\textcolor{blue}{\star_1} + \textcolor{blue}{\star_2}$ can be packaged together. To do so we notice 
\begin{itemize}
\item $\sideset{}{'}\sum_{l=N-2\lceil \frac{N}{2} \rceil +2}^{2k-N} + \sideset{}{'}\sum_{m=N-2\lfloor \frac{N}{2} \rfloor +1}^{2k-N-1} = \sum_{p=1}^{2k-N} $
\item $M_i,M_j$ and $O_p$ that enter in the sum satisfy $i+j-p = N$
\item In addition, the above $i$ and $j$ satisfy $N+p-k \le i \le j \le k$
\item All terms satisfying the above 3 criteria are present in the sums
\end{itemize}
Therefore $\textcolor{blue}{\star_1} + \textcolor{blue}{\star_2}$ can be written
\be 
\sum_{p=1}^{2k-N} \, \sum_{N+p-k \le i \le j \le k} \, \e_5 \, \left[ M_i M_j O_{p} \right] \de_{i+j-p,N} \nn 
\ee 
So the last version of the superpotential is
\begin{align}
&W_k= H(-p_\star) \sum_{p=1}^{k} \Bigg( \sum_{i=p}^{k} \e_{2N-2k-2i+2p} \, \e_5 \, \left[ A_k^{N-k-i-1+p} \, Q_k^2 M_i O_p \right]  \nn \\
& + \e_{2N-4k+2p} \, \e_5 \, \left[ A_k^{N-2k-2+p} \, Q_k^4 O_p \right] \Bigg) + H(p_\star-\e) \sum_{p=p_\star}^{k} \Bigg( \e_{2N-4k+2p} \, \e_5 \, \left[ A_k^{N-2k-2+p} \, Q_k^4 O_p \right] \nn \\
& +\sum_{i=p}^{k} \e_{2N-2k-2i+2p} \, \e_5 \, \left[ A_k^{N-k-i-1+p} \, Q_k^2 M_i O_p \right] \Bigg) \nn \\
&+H(p_\star-1-\e)\left(\e_5 \, \left[ M_{k} Q_{k}^2 O_{p_\star-1} \right] + \sum_{i=p_\star-1}^{k-1} \e_{2k+2-2i} \, \e_5 \, \left[ A_{k}^{k-i} \, Q_{k}^2 M_i O_{p_\star-1} \right] \right) \nn \\ 
&+H(p_\star-2-\e) \sum_{p=1}^{2k-N} \left( \sum_{i=p}^{N+p-k-1} \e_{2N+2p-2i-2k} \, \e_5 \, \left[ A_{k}^{N+p-k-i-1} \, Q_{k}^2 M_i O_{p} \right] \right. \nn \\
&\hspace{4cm} \left.+ \sum_{N+p-k \le i \le j \le k} \, \e_5 \, \left[ M_i M_j O_{p} \right] \de_{i+j-p,N} \right)
\end{align}
with $p_\star = 2k+2-N$ and $H(x) = \begin{cases} 1 & \text{if $x\ge0$} \\
0 & \text{if $x<0$}\end{cases}$ \\
With $W_k$ we can write the quiver of the theory in the frame $\CT_k$
\be \label{UspTk} \bpic[node distance=2cm,gSUnode/.style={circle,red,draw,minimum size=8mm},gUSpnode/.style={circle,blue,draw,minimum size=8mm},fnode/.style={rectangle,draw,minimum size=8mm}]   
\node at (-1,2) {$\CT_k:$};
\node[gUSpnode] (G1) at (0,0) {$2N-2k$};
\node[fnode,minimum size=1cm] (F1) at (3,0) {$5$};
\node[fnode,red] (F2) at (0,-2.5) {$1$};
\node[fnode] (F3) at (2.1,-2.5) {$1$};
\node at (3.1,-2.5) {$\dots$};
\node[fnode] (F4) at (3.9,-2.5) {$1$};
\draw (G1) -- pic[pos=0.7,sloped]{arrow} (F1) node[midway,above] {$Q_{k}$};
\draw (G1) -- (F2) node[midway,left] {$F_{k}$};
\draw (F1) -- pic[pos=0.3,sloped]{arrow} (F3) node[midway,left] {$O_1$};
\draw (F1) -- pic[pos=0.4,sloped,very thick]{arrow=latex reversed} (F4) node[midway,right] {$O_{k}$};
\draw (0.5,0.8) to[out=90,in=0]  (0,1.4) to[out=180,in=90] (-0.5,0.8);
\draw (3.5,0.2) to[out=0,in=90] pic[pos=0.4,sloped,very thick]{arrow=latex reversed} (3.9,0) to[out=-90,in=0] pic[pos=0.6,sloped,very thick]{arrow=latex reversed} (3.5,-0.2);
\node at (4.3,0) {$\dots$};
\draw (3.5,-0.4) to[out=0,in=-90] pic[pos=0.4,sloped,very thick]{arrow=latex reversed} (4.7,0) to[out=90,in=0] pic[pos=0.6,sloped,very thick]{arrow=latex reversed} (3.5,0.4);
\node[right] at (5.7,-1) {$\CW = \sum_{i=2}^{N-k} \, \a_i \, \tr(A_k^i) + W_k$};
\node at (0.8,1.3) {$A_k$};
\node at (4.6,0.9) {$M_1, \dots, M_{k}$};
\epic \ee
To finish, when $k=N-1$ then $p_\star=N$ and it's easy to see that \eqref{UspTk} becomes \eqref{UspTN-1}.

\section{Unitary presentation of the S-confining dualities}\label{unitary}
The S-confining dualities discussed in this paper are not written in a unitary fashion. This is not by itself a problem, since the gauge theories can be part of larger gauge theory, where no unitary violation is present. Nevertheless, for completeness, we give a presentation of the confining dualities flipping all the operators that violate unitarity. In the Wess-Zumino there can only be cubic superpotential terms (higher order terms automatically imply that at least one field has $R<\frac{2}{3}$). So all surviving operators have R charge precisely $\frac{2}{3}$ and the WZ superpotential is cubic.

 Obviously, flipping some chiral ring generator $\mathcal{O}$, removes the operator $\mathcal{O}$ from the chiral ring. If $R[\mathcal{O}]>2/3$, the flipper field appears in the chiral ring of the new theory, but notice that if $R[\mathcal{O}]<2/3$ the flipper field is not a chiral ring generator of the new theory. This is due to quantum effects: if the flipper takes a vev, the theory develops a quantum generated superpotential that lifts the supersymmetric vacua at the origin \cite{Benvenuti:2017lle}.
 
The unitary dualities look somewhat trivial sometimes, having very few fields left in the chiral ring. Let us start from the very simple examples of $SU(N)$ and $Usp(2N)$ with only fundamentals. The case of $SU(2)=Usp(2)$ has cubic dual Wess-Zumino and it is unitary.

\paragraph{$Usp(2N)$ with $2N+4$ fundamentals, $N>2$}
The mesons violate the unitarity bound, so the duality becomes unitary upon flipping  the mesons:
\be \ba{ccc}
\ba{c} Usp(2N) \text{ with $2N+4$ fundamentals} \, Q \\
 \CW= \sigma  tr(Q Q) \ea 
    &\quad \Longleftrightarrow \quad & \quad
\ba{c} \text{Trivial WZ with $0$ chirals} \,.\ea 
   \ea \ee

\paragraph{$SU(N)$ with $(N+1, N+1)$ flavors, $N>2$}
Also in this case we can make the duality unitary flipping the mesons:
\be \ba{ccc}
\ba{c} SU(N) \text{ with}\\
\text{ $(N+1, N+1)$ flavors } \, (Q, \Qt) \\
 \CW= \sigma tr(Q \Qt) \ea 
    &\quad \Longleftrightarrow \quad & \quad
\ba{c} \text{Free WZ with $2(N+1)$ chirals} \,\,\\
 \{B, \Bt\}  \lra \{Q^N, \Qt^{N}\}\\
 \CW=0 \;.\ea 
   \ea \ee
Let us emphasize a peculiar aspect of this duality. On the l.h.s. the UV non anomalous R-symmetry is $R[Q]=R[\Qt]=1-\frac{N}{N+1}$, from which $R[Q\Qt] = \frac{2}{N+1}$ and $R[Q^N]=R[\Qt^N]=\frac{N}{N+1}$. So there is a mismatch: on the l.h.s. the baryons have $R=\frac{N}{N+1}$, on the r.h.s. their image have $R=\frac{2}{3}$. The resolution of this apparent paradox is that on the l.h.s there is a additional $U(1)$ global symmetry emerging in the IR and mixing with the superconformal R-symmetry. A similar phenomenon will appear in the following for the S-confining theory $SU(6)$ with 2 antisymmetrics and $(1,5)$ flavors.

\paragraph{$Usp(2N)$ with antisymmetric  and 6 fundamentals}
In this case, since all the powers of the antisymmetric are flipped, the duality \eqref{UspT0} $\lra$ \eqref{UspTDec} is already unitary, with a cubic superpotential on the confined side:
\be \ba{ccc}
\ba{c} Usp(2N) \text{ with antisymmetric $A$}\\
 \text{ and 6 fundamentals } \,\,Q \\
 \CW=  \sum_{i=2}^{N} \, \a_i \, \tr(A^i) \ea 
    &\quad \Longleftrightarrow \quad & \quad
\ba{c} \text{WZ with} \,\,15 N\\
 \text{chiral fields }  \mu_i \lra tr(Q A^{i-1} Q) \\
 \CW= \sum_{N\ge i\ge j\ge k\ge 1} \, \e_{6} \, \left( \mu^i \, \mu^j \,  \mu^k \right) \, \de_{i+j+k,2N+1} \;.\ea 
   \ea \ee
It is easy to check that a-maximization in one variable leads to $R[A]=0, R[Q]=\frac{1}{3}$, hence $R[Q A^{i} Q]=\frac{2}{3}$.

\paragraph{$SU(2N+1)$ with antisymmetric, $\overline{\text{antisymmetric}}$  and $(3,3)$ flavors}
Duality \eqref{SU2N+1antisym'T1Flip} $\lra$ \eqref{SU2N+1antisym'T4Flip} becomes unitary upon flipping $A^NQ, \At^N\Qt$:
\be \ba{ccc}
\ba{c} SU(2N+1) \text{ with  $A, \At$}\\
 \text{ and $(3,3)$ flavors } \,\,(Q,\Qt) \\
 \CW=  \sum_{i=1}^{N} \a_i  \tr(A  \At)^i+\\
  \b  A^{N-1}Q^3 + \tilde \b  \At^{N-1}\Qt^3 +\\
   \gamma A^NQ+ \tilde \gamma \At^N\Qt \ea 
    &\quad \Longleftrightarrow \quad & \quad
\ba{c} \text{WZ with $15N$ chirals} \,\,\\
\{ M_i, \Ht_i, H_{i+1} \} \lra \{ Q (A\At)^{i} \Qt, A \Qt^2 (A\At)^{i}, \At Q^2 (A\At)^{i} \}\\
 i = 0, \dots, N-1 \\
  \CW= \sum_{i,j,k=1}^{N-1} \, \left(\Ht_i \, M_j \, H_k + M_i \, M_j \, M_k \right) \, \de_{i+j+k,2N-1}\\
 \hspace{-0.4cm} + \Ht_{N-1} M_0 H_N + \Ht_0 M_{N-1} H_N + \sum_{l=0}^{N-3} \, \left( \Ht_{l+1} M_{N-2-l} H_N \right)  \;.\ea 
   \ea \ee
A-maximization in one variable leads to $R[A,\At]=0, R[Q]=R[\Qt]=\frac{1}{3}$, hence\\
 $R[Q (A\At)^j\Qt, A (A\At)^j \Qt^2, \At (A\At)^j Q^2]=\frac{2}{3}$.

\paragraph{$SU(2N)$ with antisymmetric, $\overline{\text{antisymmetric}}$  and $(3,3)$ flavors}
Duality \eqref{SU2Nantisym'T1Flip} $\lra$ \eqref{SU2Nantisym'T4Flip} is already unitary. On the gauge theory side we flip $tr((A \At)^j)$ and $A^N, \At^N$:
\be \ba{ccc}
\ba{c} SU(2N) \text{ with  $A, \At$}\\
 \text{ and $(3,3)$ flavors } \,\,(Q,\Qt) \\
 \CW=  \sum_{i=1}^{N-1} \, \a_i \, \tr(A \, \At)^i+\\
 + \b \e_{2N} \, A^{N}  + \tilde \b  \e^{2N} \, \At^{N}  \ea 
    &\quad \Longleftrightarrow \quad & \quad
\ba{c} \text{WZ with $15N$ chirals} \,\, \\
M_{k+1} \lra Q (A \At)^k \Qt, \, k=0, \dots, N-1 \\
\scalebox{0.9}{ \{$ \Ht_{m+2}, H_{m+1} \} \lra \{ A \Qt^2 (A\At)^{m}, \At Q^2 (A\At)^{m} \}, \, m=0, \dots, N-2$} \\
\{\Bt_2, B_2 \} \lra \{\At^{N-1} \Qt^2, A^{N-1} Q^2 \} \\
 \CW=  \Bt_2 B_2 M_N +\\ 
  \left.\sum_{i,j,k=1}^{N} \, \e_3 \, \e^3 \, \left(\Ht_i \, M_j \, H_k + M_i \, M_j \, M_k \right) \de_{i+j+k,2N+1}\right|_{H_N = \, \Ht_1 \, =0} \;.\ea 
   \ea \ee
A-maximization in one variable leads to $R[A,\At]=0, R[Q]=R[\Qt]=\frac{1}{3}$, hence
 $R[Q (A\At)^j\Qt,$ $A (A\At)^j \Qt^2, \At (A\At)^j Q^2]=\frac{2}{3}$.

\paragraph{$SU(2N+1)$ with antisymmetric  and $(4,2N+1)$ flavors}
We can make the duality \eqref{SU2N+11antisymT1} $\lra$ \eqref{SU2N+11antisymT3} unitary flipping  $Q \Qt$ and $A \Qt \Qt$:
\be \ba{ccc}
\ba{c} SU(2N+1) \text{ with antisymmetric $A$}\\
 \text{ and $(4,2N+1)$ flavors } \, (Q, \Qt) \\
 \CW= \sigma tr(A \Qt^2) + \eta tr(Q \Qt) \ea 
    &\quad \Longleftrightarrow \quad & \quad
\ba{c} \text{WZ with $9$ chirals} \,\,\\
 \{\Bt, B_1, B_3\}  \lra \{\Qt^{2N+1}, A^N Q, A^{N-1}Q^3\}\\
 \CW= \Bt B_1 B_3 \;.\ea 
   \ea \ee
 A-maximization in two variables leads to $R[A]=\frac{4}{3+6N}, R[Q]=R[\Qt]=\frac{2}{3+6N}$, hence $R[A^N Q, \Qt^{2N+1}, A^{N-1}Q^3]=\frac{2}{3}$.

\paragraph{$SU(2N)$ with antisymmetric  and $(4,2N)$ flavors}
We can make the duality \eqref{SU2N1antisymT1} $\lra$ \eqref{SU2N1antisymT4} unitary if we flip $A \Qt \Qt$ and $Q\Qt$:
\be \ba{ccc}
\ba{c} SU(2N) \text{ with antisymmetric $A$}\\
 \text{ and $(4,2N)$ flavors } \,\,(Q,\Qt) \\
 \CW=  \sigma tr(A \Qt^2) + \eta tr(Q\Qt) \ea 
    &\quad \Longleftrightarrow \quad & \quad
\ba{c} \text{WZ with $9$ chirals} \,\,\\
 \{\Bt, B_0, B_2, B_4 \}  \lra \{\Qt^{2N}, A^N, A^{N-1}Q^2, A^{N-2}Q^4\}\\
 \CW= \Bt B_0 B_4 + \Bt B_2^2 \;.\ea 
   \ea \ee
A-maximization in two variables leads to $R[A]=\frac{2}{3N}, R[Q]=R[\Qt]=\frac{1}{3N}$, hence $R[\Qt^{2N}, A^N, A^{N-1}Q^2, A^{N-2}Q^4]=\frac{2}{3}$.

\paragraph{$SU(7)$ with 2 antisymmetrics  and $(0,6)$ flavors} Performing a-maximization in the duality \eqref{SU7T1} $\lra$ \eqref{SU7T4}, one finds that both $A\Qt^2 \lra \Ht$ and $A^4\Qt \lra N$ have $R < \frac{2}{3}$. We can obtain a unitary theory flipping either $A\Qt^2$ or $A^4\Qt$. In both cases the dual WZ model becomes free:
\be \ba{ccc}
\ba{c} SU(7) \text{ with 2 antisymmetric $A$}\\
 \text{ and 6 antifundamentals } \,\,\Qt \\
 \CW=  \eta A\Qt^2 \ea 
    &\quad \Longleftrightarrow \quad & \quad
\ba{c} \text{Free WZ with $18$ chirals} \,\, \\
 A^4\Qt \lra N \\
 \CW=0 \;,\ea 
   \ea \ee
(A-maximization in one variable leads to $R[A]=\frac{1}{7}, R[\Qt]=\frac{2}{21}$, hence 
 $R[A^4\Qt]=\frac{2}{3}$), or
\be \ba{ccc}
\ba{c} SU(7) \text{ with 2 antisymmetric $A$}\\
 \text{ and 6 antifundamental } \,\,\Qt \\
 \CW=  \sigma A^4\Qt \ea 
    &\quad \Longleftrightarrow \quad & \quad
\ba{c} \text{Free WZ with $30$ chirals }\,\, \\
 A\Qt\Qt \lra \Ht \\
 \CW=0 \;.\ea 
   \ea \ee
(A-maximization in one variable leads to $R[A]=0, R[\Qt]=\frac{1}{3}$, hence 
 $R[A\Qt^2]=\frac{2}{3}$).

\paragraph{$SU(6)$ with 2 antisymmetrics  and $(1,5)$ flavors} Performing a-maximization in the duality \eqref{SU6T1} $\lra$ \eqref{SU6T3}, one finds that $A\Qt^2$, $Q\Qt$ and $A^3$ have $R < \frac{2}{3}$. We obtain a unitary theory flipping  $A^3$:\be \ba{ccc}
\ba{c} SU(6) \text{ with 2 antisymmetric $A$}\\
 \text{ and (1,5) flavors } \,\,Q,\Qt \\
 \CW=  \eta A^3 \ea 
    &\quad \Longleftrightarrow \quad & \quad
\ba{c} \text{WZ with $45$ chiral fields }\\
  \{A\Qt^2, Q\Qt, A^3 Q\Qt, A^4\Qt^2\} \lra \{\Ht, M, D, \Phi\} \\
 \CW=\Phi \Ht D + M \Phi^2 \;.\ea 
   \ea \ee
(A-maximization in two variables leads to $R[A]=0, R[Q,\Qt]=\frac{1}{3}$, hence 
$R[A\Qt^2, Q\Qt,$ $ A^3Q\Qt, A^4\Qt^2]=\frac{2}{3}$) or flipping $A\Qt^2$:
\be\label{SU6unitary} \ba{ccc}
\ba{c} SU(6) \text{ with 2 antisymmetric $A$}\\
 \text{ and (1,5) flavors } \,\,Q,\Qt \\
 \CW=  \eta A \Qt^2 \ea 
    &\quad \Longleftrightarrow \quad & \quad
\ba{c} \text{WZ with $29$ chiral fields }\\
  \{Q\Qt, A^3, A^3 Q\Qt, A^4\Qt^2\} \lra \{M, S, D, \Phi\} \\
 \CW= M \Phi^2 \;.\ea 
   \ea \ee
In this case A-maximization in two variables leads to $R[A]=\frac{2}{9}, R[Q]=\frac{7}{9}, R[\Qt]=-\frac{1}{9}$, hence  $R[Q\Qt, A^3, A^4\Qt^2]=\frac{2}{3}$ and $R[A^3Q\Qt]=\frac{4}{3}$. We interpret this result as follows: there is an emergent $U(1)$ global symmetry in the gauge theory side, which mixes with the $U(1)_R$ symmetry and shifts the actual R-charges of $A^3Q\Qt$ to $\frac{2}{3}$. Notice indeed that on the dual WZ side, in \eqref{SU6unitary} there is an additional $U(1)$ global symmetry, w.r.t. to the gauge theory side.

\paragraph{$SU(5)$ with 2 antisymmetrics  and $(2,4)$ flavors} Performing a-maximization in the duality \eqref{SU52antisymT1} $\lra$ \eqref{SU52antisymT4}, one finds that $A\Qt^2$, $Q\Qt$, $A^2Q$ and $A^3 \Qt$ have $R < \frac{2}{3}$. We obtain a unitary theory flipping  $Q\Qt$ and $A^2 Q$:\be \ba{ccc}
\ba{c} SU(5) \text{ with 2 antisymmetric $A$}\\
 \text{ and (2,4) flavors } \,\,Q,\Qt \\
 \CW=  \eta Q\Qt +\delta A^2Q \ea 
    &\quad \Longleftrightarrow \quad & \quad
\ba{c} \text{WZ with $24$ chiral fields }\\
  \{A\Qt^2, A^3 \Qt, A^2 Q^2 \Qt\} \lra \{\Ht, N, f\} \\
 \CW= \Ht N f  \;.\ea 
   \ea \ee
(A-maximization in two variables leads to $R[A]=\frac{2}{15}, R[Q]=\frac{1}{15}, R[\Qt]=\frac{4}{15}$, hence 
 $R[A\Qt^2, A^3 \Qt, A^2 Q^2 \Qt]=\frac{2}{3}$).

\newpage

\paragraph{$SU(5)$ with 3 antisymmetrics  and $(0,3)$ flavors} Performing a-maximization in the duality \eqref{SU53antisymT1} $\lra$ \eqref{SU53antisymT5}, one finds that $R[A]=\frac{2}{15}, R[\Qt]=\frac{4}{15}$, hence 
 $R[A\Qt^2, A^3 \Qt, A^5]=\frac{2}{3}$, and the duality is already unitary:
\be \ba{ccc}
\ba{c} SU(5) \text{ with 3 antisymmetric $A$}\\
 \text{ and (0,3) flavors } \,\,\Qt \\
 \CW=  0 \ea 
    &\quad \Longleftrightarrow \quad & \quad
\ba{c} \text{WZ with $39$ chiral fields }\\
  \{A\Qt^2, A^3 \Qt, A^5\} \lra \{\Ht, O, S\} \\
 \CW=  O S \Ht  + O^3  \;.\ea 
   \ea \ee

\bibliographystyle{ytphys}
\bibliography{refs}
\end{document}